\documentstyle[epsfig,times]{mn}
\def\spose#1{\hbox to 0pt{#1\hss}}

\def\lta{\mathrel{\spose{\lower 3pt\hbox{$\mathchar"218$}}\raise 2.0pt\hbox{$\mathchar"13C$}}}
\def\gta{\mathrel{\spose{\lower 3pt\hbox{$\mathchar"218$}}\raise 2.0pt\hbox{$\mathchar"13E$}}}
\def\arcsec{$^{\prime\prime}$}

\title[Excess emission in $z \sim 1$ radio galaxies]
{Deviations from passive evolution - Star formation and the UV excess
  in $z \sim 1$ radio galaxies}
      
\author[K.\,J.\, Inskip, P. N. Best \& M.S. Longair]
{K.\,J.\, Inskip$^{1,2}$\footnotemark, P.\,N.\,Best$^3$ and
M.\,S. Longair$^2$ \\
$^1$ Department of Physics \& Astronomy, University of Sheffield,
Sheffield S3 7RH,\\$^2$ Cavendish Laboratory, Madingley Road, Cambridge, CB3 0HE,\\ 
$^3$
Institute for Astronomy, Royal Observatory Edinburgh, Blackford Hill,
Edinburgh, EH9 3HJ}

\date{}

\pagerange{\pageref{firstpage}--\pageref{lastpage}}

\pubyear{2006}

\begin{document}

\label{firstpage}

\maketitle

\begin{abstract}

Galaxy colours are determined for two samples of 6C and 3CR radio
sources at $z \sim 1$, differing by a factor of $\sim 6$ in radio
power.  Corrections are made for emission line contamination and the
presence of any nuclear point source, and the data analysed as a
function of both redshift and the radio source properties.  The
galaxy colours are remarkably similar for the two populations, and the 
UV excess evolves with radio source size similarly in both samples,
depsite the fact that the alignment effect is more extensive for the
more powerful 3CR radio galaxies. 
These results seem to suggest that the alignment effect at these
redshifts does not scale strongly with radio power, and is instead
more closely dependent on galaxy mass (which is statistically comparable for the two
samples).   However, it is likely that the presence of relatively
young ($\lta$ several $10^8$ years old) stellar
populations has considerably contaminated the $K-$band flux of these
systems, particularly in the case of the more powerful 3CR sources, which are $\sim
0.5$mag more luminous than the predictions of passive evolution models
at $z \sim 1$.  The higher luminosity of the 3CR alignment effect is
balanced by emission at longer wavelengths, thereby leading to
comparable colours for the two samples.

\end{abstract}


\section{Introduction}
\footnotetext{E-mail: k.inskip@shef.ac.uk}
The properties of powerful radio sources vary considerably with both
radio power and redshift.  Up to redshifts of $z \sim 1$, radio
sources are usually hosted by massive elliptical galaxies, with scale
sizes of $\sim 10-15$kpc (e.g. Inskip et al 2005; McLure et al 2004).
Whilst the average galaxy size does not appear to vary greatly, the
host galaxies of the higher redshift radio sources within this
redshift range ($z \gta 0.6$) are intrinsically more luminous than
those at lower redshifts (Inskip et al 2002a, 2005), to an extent
which implys that factors other than simple passive evolution are
involved.  At larger redshifts ($z > 2$), the host galaxies of
the most distant radio sources are clearly still in the process of
formation (e.g. van Breugel et al 1998, Pentericci et al 2001).

Redshift evolution is also observed in the emission surrounding the
host galaxies.   At $z > 0.3$, the host galaxies are
often seen to be surrounded by considerable excess
rest-frame UV emission; at higher redshifts ($z \gta 0.6$) this
emission is usually more extensive, is generally observed to be
closely aligned  
with the radio source axis (Chambers, Miley \& van Breugel 1987;
McCarthy et al. 1987; Allen et al 2002), and is known as the {\it
  Alignment Effect}.  Both the alignment 
effect and the properties of the extended emission line regions
surrounding these sources are seen to be more extreme (in terms of
luminosity, alignment with the radio axis, physical extent and gas kinematics) both for more
powerful radio sources, and also for the smaller radio sources
(Best et al 1996, 2000b; Inskip et al 2002c, 2003, 2005).
Several different mechanisms are thought to be responsible for
producing these regions of extensive aligned emission.  These include:
extended line emission and nebular continuum radiation (Dickson et al
1995), scattering of the UV continuum from the AGN (e.g. Tadhunter et
al 1992; Cimatti et al 1993) and young stars (e.g. Chambers \&
McCarthy 1990) potentially produced in a radio jet
induced starburst (McCarthy et al 1987).  Merger-induced starbursts
may also be responsible for the presence of a relatively young stellar
population in/around the host galaxy (Tadhunter et al 2005 and
references therein).

McCarthy, Spinrad \& van Breugel (1995) found that extended line
emission can be observed around the majority of 3CR galaxies at $z
\gta 0.3$. The relative contribution of emission lines to the total 
aligned emission varies from source to source, although both the
alignment effect and emission line flux are more extreme for smaller
radio sources.  Typically, line emission provides from 2\% up to 30\%
of the total rest-frame UV aligned emission for 3CR galaxies at $z
\sim 1$ (although this does depend on the distribution of emission lines
within the wavelength range of the observed filter), with similar
proportions found for the less powerful 6C radio sources at the same
redshifts (Best 1996; Inskip et al 2003).
In addition to line emission, nebular continuum 
radiation is also produced due to other radiative processes associated
with the ionized gas.
Although a significant process for some sources, the total flux
provided by line emission and nebular continuum emission alone cannot
account for all of the excess emission forming the alignment effect
(e.g. Inskip et al 2003, Tadhunter et al 2002). 

Although the active nucleus of a radio galaxy may be obscured from view, the
powerful UV continuum emitted by the AGN may be scattered towards an
observer by dust or 
electrons in the extended structures surrounding the galaxy.
Emission polarised perpendicularly to the direction of emission from
the AGN due to scattering has been observed from these extended regions, and is consistent with the
orientation--based unification scheme for radio galaxies and quasars
(e.g. Cimatti it et al 1993, Tran et al 1998). 
However, while polarized emission is frequently observed, the emission from
the aligned structures surrounding many radio galaxies lacks the high
levels of polarisation expected if scattering of the UV emission from
the obscured quasar nucleus were the only mechanism occurring. A
recent study of $0.15 < z < 0.7$ radio galaxies by Tadhunter et
al (2002) found that scattering contributed a significant proportion
of the UV excess in many cases, but was very rarely the dominant
factor.  Further
to this, some 3CR sources do not exhibit any polarization of 
their extended structures (e.g. Wills et al 2002). 

Finally, one of the first explanations proposed for the excess UV
continuum was that it was due to emission from young stars, whose
formation was triggered by the passage of the expanding radio source
(e.g. McCarthy et al 1987, Chambers, Miley \& van Breugel 1987).
The emission from such a population of hot, young stars would
dominate the UV emission of the galaxy, and produce a fairly flat
spectral shape, whilst at near infrared wavelengths, the emission would still be
predominantly due to the old stellar population of the host galaxy.
Any young stellar population would quickly (within $\lta 10^7$years)
evolve, accounting for the rapid evolution with radio size seen in the UV aligned
structures at $z \sim 1$ (Best, Longair \& R\"{o}ttgering 1996).
In order to account for the
excess emission, the mass of stars formed in the interactions with the
radio source is typically required to be only a few $10^8 M_\odot$
(e.g. Best, Longair \& R\"{o}ttgering 1997a). 
  It is questionable how easily jet-induced star
formation can occur.  Numerical simulations in the literature
often disagree on whether clouds will be compressed or
shredded/dissipated (e.g. Rees 1989; Begelman \& Cioffi 1989; Klein,
McKee \& Colella 1994; Icke 1999;
Poludnenko, Frank \& Blackman 2002), although recent work including the effects of cooling
(e.g. Mellema, Kurk \& R\"{o}ttgering 2002) suggests that such triggered star formation
is indeed plausible. Observational evidence for star formation triggered by the radio source jets is
seen in isolated objects: 3C 34 (Best, Longair \& R\"{o}ttgering
1997a), Minkowski's object (van Breugel et al 1985), 3C285 (van
Breugel \& Dey 1993).

Evidence that shocks strongly influence the ionization and kinematics
of the emission line gas has been observed in the spectra
of many distant radio sources (e.g. Best, R\"{o}ttgering \& Longair
2000b; Sol\'{o}rzano-I\~{n}arrea, Tadhunter \& Axon 2001; Inskip et al
2002b), particularly in the case of smaller radio
sources, i.e. those with a projected physical size of $< 120$kpc.  We
also find that the sources in which shocks have the greatest impact on 
the emission line gas properties  are those with the most extensive, luminous
alignment effects (Inskip et al 2005, 2002c).
The shocks associated with an expanding radio source can greatly influence the
alignment effect.  Ionizing photons associated
with the shocks may boost certain emission lines, and also lead to an
increase in nebular continuum emission.  Star formation induced by
the passage of radio source shocks through the cool dense gas clouds
is also an obvious mechanism by which the alignment effect
may be enhanced.  In addition, the passage of a fast shock can
potentially cause the break-up of optically thick clouds (Bremer, Fabian \&
Crawford 1997), increasing the covering factor for scattering
of the UV flux from the AGN.  The more numerous, smaller clouds will
also have a larger cross section for ionization by the AGN, leading to
an increase in the total flux of line emission.  

Despite the compelling evidence for each of the alignment effect mechanisms
outlined above, their relative balance is still poorly understood.
A wide range of galaxy colours provides a useful means of probing the 
contributions from different physical processes, and their dependence
on the properties of the radio source population (power, size, epoch).
This is of paricular interest, since the dependence of different
mechanisms on each of these parameters varies significantly. For
example, line
emission is known to be closely linked to both radio source power and
size. Star formation, on the other hand, may be independent of radio
power, despite evolving quickly with age.  It is necessary that these
processes are better understood before we can interpret the clear
redshift evolution of the alignment effect.

One problem with studies of 3CR radio galaxies is that the radio power
of sources in a flux-limited sample such as this increases with
redshift, leading to a degeneracy between redshift and radio power.
The less powerful 6C sample provides a population of radio galaxies
ideally suited for breaking this degeneracy.  The factor of $\sim 6$ difference
in radio power between the samples is small compared to the wide range
of powers (spanning several orders of magnitude) observed for the
radio source population as a whole.  However, it is comparable to the
difference in power between 3C sources at low ($z \sim 0.1-0.5$) and high
($z \sim 1$) redshifts, and it is the evolution within this range that
we hope to explain.   To this end, we have carried out a program of
multiwavlength imaging and spectroscopic observations of a subsample
of 11 6C radio sources at $z \sim 1$ (Inskip et al 2003; Best et al
1999; Inskip et al 2002b), which are well matched to the
3CR subsample previously studied by Best et al (1997b, 2000a).
Having already analysed the spectroscopic (Inskip et al 2002c) and
morphological properties of these systems (Inskip et al 2005, which
included an analysis of the variations in host galaxy size), we now
turn our attention to investigating the effect 
of radio power on the galaxy colours, and the nature of
the excess UV emission (including the relative
contributions of line emission and nebular continuum).
The structure of the paper is as follows.  In section 2, we briefly outline
the sample selection and observations. Colours are determined for the
two matched samples, including emission line corrections.  The results
are presented in section 3, and analysed in more depth in section 4,
where we consider the influence of radio source size and power.  We
consider the influence of the host galaxy stellar populations in
section 5, and present our
conclusions in section 6. Values for the cosmological
parameters $\Omega_0=0.3$, $\Omega_\Lambda=0.7$ and
$H_{0}=65\,\rm{km\,s^{-1}\,Mpc^{-1}}$ are assumed throughout this
paper.

\begin{table*}
\caption{Observed (roman) and calculated (italics) magnitudes for the 6C and 3CR
sources at $z \sim 1$, together with a summary of the source redshifts,
radio sizes and radio power at 178MHz.  All 
magnitudes for the sources in both samples were determined within a 4\arcsec\
diameter aperture and have been corrected for galactic
extinction. The 6C data was previously presented in Inskip et al
(2003).  The 3CR HST, $J$ and $K$ band data were initially analysed in 5\arcsec\ and
9\arcsec apertures (Best et al 1997), but magnitudes have been
re-extracted in 4\arcsec\
diameter apertures for the purposes of this paper.  The $H$ band 3CR
data were obtained via the UKIRT service program.
For 6C1256+36, the flux due to the unresolved companion
object has been modelled (Inskip et al 2005, Paper 2) and removed from the F702W and
K-band data (magnitudes which remain contaminated by this
object are marked by a '*').  No
corrections have been made for flux contamination in the HST data by
adjacent objects, as
it is usually impossible to disentangle adjacent
galaxies from any aligned line/continuum emission.   
 } 
\begin{center} 
\begin{tabular} {ccccr@{$\pm$}lr@{$\pm$}lr@{$\pm$}lr@{$\pm$}lr@{$\pm$}l}\\
Source    & Redshift & L$_{178}$ (log$_{10}$W\,Hz$^{-1}$)&D$_{rad}$ (kpc) &\multicolumn{2}{c}{F702W} &
\multicolumn{2}{c}{F814W} & \multicolumn{2}{c}{$J$}&
\multicolumn{2}{c}{$H$}& \multicolumn{2}{c}{$K$}\\\hline
6C0825+34 & 1.467 &28.31 &64  & {\it 22.59}& {\it 0.31}&22.10&0.17 &19.86 &0.17 &19.68 &0.29 & 19.12& 0.12 \\
6C0943+39 & 1.035 &28.07 &92  & 22.08&0.14 & {\it 21.55}& {\it 0.20}&19.70&0.12&19.27&0.14&18.09&0.07\\
6C1011+36 & 1.042 &28.02 &444 & 21.73&0.07 & {\it 21.23}&{\it 0.15} &19.68&0.20&18.63&0.10&17.83&0.06\\
6C1017+37 & 1.053 &28.10 &65  & 21.77&0.06 & {\it 21.10}&{\it 0.14} &19.89&0.16&19.54&0.15&18.57&0.09\\
6C1019+39 & 0.922 &28.03 &67  & {\it 21.04}&{\it 0.29} & 20.07&0.04 &18.47&0.07&17.71&0.06&16.80&0.04\\
6C1100+35 & 1.440 &28.32 &119 & {\it 22.31}& {\it 0.26}& 21.80&0.06 &19.59&0.12&19.02&0.11&17.99&0.07\\
6C1129+37 & 1.060 &28.03 &141 & 22.02&0.07 & {\it 21.64}&{\it 0.16} &19.35&0.11&18.31&0.09&17.81&0.07\\
6C1204+35 & 1.376 &28.45 &158 & {\it 21.84}&{\it 0.28} & 21.30&0.10 &19.31&0.11&18.76&0.11&18.01&0.07\\
6C1217+36 & 1.088 &28.12 &38  & {\it 21.91}&{\it 0.25} & 20.89&0.05 &19.50&0.12&18.51&0.07&17.55&0.06\\
6C1256+36 & 1.128 &28.20 &155 & 22.86&0.09 & {\it 22.60}&{\it 0.16*} &19.74&0.20*&18.58&0.09*&18.14&0.06\\
6C1257+36 & 1.004 &28.06 &336 & {\it 21.55}&{\it 0.30} & 20.98&0.05 &19.39&0.11&18.27&0.07&17.50&0.05\\\\

3C13      & 1.351 &29.20 & 259 & {\it 21.16}&{\it 0.25}&20.57 & 0.03&18.74 &0.13 &\multicolumn{2}{c}{-}&17.47 &0.11\\
3C22 	  & 0.938 &28.77 & 209 & {\it 20.00}&{\it 0.15}&19.29 &0.02 &17.53 &0.08 & 16.95&0.09 &15.66&0.05\\
3C34 	  & 0.690 &28.51 & 359 & {\it 20.70}&{\it 0.25} &{\it 20.03} &{\it 0.25} &18.30 &0.11 &\multicolumn{2}{c}{-} &16.46 & 0.07\\
3C41 	  & 0.795 &28.48 & 194 & {\it 20.42}&{\it 0.20}&{\it 19.79} &{\it 0.20} &18.78 &0.14 & \multicolumn{2}{c}{-}&15.89 & 0.05\\
3C49 	  & 0.621 &28.25 & 7   & {\it 19.97}&{\it 0.15} &19.33 &0.05 &\multicolumn{2}{c}{-}&\multicolumn{2}{c}{-} &16.25 & 0.06\\
3C65 	  & 1.176 &29.09 & 160 & {\it 22.08}&{\it 0.16} &21.07 &0.03 &18.93 &0.14 &\multicolumn{2}{c}{-} &17.19 & 0.10\\
3C68.2 	  & 1.575 &29.33 & 218 & {\it 22.37}&{\it 0.30} &{\it 21.87 }& {\it 0.30}&19.78 &0.22 & \multicolumn{2}{c}{-}&18.18 & 0.16\\
3C217 	  & 0.897 &28.69 & 110 & {\it 20.81}&{\it 0.17} &20.27 & 0.02&18.81 &0.13 &18.99 &0.23 &17.88&0.13 \\
3C226 	  & 0.820 &28.75 & 259 & {\it 20.46}&{\it 0.18} &{\it 19.64} &{\it 0.16} &18.46 &0.11 &17.82 &0.13 &16.83
&0.08\\		    		    
3C239 	  & 1.781 &29.60 & 111 & {\it 21.60}&{\it 0.20} &21.16 &0.03 &19.01 & 0.14&\multicolumn{2}{c}{-} &17.90 &0.13 \\
3C241 	  & 1.617 &29.39 & 8   & {\it 22.19}&{\it 0.20} &21.62 &0.04 &19.19 & 0.16&\multicolumn{2}{c}{-} &17.82 &0.13 \\
3C247 	  & 0.749 &28.44 & 110 & {\it 20.35}&{\it 0.24} &19.46 &0.02 &\multicolumn{2}{c}{-}&\multicolumn{2}{c}{-} &16.04 &0.06 \\
3C252 	  & 1.105 &28.98 & 501 & {\it 21.10}&{\it 0.13} &20.58 &0.03 &\multicolumn{2}{c}{-}&\multicolumn{2}{c}{-} &17.54 & 0.11\\
3C265 	  & 0.811 &28.88 & 636 & {\it 19.51}&{\it 0.24}&{\it 19.06} &{\it 0.30} &17.81 &0.08 &17.52 &0.11 &16.39
&0.07\\		    		    
3C266 	  & 1.272 &29.13 & 41  & 21.22&0.03 &20.55 &0.03 &\multicolumn{2}{c}{-}&\multicolumn{2}{c}{-} &17.99 &0.14\\
3C267 	  & 1.144 &29.11 & 339 & 21.50& 0.04&{\it 20.95} &{\it 0.10} &19.20 &0.16 &\multicolumn{2}{c}{-} &17.46 &0.11 \\
3C277.2   & 0.766 &28.62 & 422 & {\it 20.16}&{\it 0.20} &19.82 &0.02 &18.59 &0.12 &\multicolumn{2}{c}{-} &17.32 & 0.10\\
3C280     & 0.996 &29.13 & 118 & {\it 20.92}&{\it 0.06} &20.29 &0.02 &18.48 &0.11 &18.10 &0.15 &17.05
&0.09\\		    		    
3C289 	  & 0.967 &28.81 & 90  & {\it 21.49}&{\it 0.07}&20.50 &0.03 &18.66 &0.12 &17.97 &0.14 &17.06
&0.09\\		    		    
3C324 	  & 1.206 &29.18 & 100 & 21.49&0.03&{\it 20.84} &{\it 0.20} &19.18 &0.17 &18.30 &0.16 &17.33
&0.10\\		    		    
3C337 	  & 0.635 &28.32 & 326 & {\it 20.79}&{\it 0.25} &19.94 &0.02 &18.45 &0.11 &\multicolumn{2}{c}{-} &16.84 &0.08 \\
3C340     & 0.775 &28.48 & 363 & {\it 20.88}&{\it 0.18} &{\it 19.71} &{\it 0.21} &18.52 &0.12 &17.89 &0.12 &17.08
&0.09\\		    		    
3C352 	  & 0.806 &28.61 & 100 & {\it 20.63}&{\it 0.08} &20.14 & 0.02&\multicolumn{2}{c}{-}&\multicolumn{2}{c}{-} &17.09 & 0.09\\
3C356 	  & 1.079 &28.96 & 638 & {\it 21.22}&{\it 0.09} &20.60 &0.03 &\multicolumn{2}{c}{-}& \multicolumn{2}{c}{-}&17.54 &0.11 \\
3C368 	  & 1.132 &29.17 & 75  & 20.47&0.02 &{\it 19.55 }&{\it 0.08} &19.01 &0.11 & 17.83& 0.13&17.17
&0.10\\		    		    
3C437 	  & 1.480 &29.32 & 339 & {\it 22.38}&{\it 0.30} &{\it 21.86} & {\it 0.30}&\multicolumn{2}{c}{-}&\multicolumn{2}{c}{-} &18.23 & 0.16\\
3C441     & 0.708 &28.51 & 257 & {\it 20.07}&{\it 0.15} &{\it 19.28} & {\it 0.15}&18.15 &0.10 &\multicolumn{2}{c}{-} &16.49 &0.07 \\
3C470 	  & 1.653 &29.20 & 228 & {\it 22.79}&{\it 0.46} &{\it 22.35} &{\it 0.46} &\multicolumn{2}{c}{-}&\multicolumn{2}{c}{-} & 18.20&0.15 \\

\end{tabular}		           
\end{center}		           
\end{table*}

\section{Galaxy Colours for 3CR and 6C radio sources}
\subsection{Sample Selection and Observations}
This paper focuses on observations of galaxies selected from two well
matched radio galaxy subsamples at $z \sim 1$, which have been 
extensively studied using the Hubble Space Telescope (HST), the Very
Large Array (VLA) and the United Kingdom InfraRed Telescope (UKIRT).
The 3CR subsample 
consists of 28 galaxies with $0.6 < z < 1.8$ (see Best et al 1997b, 1998 for
a description of the sample and the imaging observations \& results). The 6C
comparison sample
consists of 11 radio galaxies at $0.85 < z < 1.5$, which are
approximately 6 times less powerful radio sources 
than 3CR sources at the same redshift.  Full selection criteria and
details of the observations of these 6C sources can be found in Papers I and II of this
series (Inskip et al 2003, 2005) and also Best et al (1999).  Deep
spectroscopic observations are also available for a subset of 14 3CR
galaxies with $0.7 < z < 1.25$ and the 8 6C sources with $0.85 < z <
1.3$ (Best et al 2000a, 2000b; Inskip et al 2002b). 

\subsection{Calculating the colours}

The HST/WFPC2 and UKIRT magnitudes previously presented for the 3CR
and 6C sources (Best et al 1997b; Inskip et al 2003) were determined
through 9\arcsec\ 
diameter apertures.  The advantage of this approach was that such an
aperture includes essentially all the light emitted from the galaxy.
However, for several sources the 9\arcsec\ diameter apertures are
contaminated by flux from other nearby objects.  For the current
analysis, we use galaxy magnitudes determined within a 4\arcsec\
diameter aperture (see Table 1), where removal of flux from nearby companion
objects is not usually required. The exception to this rule is an unresolved
point source in close proximity to the host galaxy of 6C1256+36, the
flux from which (including associated errors) has been modelled and removed from the $K-$band and
F702W data (see papers 1 \& 2 of this series, Inskip et al 2003,
2005).   The data for the 6C sources was presented in Paper 1 (Inskip et
al 2003, which also includes full details of our data reduction
methods); the 3CR 4\arcsec\ diameter aperture 
magnitudes presented in the current paper were determined using an identical data
reduction methodology.  With our assumed cosmology, the angular scale
at $z \sim 1$ is $\sim 8.6$ kpc/arcsec; this varies by less than $\sim
10$\% over the redshift range of our data, and we therefore study
essentially the same physical aperture size at all redshifts.  This
4\arcsec\ aperture is also large enough that we rarely exclude any
emission due to the alignment effect, and that the effects of the
seeing remain negligible. The data of Best et al (1997b) have been
supplemented by $H$ band observations of 9 3CR sources, obtained via
the UKIRT service observing programme; these data were analysed
identically to our other infrared observations, and the resulting
4\arcsec\ magnitudes are also presented in Table 1.
  
\begin{table*}
\caption{This table presents the emission line and unresolved point
source percentages for each filter (where known). The emission line data
were previously presented for these sources in Inskip et al (2002b),
and Best et al (2000a); analysis of the emission line contributions in
the directly observed
filters can be found in Inskip et al (2003) and Best et al (1997) (see
text for details of the emission line corrections for the remaining data).
Our analysis of the nuclear point source contributions is presented in
Inskip et al (2005) and Best et al (1998).  Also listed
are the emission line corrected (i.e. continuum) galaxy 
colours, and their associated errors, derived from the magnitudes in
Table 1 and the emission line contributions listed below. The HST/WFPC2 filter
colours can be approximated to ground based UBVRI colours as F702W$ -
K$ $\approx (R - K) - 0.15$ and F814W$ - K$ $\approx (I - K) + 0.05$. }  
\begin{center} 
\scriptsize{}
\begin{tabular} {cccccccccr@{$\pm$}lr@{$\pm$}lr@{$\pm$}lr@{$\pm$}l}\\
Source    & \multicolumn{5}{c}{Emission line percentages$^1$} &
\multicolumn{3}{c}{Nuclear point source percentages}&\multicolumn{8}{c}{Emission line
  corrected (i.e. continuum) colours}\\
& F702W & F814W & $J^1$ & $H^1$ & $K^1$ & F702W & F814W &  $K$ &\multicolumn{2}{c}{F702W$-K$}&\multicolumn{2}{c}{F814W$-K$}&\multicolumn{2}{c}{$J-K$}&\multicolumn{2}{c}{$H-K$}\\\hline
6C0825+34 &$3 \pm 1$\% &$3 \pm 2$\% &$25$\% &$28$\% &5\% &- &- &- &3.44& 0.36&2.95&0.25 &0.93&0.32 &0.78& 0.42 \\
6C0943+39 &$18 \pm 5$\% &$21 \pm 4$\% &1\% &0\% &0\% &$36 \pm 4$\% &- &$41.5 \pm 3.5$\% & 4.21&0.17 &3.72 &0.21 & 1.62& 0.14& 1.18& 0.16 \\
6C1011+36 &$8 \pm 4$\% &$9 \pm 1$\% &3\% &0\% &0\% &$25 \pm 2$\% &- &$28.8 \pm 1$\% & 3.99&0.10 &3.50 &0.16 & 1.88& 0.21& 0.80& 0.12 \\
6C1017+37 &$17 \pm 4$\% &$22 \pm 3$\% &14\% &1\% &0\% &$12 \pm 3$\% &- &$4 ^{+18}_{-4}$\%  & 3.40&0.11 &2.80 &0.17 & 1.46& 0.23& 0.98& 0.18\\
6C1019+39 &$5 \pm 1$\% &$5 \pm 1$\% &2\% &0\% &0\% &- &$0 \pm 1$\% &$0 \pm 1$\% &4.30 &0.29 & 3.33&0.06 &1.69 &0.08 &0.91 &0.07   \\
6C1100+35 &$5 \pm 1$\% &$7 \pm 4$\% &$35$\% &$28$\% &3\% &- &$41 ^{+11}_{-1} 2$\% &$15 \pm 14$\%  &4.35 &0.27 & 3.86&0.10 &1.89 & 0.36& 1.27&0.30  \\
6C1129+37 &$15 \pm 5$\% &$25 \pm 4$\% &1\% &0\% &0\% &- &- &$18 \pm 1$\%  & 4.39&0.11 & 4.14& 0.17&1.55 &0.13 &0.50 &0.11 \\
6C1204+35 &$7 \pm 2$\% &$11 \pm 7$\% &$56$\% &$50$\% &8\% &- &- &$13\pm 1$\% &3.82& 0.30& 3.33&0.17 & 1.70& 0.51&1.11 &0.47  \\
6C1217+36 &$2 \pm 1$\% &$3 \pm 1$\% &0\% &0\% &0\% &- &$23 \pm 5$\% &$25 \pm 0.5$\% &4.38&0.26 & 3.37&0.08 & 1.95& 0.13& 0.96& 0.09\\
6C1256+36 &$9 \pm 4$\% &$15 \pm 3$\% &0\% &0\% &0\% &0\% &- &$0 \pm
3$\% & 5.02&0.12 &4.64 & 0.17&1.60 &0.21 &0.44 & 0.21  \\
6C1257+36 &$9 \pm 1$\% &$10 \pm 2$\% &28\% &0\% &0\% &- &$3.5 ^{+1}_{-3.5} 2$\% &$16 \pm 0.5$\%  &4.15 &0.30 & 3.59&0.07 &2.16 &0.29 &0.77 &0.09 \\\\

3C13      &$ 8 \pm 2$\%  & $19 \pm 4$\%  & 95\% &-  &17\%  &- &-&$0^{+7}_{-0}$\% &3.65 &0.32 & 3.16&0.21 &1.82& 0.76&\multicolumn{2}{c}{-} \\
3C22 	  &$10 \pm 1$\%  & $19 \pm 1$\%  & $45$\% &2\%  &0\%  &- &-
&$50^{+20}_{-10} $\%&4.44 &0.16 & 3.86&0.05&2.27 &0.41 &1.31 & 0.11 \\
3C34 	  &$ 1 \pm 1$\%  & $ 1 \pm 1$\%  & 0\% &-  &0\%  &- &- &$5 ^{+6}_{-0}$\%&4.25 &0.26 &3.58 &0.26 &1.84& 0.13 &\multicolumn{2}{c}{-} \\
3C41 	  &$15 \pm 2$\%  & $11 \pm 2$\%  & 38\% &-  &0\%  &- &- &$31 ^{+10}_{-8} $\%& 4.68&0.21 &4.01 &0.21 & 3.24&0.38&  \multicolumn{2}{c}{-}\\
3C49 	  &$17 \pm 3$\%  & $ 8 \pm 2$\%  & - &-  &0\%  &- &- &$0 ^{+2}_{-0} $\%&3.90 &0.16 &3.17&0.08 & \multicolumn{2}{c}{-}  &\multicolumn{2}{c}{-} \\
3C65 	  &$ 8 \pm 2$\%  & $ 5 \pm 1$\%  & 0\% &-  &0\%  &- &- &$0 ^{+8}_{-0} $\%& 4.96& 0.19&3.92 &0.11 &1.73  &0.17& \multicolumn{2}{c}{-} \\
3C68.2 	  &$ 4 \pm 1$\%  & $ 4 \pm 1$\%  & 38\% &-  &0\%  &- &- &-& 4.21& 0.34&3.71 &0.34 &1.94 &0.45 &  \multicolumn{2}{c}{-}\\
3C217 	  &$31 \pm 3$\%  & $20 \pm 1$\%  & 84\% &9\%  &0\%  &- &- &$0 ^{+6}_{-0} $\%&3.22 &0.21 &2.63 &0.13 &1.59 &0.69 &1.20 &0.28 \\
3C226 	  &$14 \pm 2$\%  & $10 \pm 1$\%  & $29$\% &3\%  &0\%  &- &- &$13 ^{+8}_{-13} $\%&3.77 &0.20 &2.91&0.18 & 1.91 &0.31 &1.02 &0.16 \\
3C239 	  &$11 \pm 3$\%  & $ 6 \pm 1$\%  & 3\% &-  &1\%  &- &- &-&3.81 &0.24 &3.32 &0.13 & 1.13& 0.19&  \multicolumn{2}{c}{-}\\
3C241 	  &$ 1 \pm 1$\%  & $ 8 \pm 2$\%  & 3\% &-  &0\%  &- &- &-&4.38 &0.24 &3.89 &0.14 &1.40 &0.21 &  \multicolumn{2}{c}{-}\\
3C247 	  &$11 \pm 1$\%  & $15 \pm 1$\%  & - &-  &0\%  &- &- &-&4.42 &0.25 &3.60&0.06 & \multicolumn{2}{c}{-} &  \multicolumn{2}{c}{-}\\
3C252 	  &$ 7 \pm 1$\%  & $11 \pm 1$\%  & - &-  &0\%  &- &- &$17 ^{+7}_{-17} $\%&3.63 &0.17 & 3.17&0.11 & \multicolumn{2}{c}{-} &  \multicolumn{2}{c}{-}\\
3C265 	  &$22 \pm 2$\%  & $18 \pm 2$\%  & $56$\% &8\%  &0\%  &- &- &$8 ^{+4}_{-8} $\%&3.34 &0.25&2.85 &0.31  &1.90 &0.49 &1.21 &0.15 \\
3C266 	  &$10 \pm 1$\%  & $16 \pm 1$\%  & - &-  &9\%  &- &- &$4 ^{+9}_{-4} $\%&3.25 &0.17 &2.66 &0.17 & \multicolumn{2}{c}{-}&   \multicolumn{2}{c}{-}\\
3C267 	  &$ 8 \pm 2$\%  & $12 \pm 2$\%  & 1\% &-  &1\%  &- &- &$0 ^{+5}_{-0} $\%&4.12 &0.12 &3.60 &0.15 &1.74 & 0.19&  \multicolumn{2}{c}{-}\\
3C277.2   &$23 \pm 3$\%  & $ 9 \pm 2$\%  & $20$\% &-  &0\%  &- &- &$15 ^{+6}_{-15} $\%&3.08 &0.22 &2.60 &0.10 &1.47 &0.25 &  \multicolumn{2}{c}{-}\\
3C280     &$15 \pm 2$\%  & $20 \pm 1$\%  & $51$\% &0\%  &0\%  &- &- &$0 ^{+8}_{-0} $\%&4.02 &0.11 &3.48 &0.09 &1.88 &0.47 &1.05 &0.17 \\
3C289 	  &$ 8 \pm 1$\%  & $10 \pm 1$\%  & $28$\% &0\%  &0\%  &- &- &$1 ^{+6}_{-1}$\%&4.51 &0.11 & 3.55&0.10 &1.87 &0.31 &0.91 &0.17 \\
3C324 	  &$ 7 \pm 1$\%  & $14 \pm 2$\%  & 4\% &17\%  &9\%  &- &- &$0 ^{+4}_{-0} $\%&4.15 &0.14 &3.56 &0.24& 1.80& 0.22 &1.05 &0.27 \\
3C337 	  &$ 7 \pm 1$\%  & $ 5 \pm 1$\%  & 0\% &-  &0\%  &- &- &$2 ^{+7}_{-2}$\%&4.02 &0.26 & 3.16&0.08 & 1.61& 0.14&  \multicolumn{2}{c}{-}\\
3C340     &$12 \pm 1$\%  & $18 \pm 2$\%  & 19\% &3\%  &0\%  &- &- &$0 ^{+4}_{-0} $\%&3.92 &0.20 &2.81 &0.23 &1.63 &0.24 &0.84 &0.15 \\
3C352 	  &$22 \pm 2$\%  & $13 \pm 1$\%  & - &-  &0\%  &- &- &$7 ^{+6}_{-7}$\%&3.76 &0.12 &3.20 &0.09 &  \multicolumn{2}{c}{-} & \multicolumn{2}{c}{-}\\
3C356 	  &$10 \pm 1$\%  & $13 \pm 1$\%  & - &-  &0\%  &- &- &$0 ^{+14}_{-0} $\%& 3.78&0.14 &3.21 &0.11 & \multicolumn{2}{c}{-} & \multicolumn{2}{c}{-}\\
3C368 	  &$10 \pm 1$\%  & $12 \pm 1$\%  & 3\% &2\%  &4\%  &- &- &-& 3.37& 0.11&2.46 &0.13 &1.83 &0.16 &0.64 &0.17 \\
3C437 	  &$ 3 \pm 1$\%  & $ 5 \pm 1$\%  & - &-  &3\%  &- &- &-&4.15 &0.34 &3.65 &0.34 &\multicolumn{2}{c}{-} &  \multicolumn{2}{c}{-}\\
3C441     &$ 6 \pm 1$\%  & $10 \pm 1$\%  & 0\% &-  &0\%  &- &- &$0 ^{+5}_{-0}$\%&3.64 &0.17& 2.89&0.17  & 1.66& 0.12& \multicolumn{2}{c}{-} \\
3C470 	  &$14 \pm 2$\%  & $ 8 \pm 2$\%  & - &-  &0\%  &- &- &-& 4.73&0.48 &4.23 &0.48 &\multicolumn{2}{c}{-} & \multicolumn{2}{c}{-} \\\hline
{Notes:} &\\
{[1]:}&\multicolumn{16}{l}{The infrared
emission line contributions are derived from appropriate emission line ratios and the observed fluxes of
[O\textsc{ii}]3727\AA\ and H$\beta$, and are given as a rough estimate
only,
}\\
& \multicolumn{16}{l}{ as they are only believed to be accurate to about a factor of 2 (see text for details).  An error of
50\% on the calculated emission line contribution to the $JHK$
magnitudes has been }\\
& \multicolumn{16}{l}{included in our calculation of the error on the
emission line corrected galaxy colours.}\\
\end{tabular}		           
\normalsize
\end{center}		           
\end{table*}

All eleven 6C sources have been observed with one filter of either 
F702W or F814W. 
 The majority of the $z \sim 1$ 3CR sources studied by
Best, Longair \& R\"{o}ttgering (1997b) were also observed in one of
these two filters. These broad-band filters roughly correspond to ground
based filters in the $UBVRI$ system as follows: $F814W \approx I +
0.07(V-I)$, and $F702W \approx R - 0.3(V-R)$.
In order to consistently compare the rest-frame
UV-optical 
colours of the host galaxies of 6C and 3CR radio sources at $z \sim 1$,
magnitudes through the same filters must be obtained for both full
samples.  For the galaxies which have not been observed in either the
F702W or F814W filters, magnitudes in these filters have been
estimated using the observations made in other filters.  
 
For the fourteen 3CR and eight 6C sources for which high quality
spectra are available (Best, R\"{o}ttgering \& Longair 2000a; Inskip {\it et al} 2002b), the
following method has been used to determine the F702W and F814W filter
magnitudes for cases where no observations in the relevant filter
exist.  Using the spectroscopic observations, the continuum flux 
density levels were compared at the mean wavelengths of an observed
filter and the filter for which a magnitude was to be determined.  By
comparing the relative flux levels of the continuum at these
wavelengths, and accounting for the different filter widths, the flux
in the unobserved filter can be estimated and converted to a
magnitude (including the appropriate emission line contribution for
that filter).  

This method assumes that the percentage contributions of emission 
lines, host galaxy and aligned emission does not vary significantly
between the region of sky covered by the extracted spectra (a
1.5\arcsec\ to 2.5\arcsec\ slit aligned along the radio source axis) and 
the 4\arcsec\ apertures used for our photometry.  The emission line
imaging data obtained by McCarthy, Spinrad and van Breugel (1995),
which includes a subset of our $z \sim 1$ 3CR sources, supports this
assumption.  For the sources in question, the surface profiles of the line
emission and broad band emission are generally very similar within the
aperture size in question. In addition, the bulk of the rest-frame UV
excess (both line and continuum emission) lies along the radio source
axis and therefore within our spectroscopic slit (see Inskip et al
2003 \& Best et al 1997 for details); the flux levels in
the remainder of our 4\arcsec diameter apertures are generally at a
substantially lower level, and any discrepancy will therefore not be a
major source of error.   We have analysed the spatial variation of the
emission line contribution for two of the 6C sources which displayed
particularly extensive emission line regions (6C0943+39 and 6C1129+37,
see Inskip 2003 for full details).  While the emission line percentage
in the outer regions does vary (e.g. 39.5 per cent of the flux
in the outer regions of 6C0943+39 can be accounted for by line
emission, cf. 17.5 per cent line emission in
the central 4\arcsec\ of the spectroscopic data), the overall levels of
emission outside the central regions of the slit are much lower.
Given the observed distribution of line emission around radio
galaxies, we expect that no more than 20\% of the total line flux in
the 4\arcsec\ diameter apertures lies outside the regions covered by our
long-slit spectroscopy.  
In order to account for any
variations in the percentage contribution of line emission in these
regions, we therefore assign an additional 10\% error to our calculated emission line
percentages (i.e. allowing for a worst-case scenario error of 50\% on
the emission line flux in these regions).

By considering filters as close
in wavelength to F702W or F814W as possible, any variation with
wavelength in the flux emitted by the aligned structures should be
minimised, and will not significantly detract from the accuracy of the
estimated magnitudes. 

Where sources were observed using more than one HST/WFPC2 filter, the flux
in the unobserved F702W or F814W filter was estimated using the above
method with both of the  observed filters, providing these were close
in wavelength 
to the filter in question. The two resulting F702W or F814W magnitudes
obtained in this way were very consistent (typically to within
5-10\%), and the average value was 
taken.  

In the case of the galaxies for which high signal--to--noise
spectra were not available, the relative continuum levels and emission
line fluxes of the
composite spectra presented in Inskip et al (2002b) were used instead.  Once
again, this process was repeated for all the HST/WFPC2 filters used in the
observations of a given galaxy which were close in wavelength to F702W
or F814W.   Overall, while the precision of the estimated magnitudes
is necessarily lower than that of those which were directly observed,
we are confident that this method does produce accurate values.

\subsection{Emission line corrections}

We have also endeavoured to determine emission line percentages in each
filter (Table 2).  For the 8 6C and 14 3CR sources of in our
spectroscopic samples, emission line percentages in the F702W and
F814W filters are determined directly from our deep optical
spectroscopy (Best, Longair \& R\"{o}ttgering 2000a; Inskip et al
2002c).  For other sources, we make use of lower signal to noise
spectra in the literature (see references in Best et al 1997 \& Inskip et al 2003 for
details).   Where the spectroscopic data for
a given source is of poor quality or does not cover the full wavelength
range required (in other words, where we do not have accurate relative
fluxes for the full complement of emission lines required), we couple
the known line strengths of major emission lines such as
[O\textsc{ii}]3727\AA\ with the composite emission line ratios of
Inskip et al (2002b), which were determined based on the spectra of
14 3CR and 8 6C sources at $z \sim 1$.  These composite spectra provide typical emission line
ratios (including errors) for both large and small radio sources in each sample;
by using the most appropriate line ratios for a given source we can
therefore account for the expected variations in the ionization state of the
emission line gas due to differences in radio source properties (power
and size) within the samples.  This method of correcting the galaxy
colours for emission line contamination does however assume that the distribution of
line and continuum emission within the apertures is similar (as noted
in the preceding section); any
errors introduced by such a discrepancy are likely to be small ($\sim 10$\%) and are
included in the uncertainties on each value.  This uncertainty is
combined with the inherent errors in flux calibration for the spectra,
the uncertainties in the continuum level for the filter, plus the
errors associated with any line ratios derived from composite spectra
where applicable (see Inskip et al 2002b, Best et al 2000a for further
details of the errors involved).
  
Given the evidence that emission line flux scales with radio power
(e.g. Rawlings \& Saunders 1991, Willott et al
1999, Jarvis {\it et al} 2001, Inskip {\it et al} 2002c), one might
expect the line emission from the 6C sources to be lower by a factor of
$\sim 6$ than that of the 3CR sources at the same redshift, and
therefore to account for a lower fraction of the total UV excess
observed.  However, the choice of broad-band HST filters for these observations was such that the 
wavelength of the dominant [O\textsc{ii}] emission line is at a higher filter
throughput for the 6C observations than the 3CR observations. 
As this is the most prominent emission line in the rest--frame UV
spectra of these galaxies, the emission line contribution to the HST
magnitudes for the 6C
sources, although still less on average than the 3CR emission line
contribution, is higher than might otherwise be expected (Inskip et al
2003). 

As we do not have high quality spectra at longer wavelengths for the
galaxies in our $z \sim 1$ samples, $J$,$H$,$K$ emission line
percentages have been estimated using standard optical emission line
ratios (McCarthy 1993, Osterbrock 1989), determined relative to
measured fluxes for H$\beta$ and/or [O\textsc{ii}].  While our
spectroscopic data is certainly subject to slit losses, we simultaneously
consider the total flux within the $J$,$H$,$K$ filters and F702W/F814W
filters.   This allows us to correctly balance the relative
contributions of line and continuum emission in different wavelength
ranges, in accord with our the observations, thus avoiding any
potential systematic errors due to a mismatch between line and
continuum flux in different filters.  Unfortunately, whilst being free
of most sources of systematic error, one problem remains, and our
estimation of the emission line fluxes in the observed-frame infrared based on
existing optical spectra is not likely to be particularly accurate
compared to that for the emission lines lying at shorter wavelengths. 
This is due to the fact that there is considerable variation in
emission line ratios with radio source 
size and power (e.g. Inskip et al 2002b), which cannot be adequately
accounted for using  a single set of 
line ratios for the full sample.  Therefore, based on the known variations in these line
ratios at shorter wavelengths, the uncertainties in these values are easily
of the order of a factor of 2.  Whilst this may seem to be a
significant factor,  the overall emission line
corrections in the observed $J$,$H$,$K$ wavebands are
typically small, and therefore the errors on the observed magnitudes will not be
greatly degraded in most cases. Nonetheless, in point of this fact,
throughout this paper we
display the resulting galaxy colours both with and without emission
line correction.  The resulting emission line
percentages  are presented in Table 2, together with the nuclear 
point source percentages determined in Inskip et al 2005 (Paper II,
this series) and Best, Longair \& R\"{o}ttgering (1998).  The emission
line corrected galaxy colours and their associated errors (calculated
from the errors on the 4\arcsec aperture magnitudes and the errors on
the emission line contribution in each filter) are also
presented in Table 2. 

\section{Results}
\subsection{Observed F702W-K \& F814W-K galaxy colours}

\begin{figure*}
\vspace{5.80 in}
\begin{center}
\includegraphics{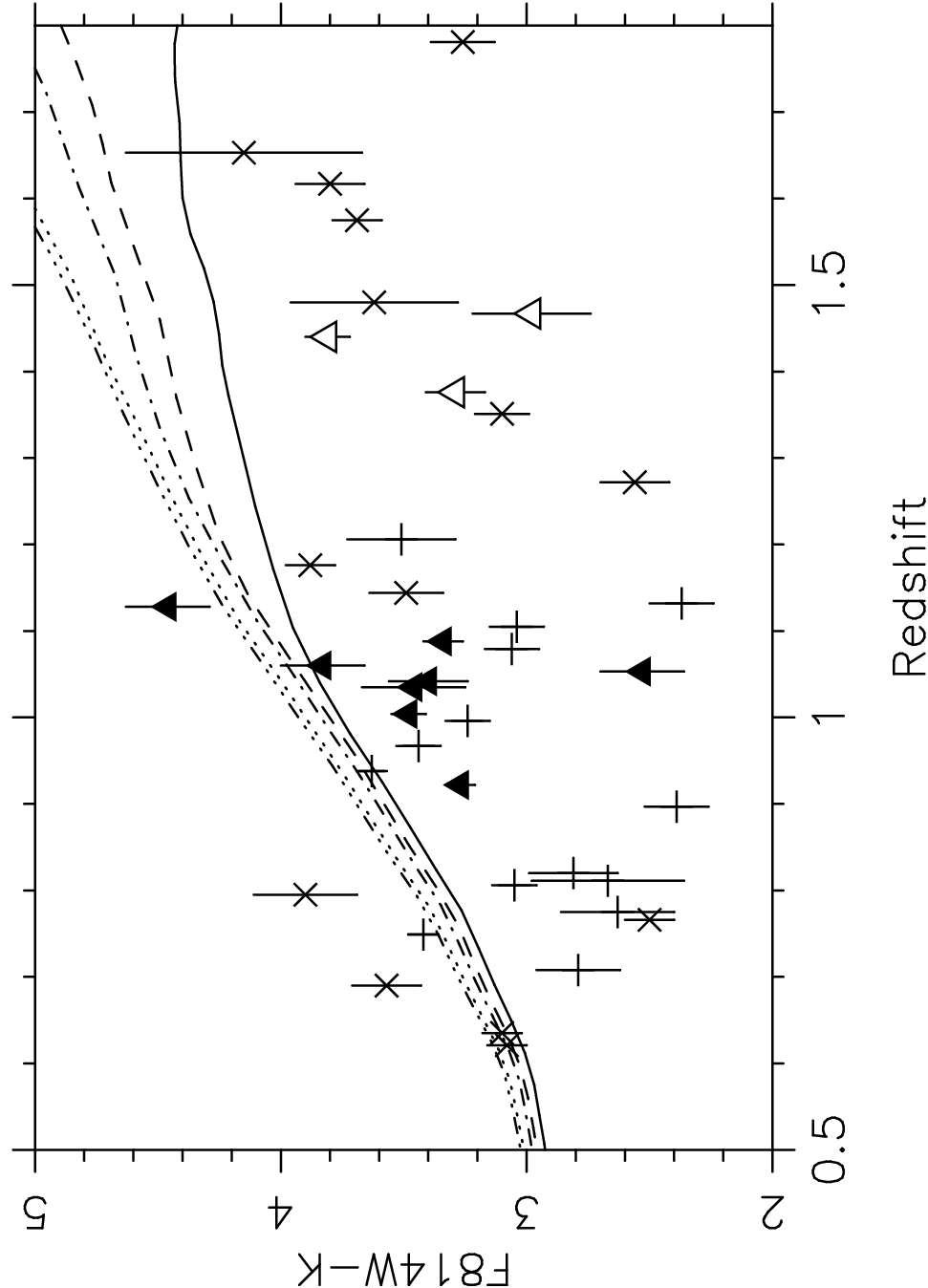}
\includegraphics{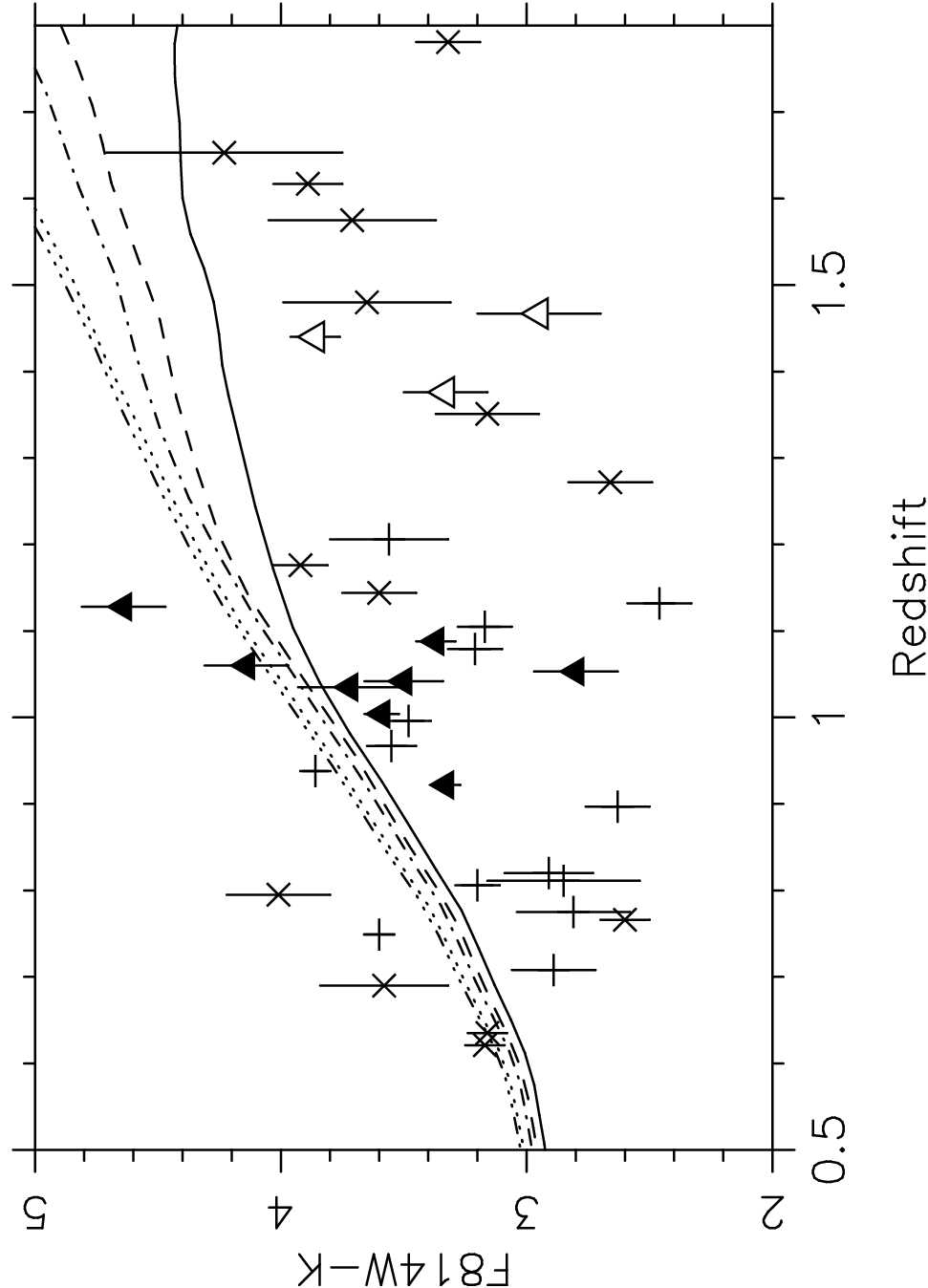}
\includegraphics{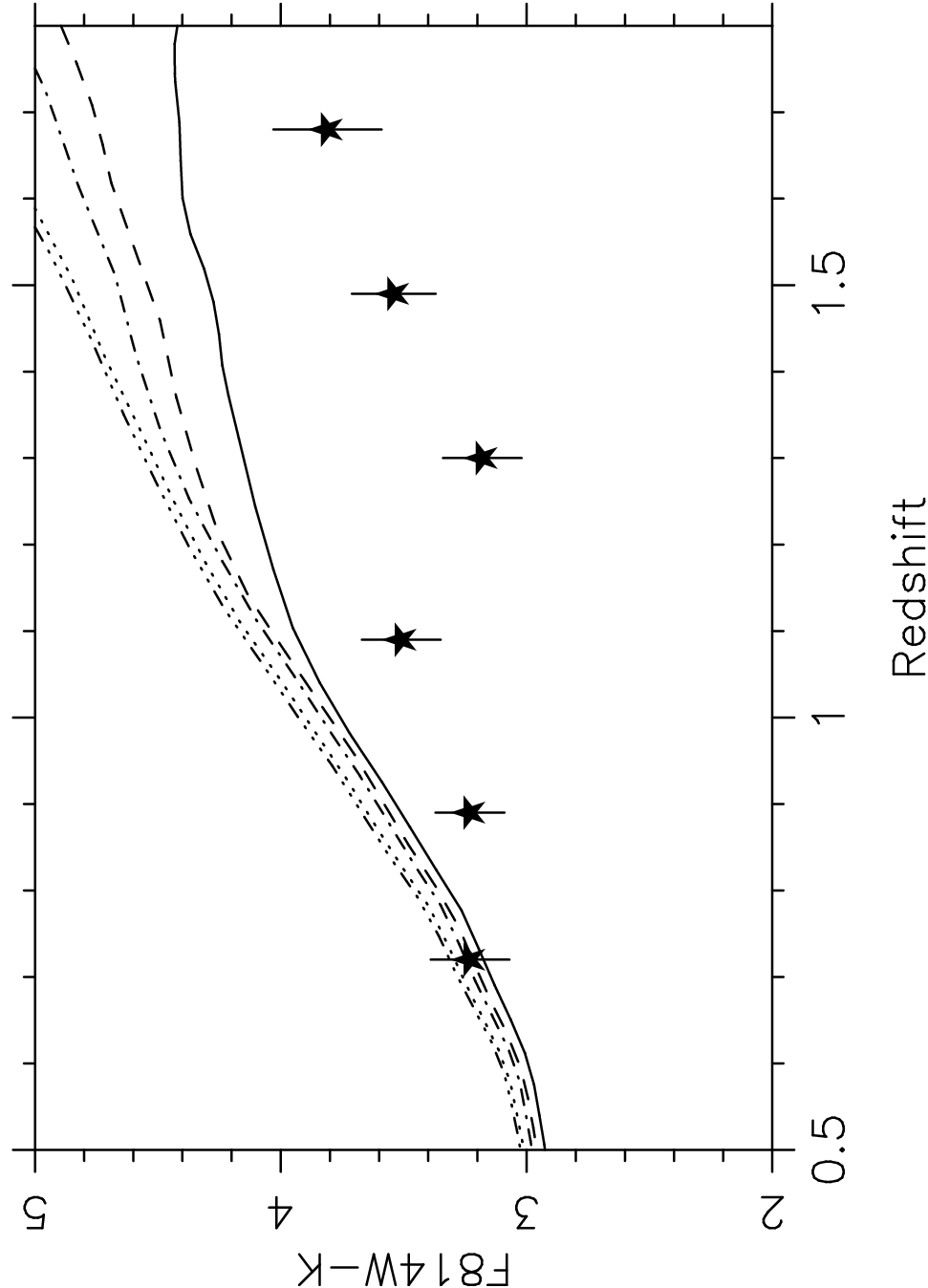}
\includegraphics{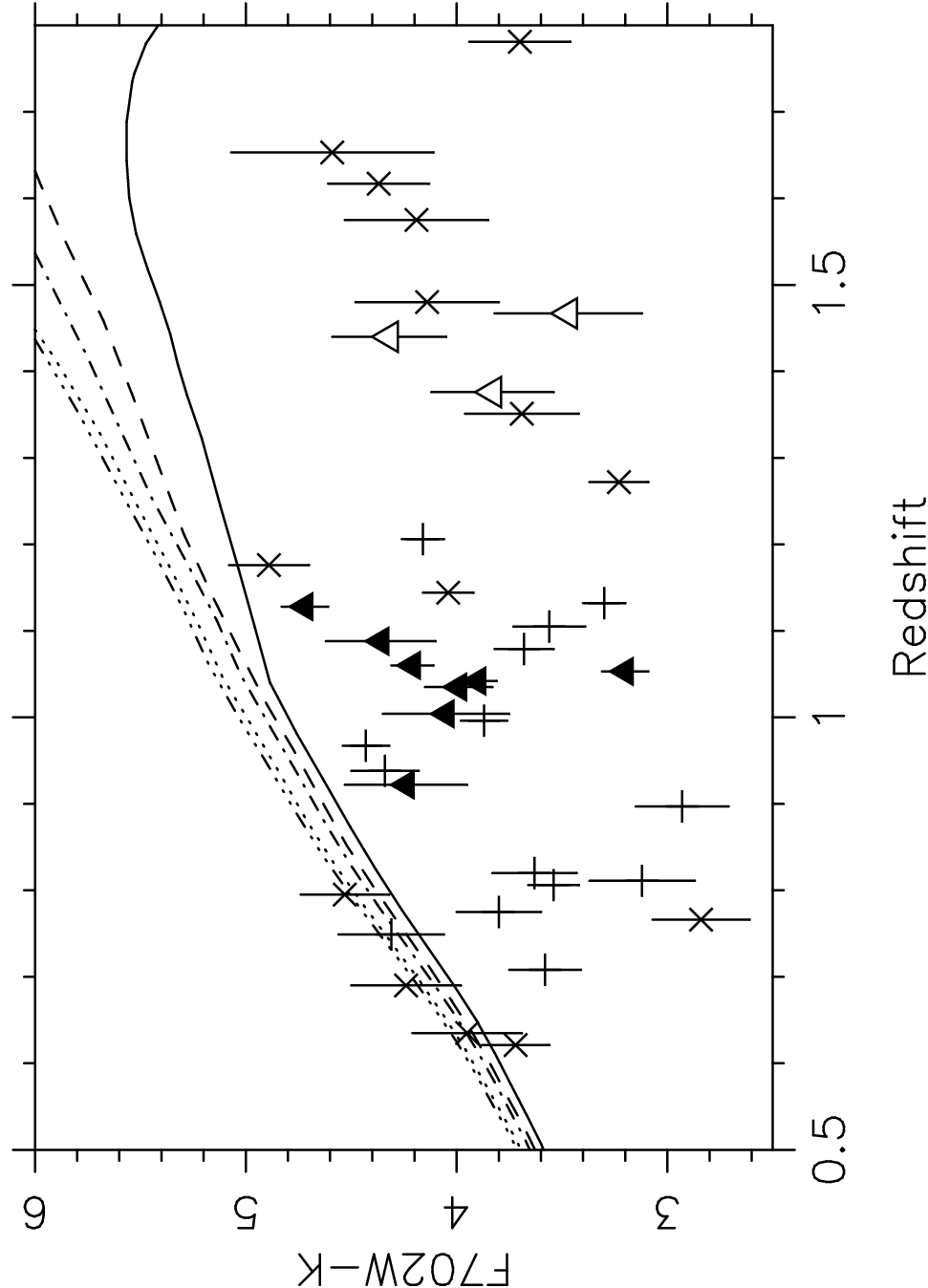}
\includegraphics{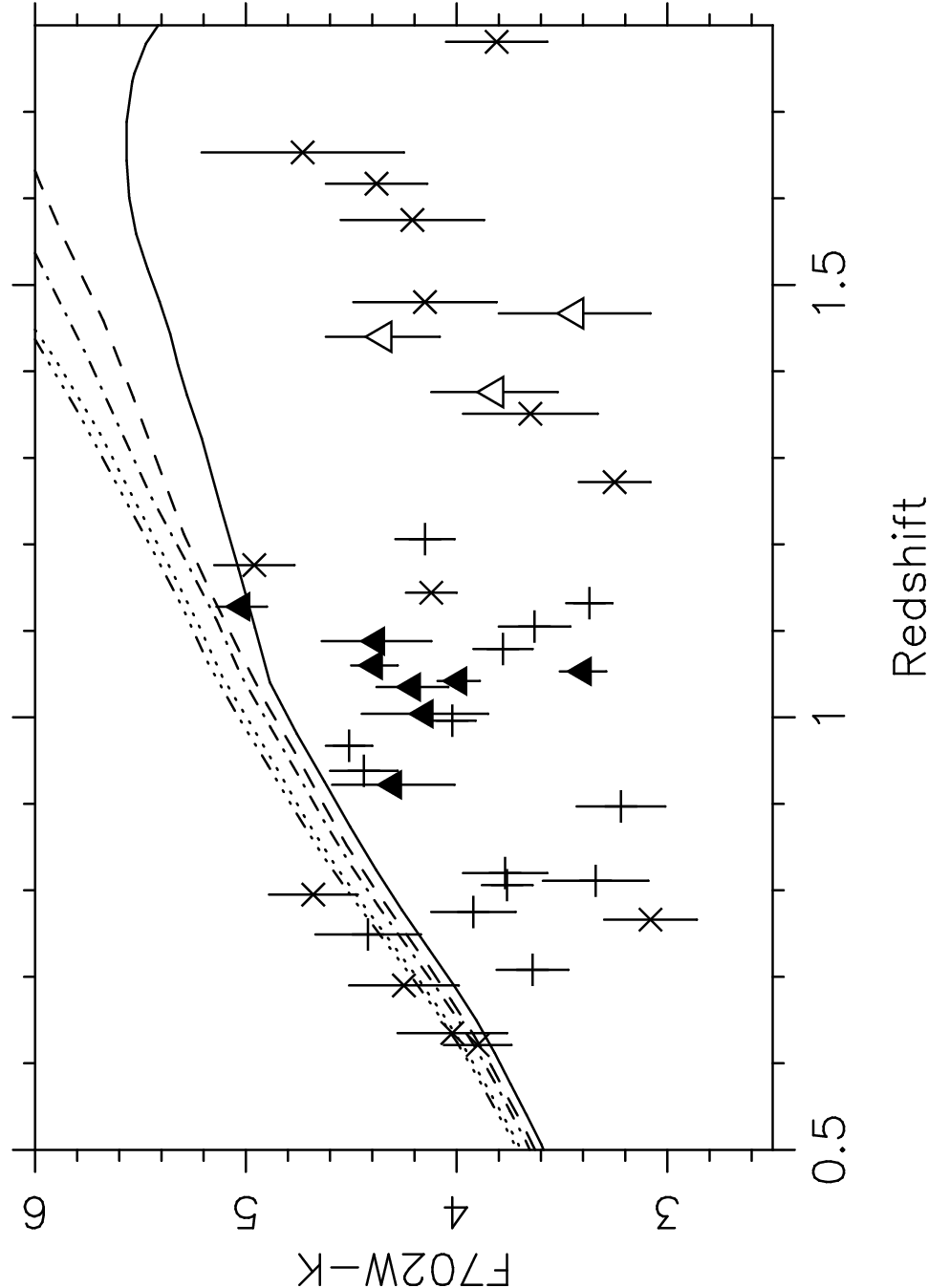}
\includegraphics{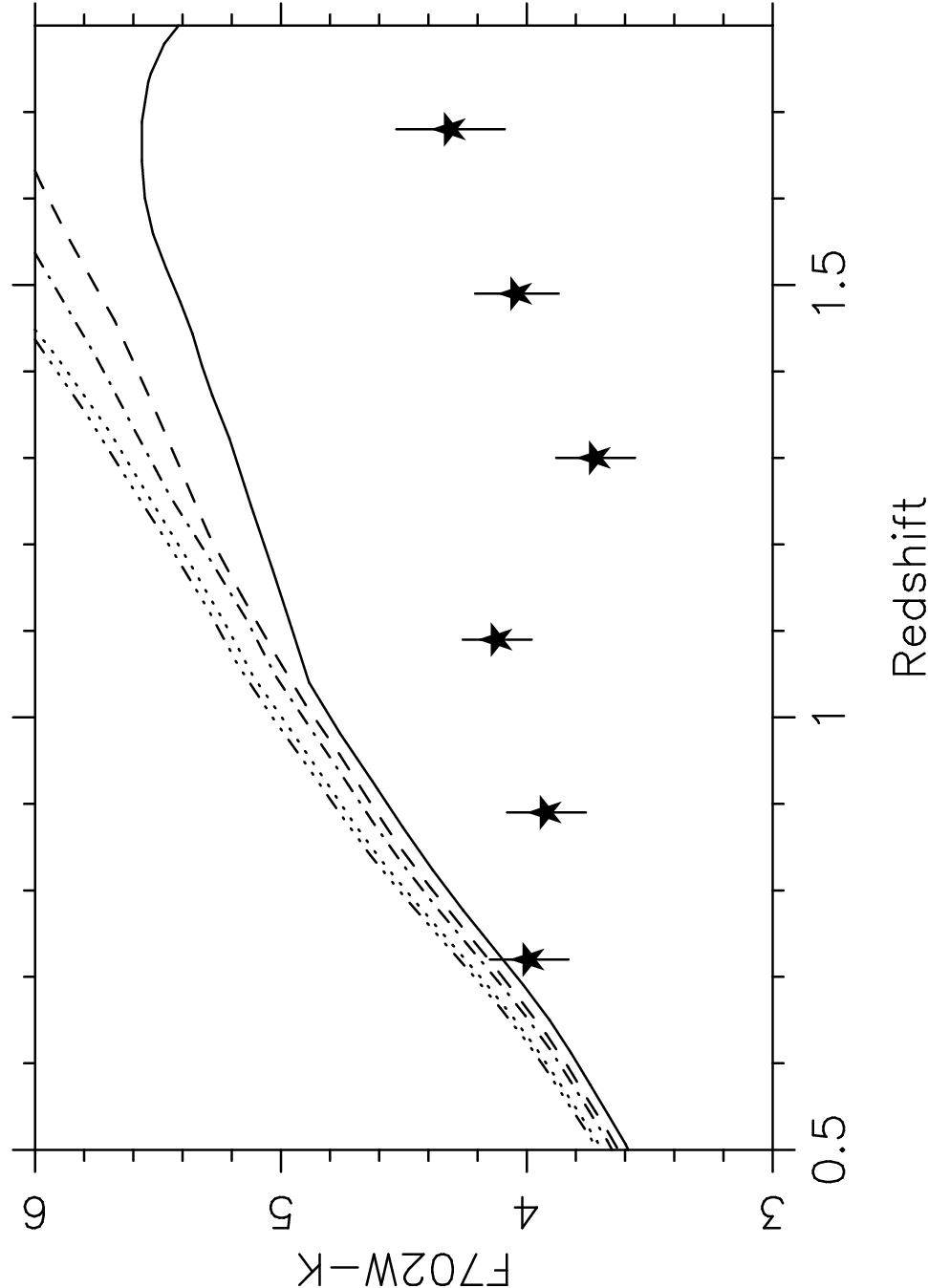}
\end{center}
\caption{Top row: Observed 4\arcsec\ F814W$-K$ (left) \& F702W$-K$ (right) colours
vs. redshift for 6C and 3CR radio galaxies. Middle row: 4\arcsec\ F814W$-K$ \&
F702W$-K$ colours after correcting for both $K-$band point source
contributions and also emission line contamination.  Bottom row: The
same (corrected) data
binned in equal redshift intervals.  6C sources are represented by the triangles: filled
triangles represent the eight sources in the $z \sim 1$ spectroscopic
sample.  3CR sources are represented by crosses: the sources in the
3CR $z \sim 1$ spectroscopic sample are marked by `$+$', with the
remaining galaxies marked by `$\times$'.  Error bars
represent the standard error on the mean value.  The tracks represent
the F814W$-K$ \& F702W$-K$ colours for passively evolving elliptical
galaxies formed at redshifts of 3 (solid line), 4 (dashed line), 5
(dot-dashed line), 10 (dotted line) and 20 (dot-dot-dot-dashed line).  
\label{Fig: 7_3}}
\end{figure*}

Fig.~\ref{Fig: 7_3} displays the uncorrected F814W$-K$ and F702W$-K$
colours (derived directly from the data in Table 1) for the 6C and 3CR
sources as a function of redshift, the 
galaxy colours after removal of the emission line contributions in each
filter (see Table 2), and the corrected data binned in equal redshift intervals. The
model tracks were created using the spectral synthesis models of
Bruzual and Charlot (2003), and represent passively evolving old
stellar populations which were formed in an instantaneous burst at
redshifts ranging from z=3 to z=20.
Correcting for line emission does not greatly alter the overall picture.  The
colours of the 6C sources with the weakest alignment effect
(e.g. 6C1019+39 at $z=0.922$ and 6C1256+36 at $z=1.128$) display
corrected colours consistent with passively evolving old stellar
populations (i.e. the tracks plotted on Fig.~\ref{Fig: 7_3}); similar
results are observed for the 3C sources with a weak 
alignment effect.  In all
other cases, the remaining sources retain their blue colours after
emission line correction, suggesting that line 
emission does not dominate the alignment effect.  The blue colour of
6C1217+36 (z=1.088), which has 
also has no obvious alignment effect, can be attributed to a clear UV
excess lying within the bounds of the host galaxy (Inskip et al 2005).

A further complicating factor is the fact that several sources are
known to have a fairly strong contribution to the 
flux in one or more filters from an unresolved nuclear point source.
Corrections for any nuclear point source contribution have not been
made, as the data are not sufficiently deep in all filters for accurate corrections for
nuclear point source emission to be carried out for all sources in all
filters.  For the sources with a measured point source contribution in
more than one filter, it is possible that a power-law model could be
applied in order to derive the point source contribution in other
filters.  However, such an approach would be fraught with error for
the sources with large errors on their point source contributions, or a
measured point source in only one filter, and would not be possible at
all for the many sources for which such measurements have not been
possible.  Additionally, it is quite plausible that any unresolved nuclear
point source may be due in part to recent star formation in the
central regions of the host galaxy (cf. the 4kpc diameter ring of young stars 
observed in Cygnus A (Fosbury et al 1999)), thus rendering any attempt at
such modelling useless at this stage.  Whilst we cannot fully correct
for any point source contribution, it is of importance to hold such
effects in consideration throughout the analysis of this data.
For the majority of sources, the point source
contributions are either relatively weak, and/or comparable in all
filters in which they were measurable.  However, stronger nuclear
emission ($> 30$\% in one or more filters) is observed for several
sources (6C0943+39, 6C1100+35, 3C22, 3C41).  For these sources, we
cannot accurately gauge the influence of nuclear point source emission
on the resulting galaxy colours. In order not to bias our results,
these four sources are excluded from any statistical
analysis, as well as the histogram plots of the galaxy colours
(Figs. 2 and 4). 

The data for both colours vary similarly with redshift.
At the lowest redshifts (where the alignment effect is generally
weaker), the galaxy colours are consistent with the 
predictions for passively evolving galaxies. Out to $z \sim 1.1-1.3$,
the passive evolution tracks become redder whilst the observed
galaxy colours remain roughly constant (as can be seen from the binned data
plotted in the bottom panels of Fig.~\ref{Fig: 7_3}). 
This reflects the increasing importance of the
excess UV emission associated with the alignment effect at
these redshifts, and indicates a clear evolution of the galaxy
properties over these redshifts. At higher redshifts, the average
colours then increase, and the galaxies become redder. 
In order to explain the variation of these colours at high
redshifts, it is useful to consider the changing rest-frame wavelengths
sampled by the HST/WFPC2 filters. At $z \sim 1$,  the rest-frame
wavelengths for the F702W and F814W filters are $\sim 3500$\AA\, and
$\sim 4000$\AA\, respectively; at $z \sim 1.5$ these become $\sim 2800$\AA\, for F702W and
$\sim 3200$\AA\, for F814W.
Although the old stellar population of a galaxy will be younger at $z
\sim 1.5$ than at $z \sim 1$ (and therefore bluer in colour), a galaxy at $z \sim 1.5$ is far less
luminous at the rest-frame wavelength of the observations than a
galaxy at $z \sim 1$: emission longwards of
the 4000\AA\, break
passes beyond the F702W filter at $z \sim 1$, and the F814W filter at
$z \sim 1.3$.  Given that the relative change in $K$--band flux is much
less, this explains the observed increase in the F814W$-K$ and
F702W$-K$ colours at $z \gta 1.1-1.3$. 

\begin{figure}
\vspace{2.35 in}
\begin{center}
\includegraphics{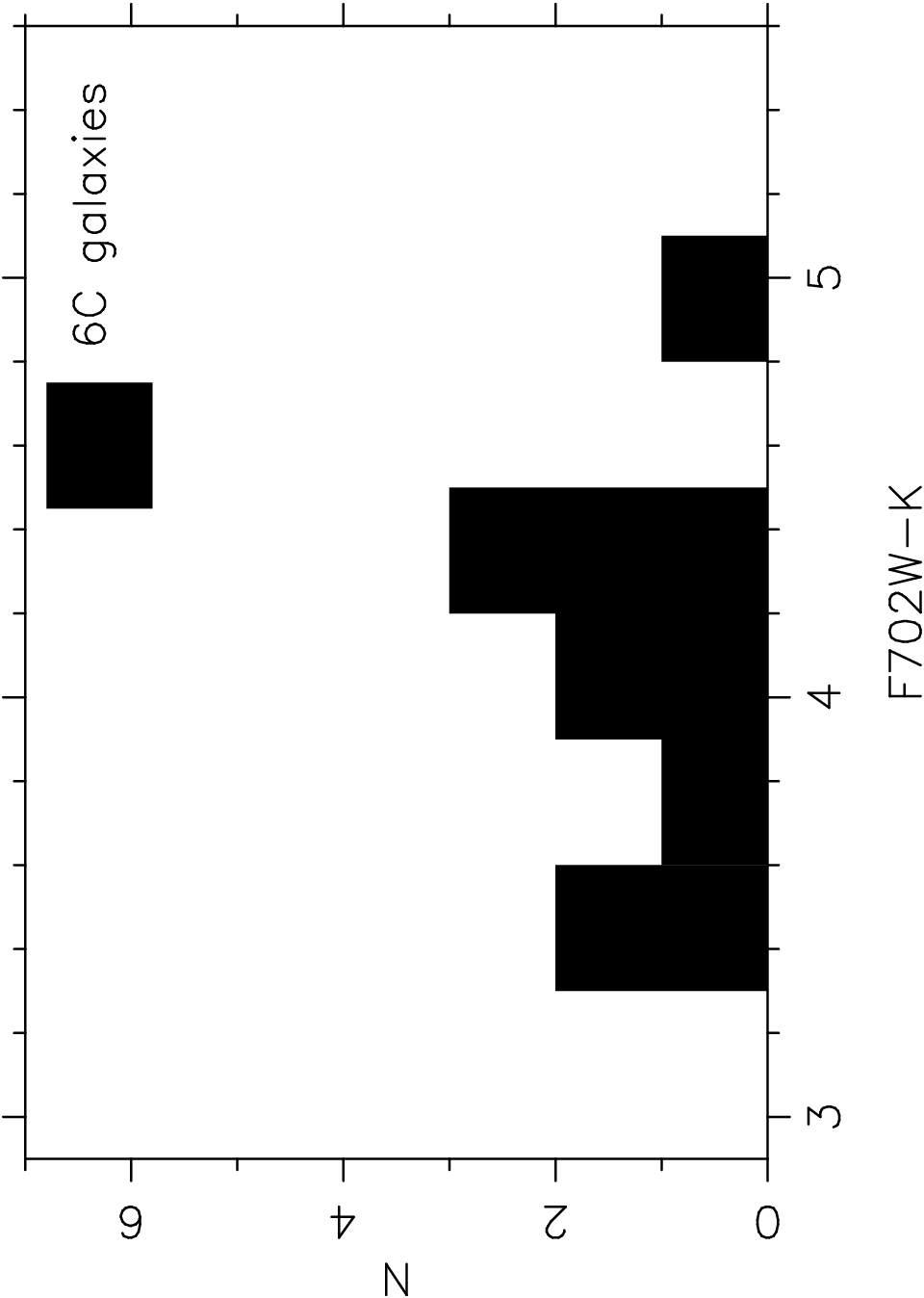}
\includegraphics{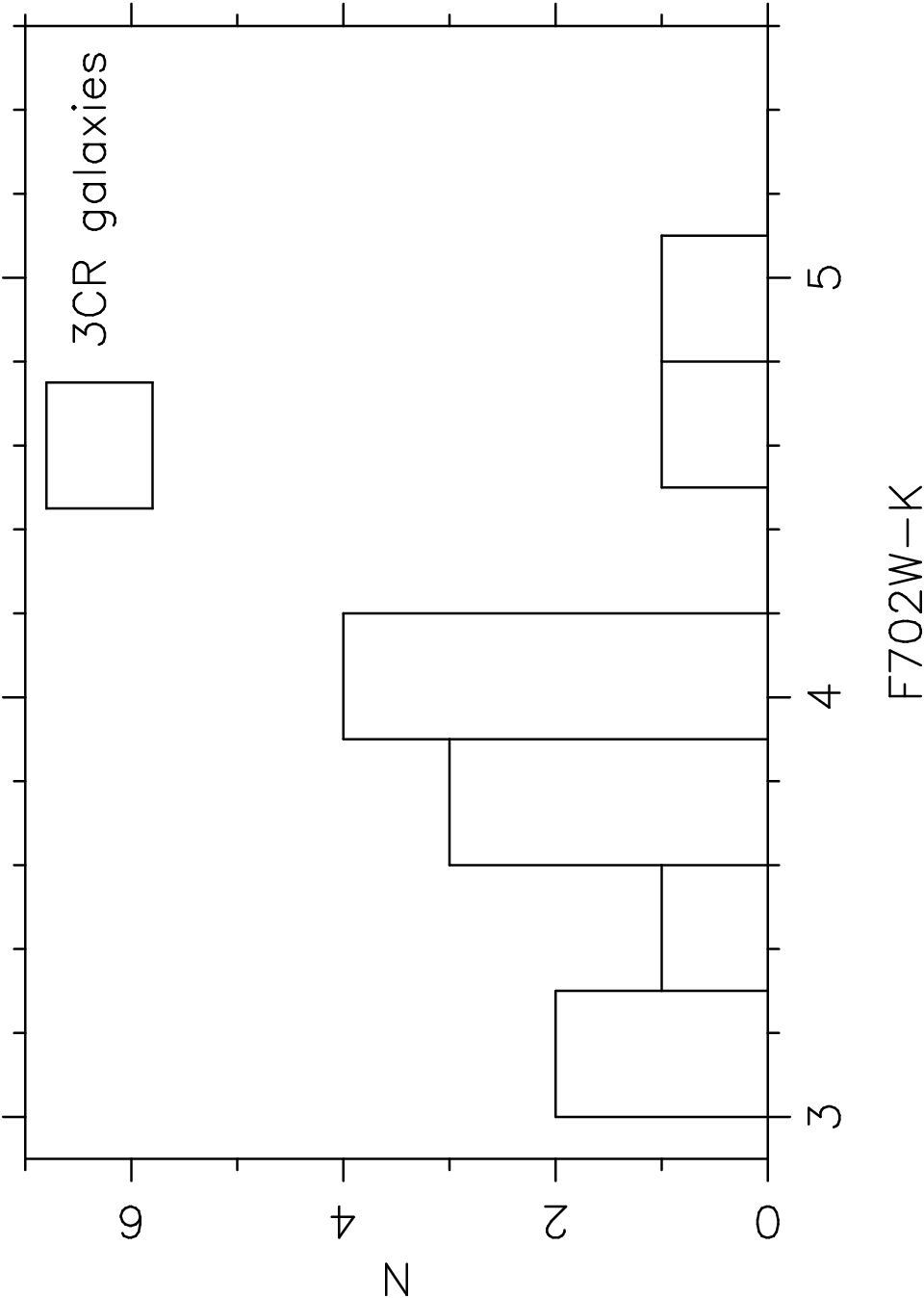}
\includegraphics{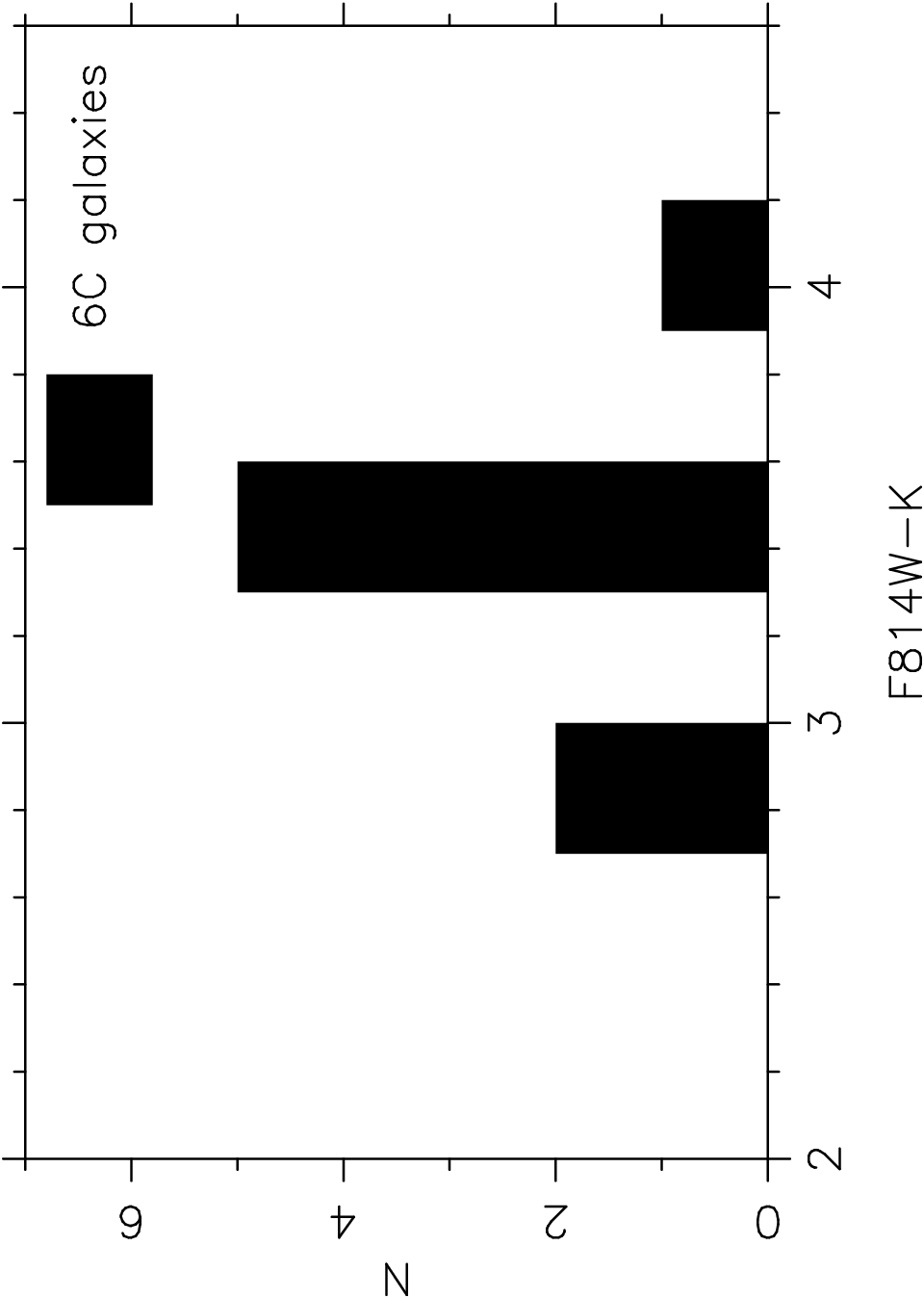}
\includegraphics{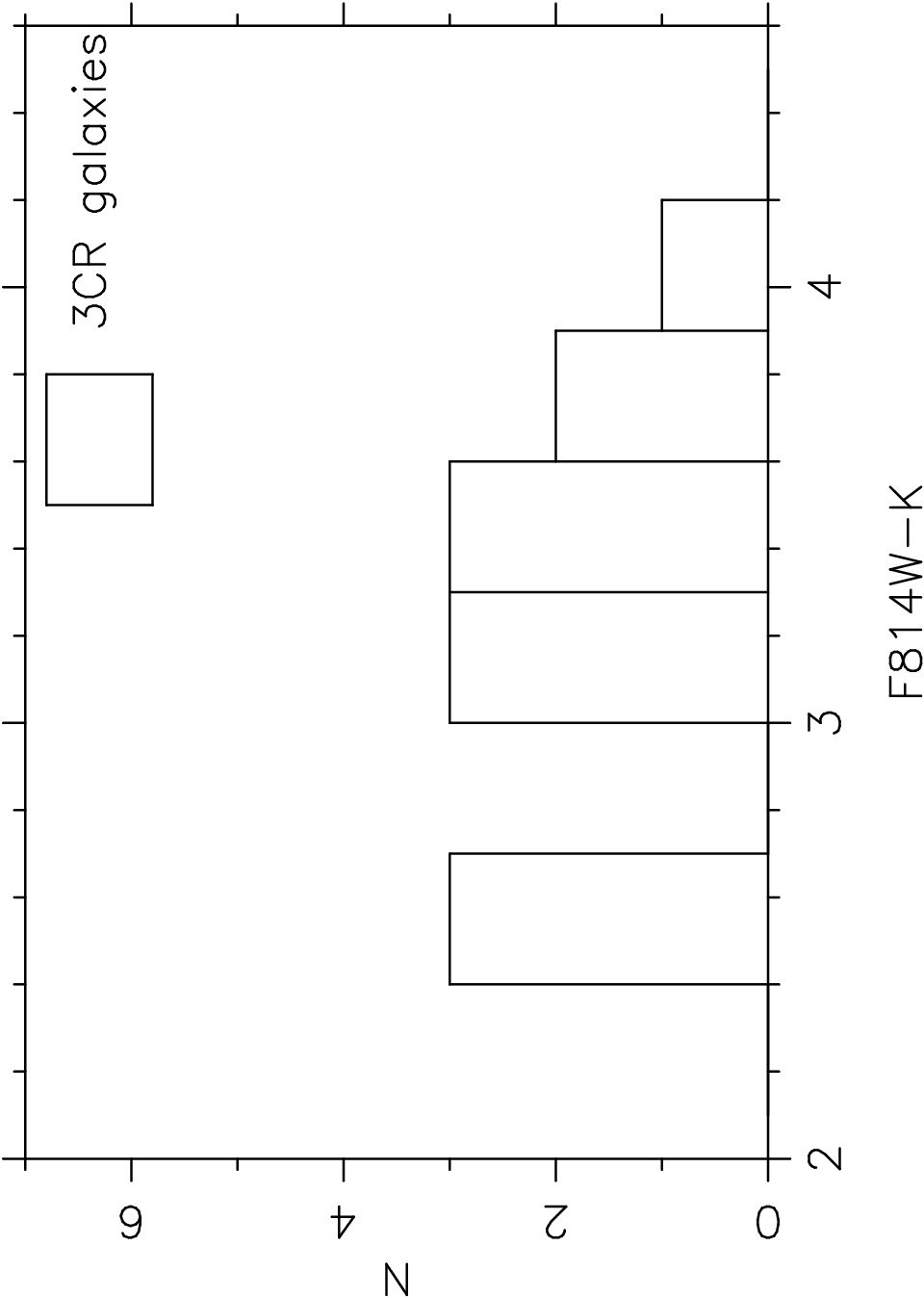}
\end{center}
\caption{Histogram of observed 4\arcsec\ F702W$-K$ (top) and F814W$-K$
(bottom) colours for the 6C (left) and 3CR (right) 
galaxies in the redshift range $0.85 < z < 1.5$.  The 6C sources are
represented by the black boxes, and the 3CR sources by the white
boxes. 
\label{Fig: 7_4}}
\end{figure}

Unlike the 3CR sources, the 6C galaxies are restricted to a narrower
redshift range closer to $z = 1$.  With the observed variation in
galaxy colours with redshift, any comparison between both full samples
would automatically lead to bluer colours on average for the 6C
sources.  To avoid any confusion, further analysis of the 3CR sources is restricted to
in the redshift range overlapping with the 6C subsample, $0.85 < z < 1.5$.
Fig.~\ref{Fig: 7_4} displays the distribution of the corrected F814W$-K$ and 
F702W$-K$ colours for the galaxies in both samples in this redshift
range as two histograms. 
 
The distribution of the F814W$-K$ colours appears to be similar for
the galaxies in both subsamples.  The mean value
of F814W$-K$ is slightly redder for the 6C sample by $\sim 0.25$
magnitudes.  The F702W$-K$ colours are almost identical for both
samples, with the mean 6C colour being $\sim 0.2$ mags redder than
that for the 3CR sources.  Kolmogorov-Smirnov tests show that the
typical galaxy colours of the two samples are not significantly different.
The fact that the two samples display such similar colours is
particularly surprising.  With so many alignment effect 
mechanisms closely linked to the properties of the radio source, the luminosity
of the extended structures at the wavelengths of the HST observations
might be expected to be greater relative to 
the host galaxy emission observed in the $K-$band for the more powerful 3CR radio sources,
leading to bluer F702W$-K$ and F814W$-K$ colours.   The
similarity in colours between the two samples suggests that this is
not the case, and that a more detailed consideration of the effects of
the radio source properties on the different types of aligned emission features is
required.  

\subsection{Galaxy colours in the infrared}

\begin{figure*}
\vspace{6.00 in}
\begin{center}
\includegraphics{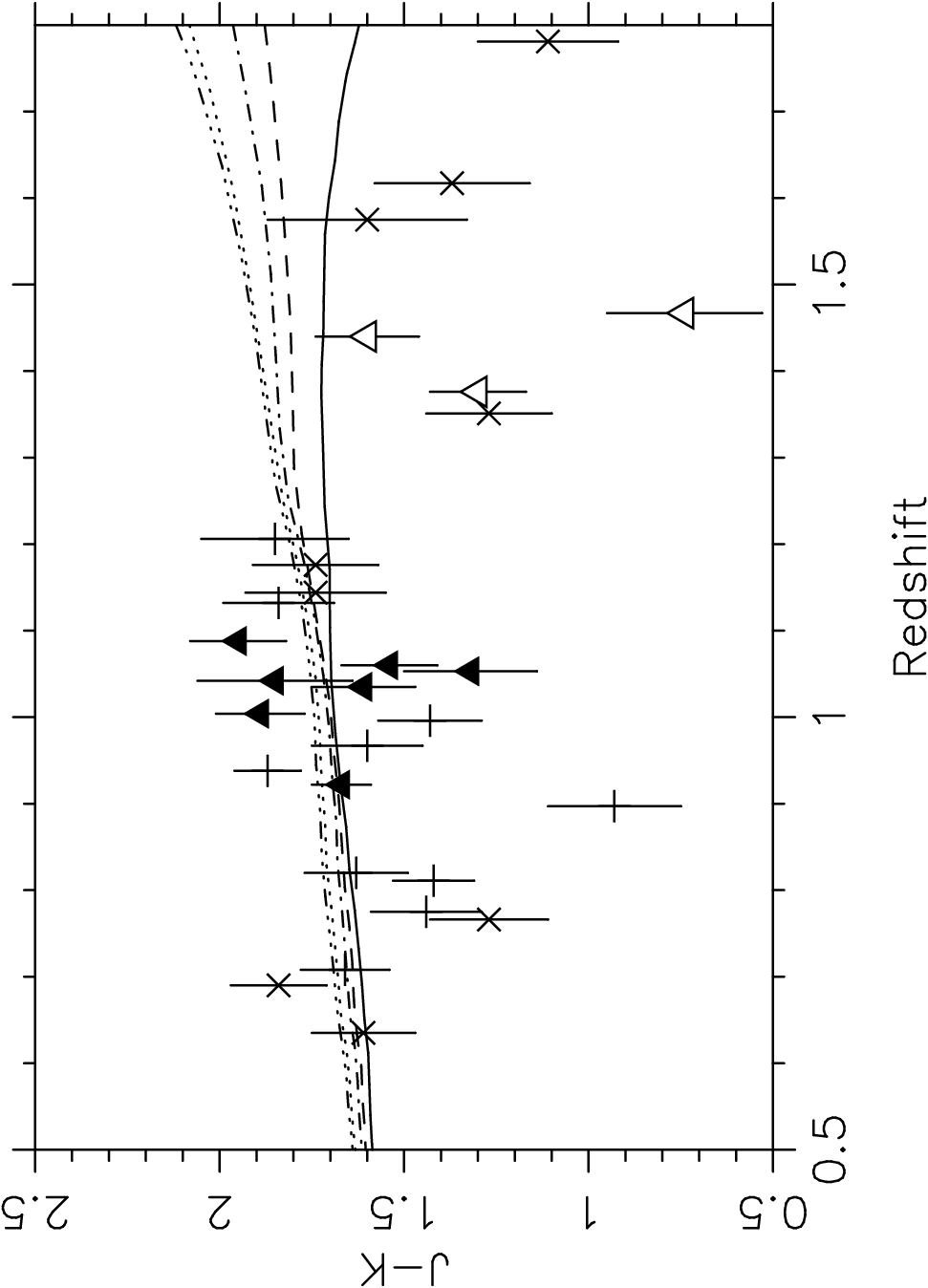}
\includegraphics{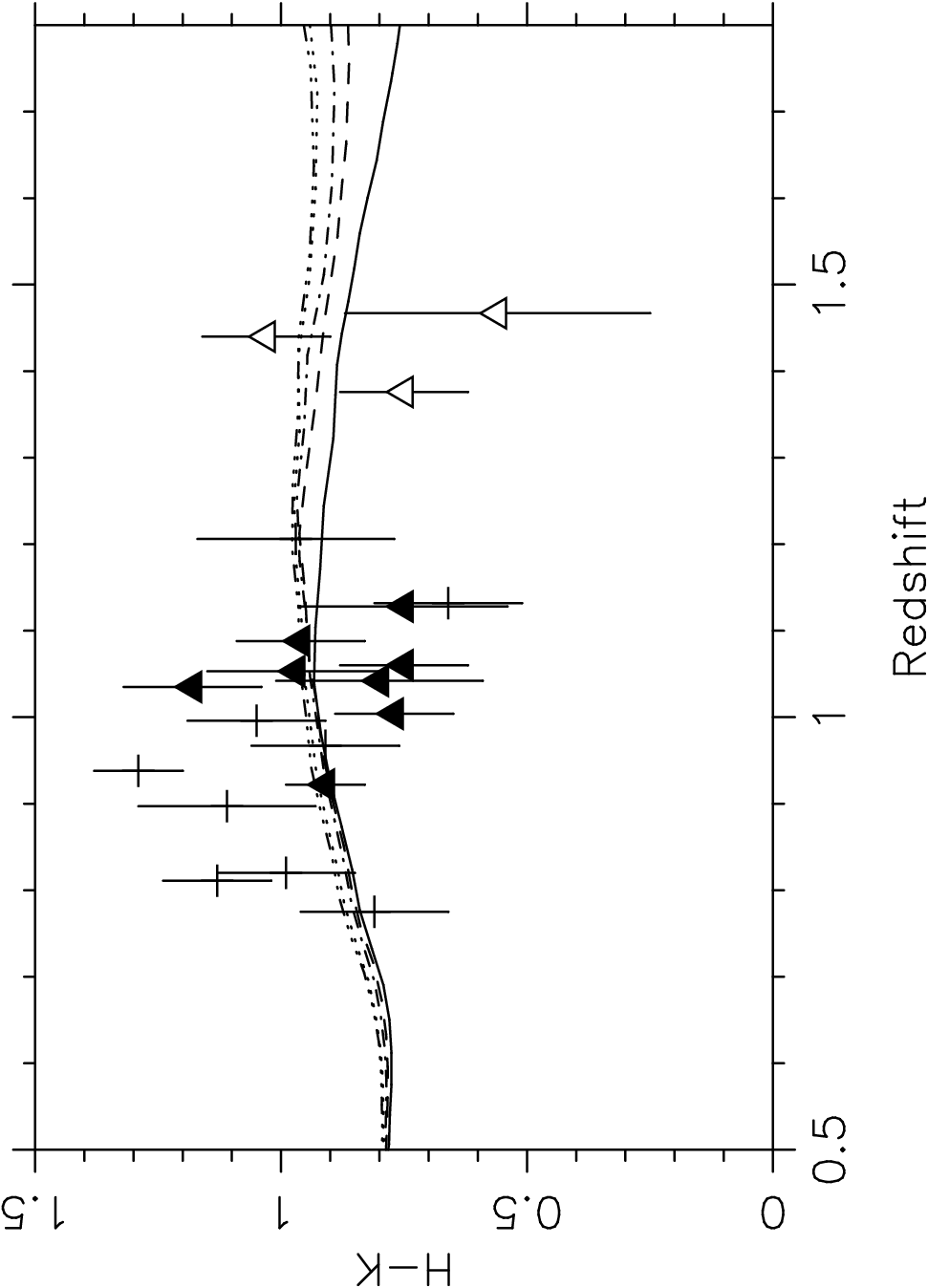}
\includegraphics{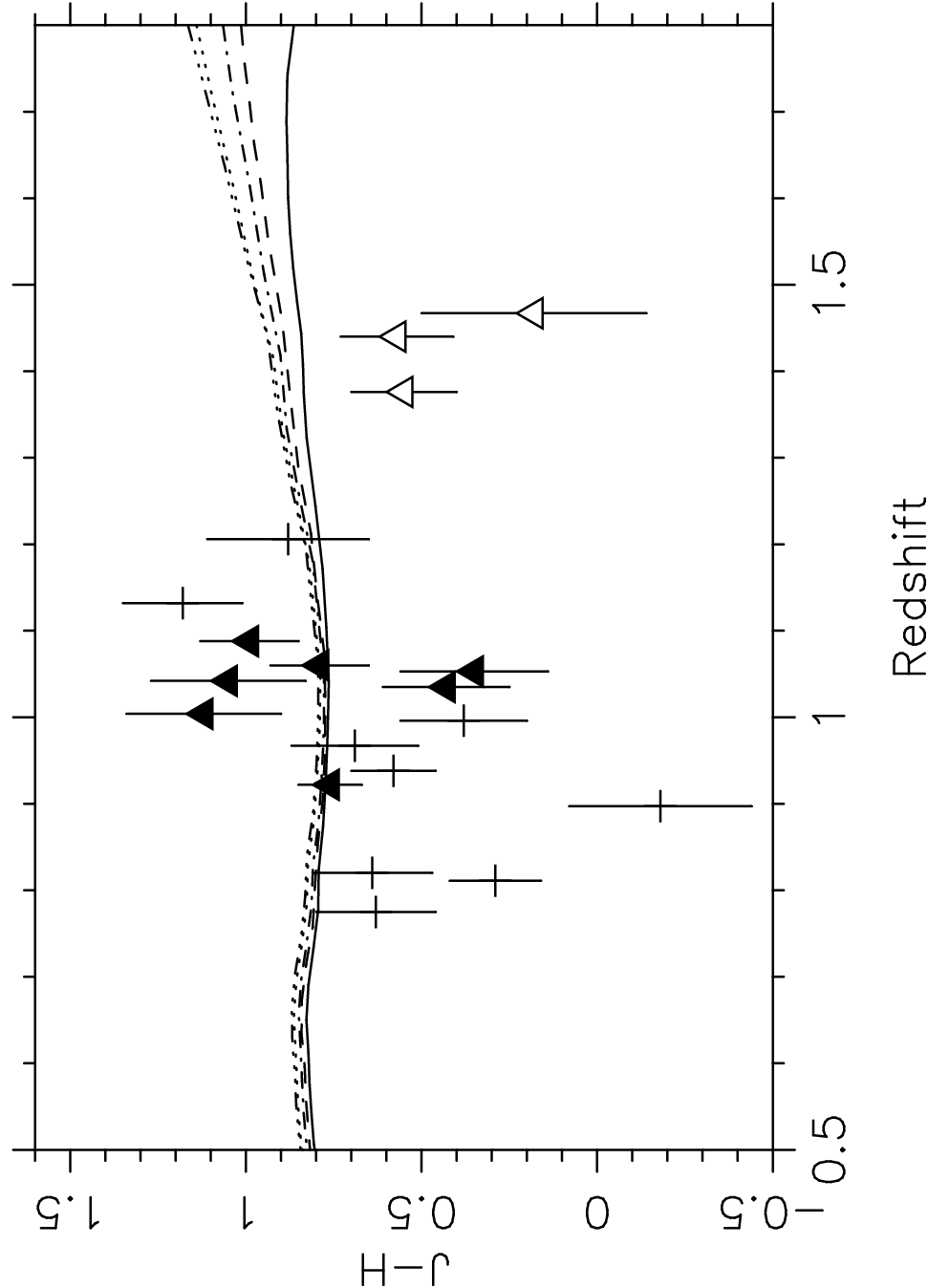}
\includegraphics{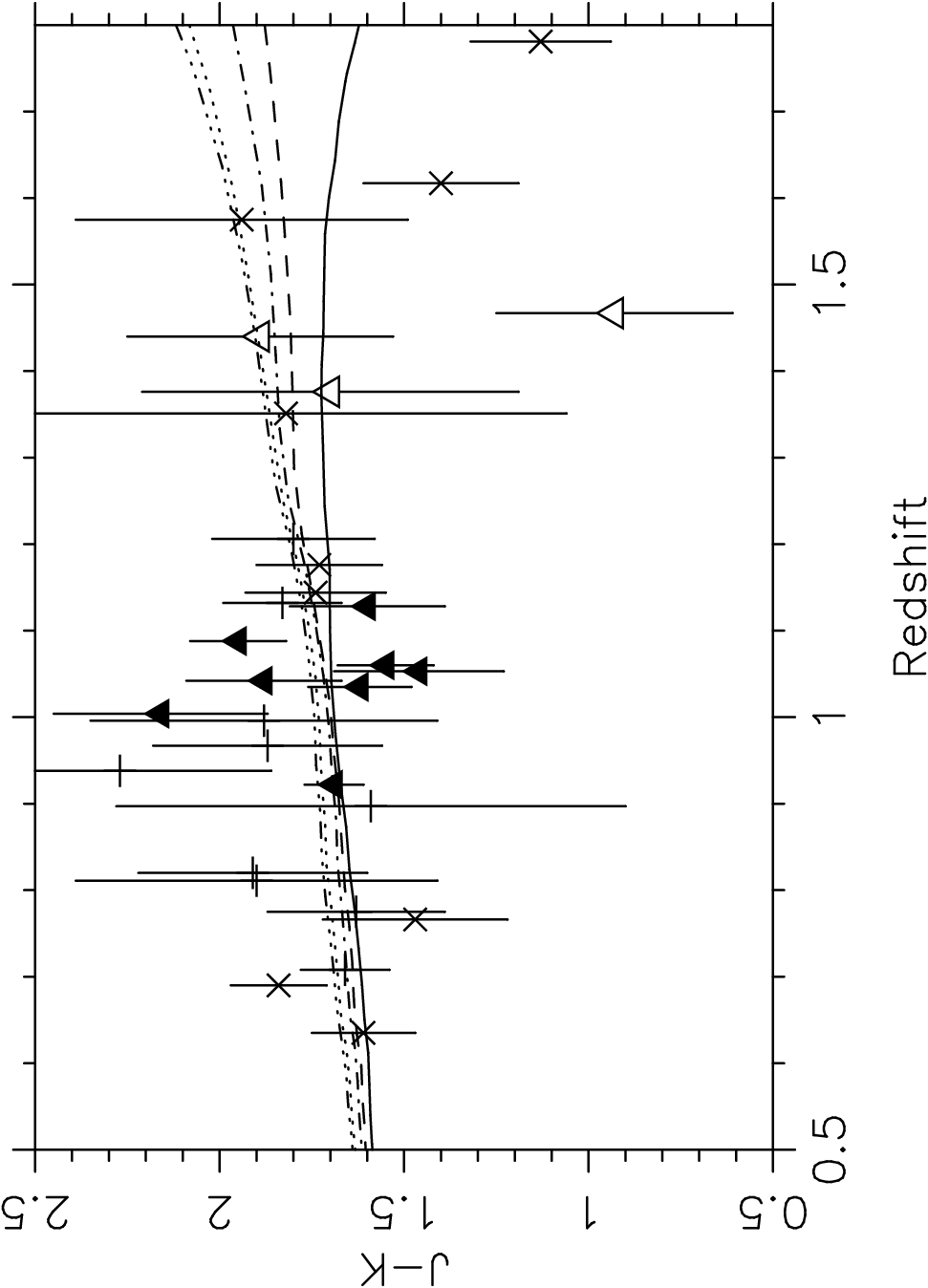}
\includegraphics{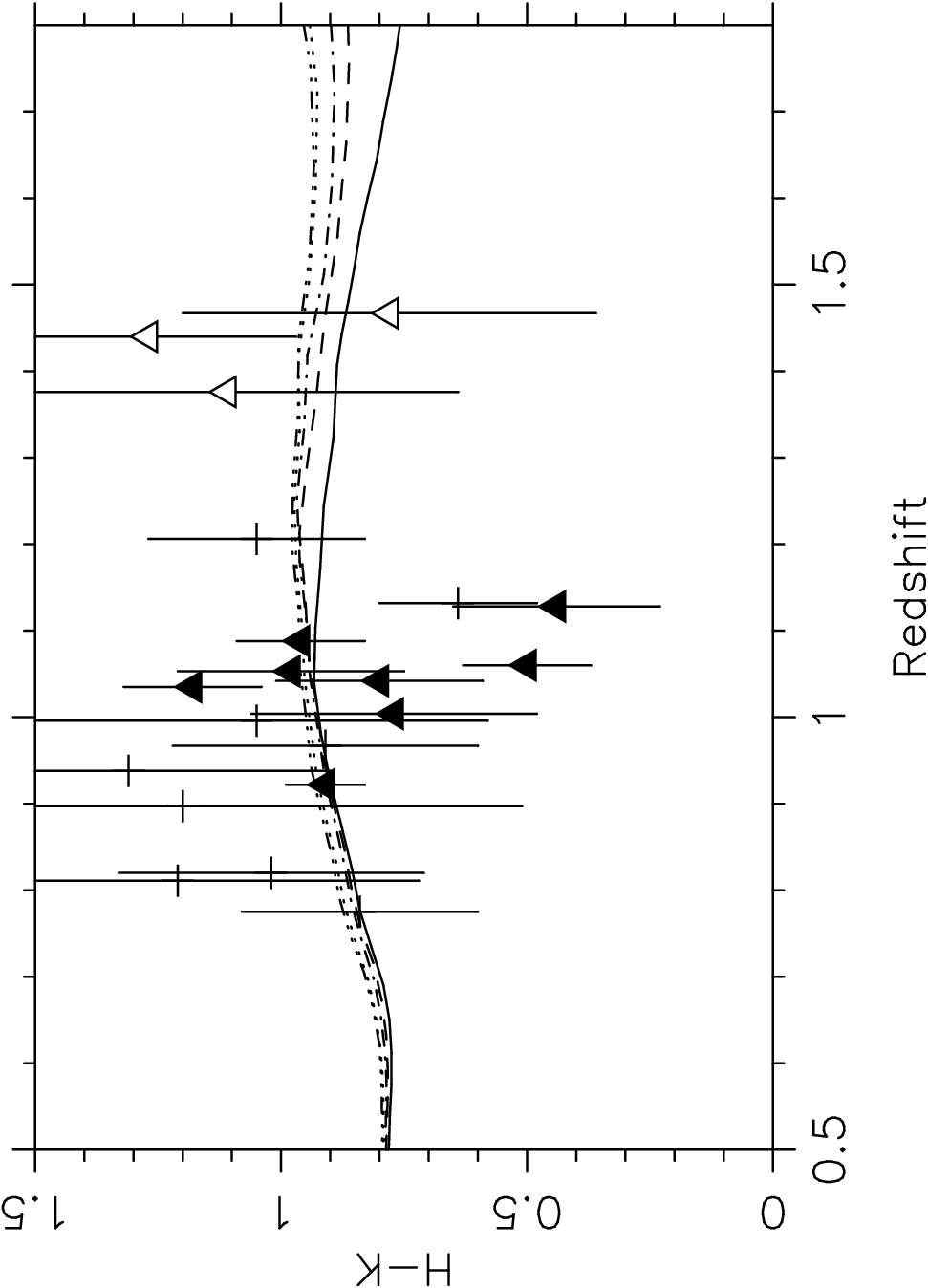}
\includegraphics{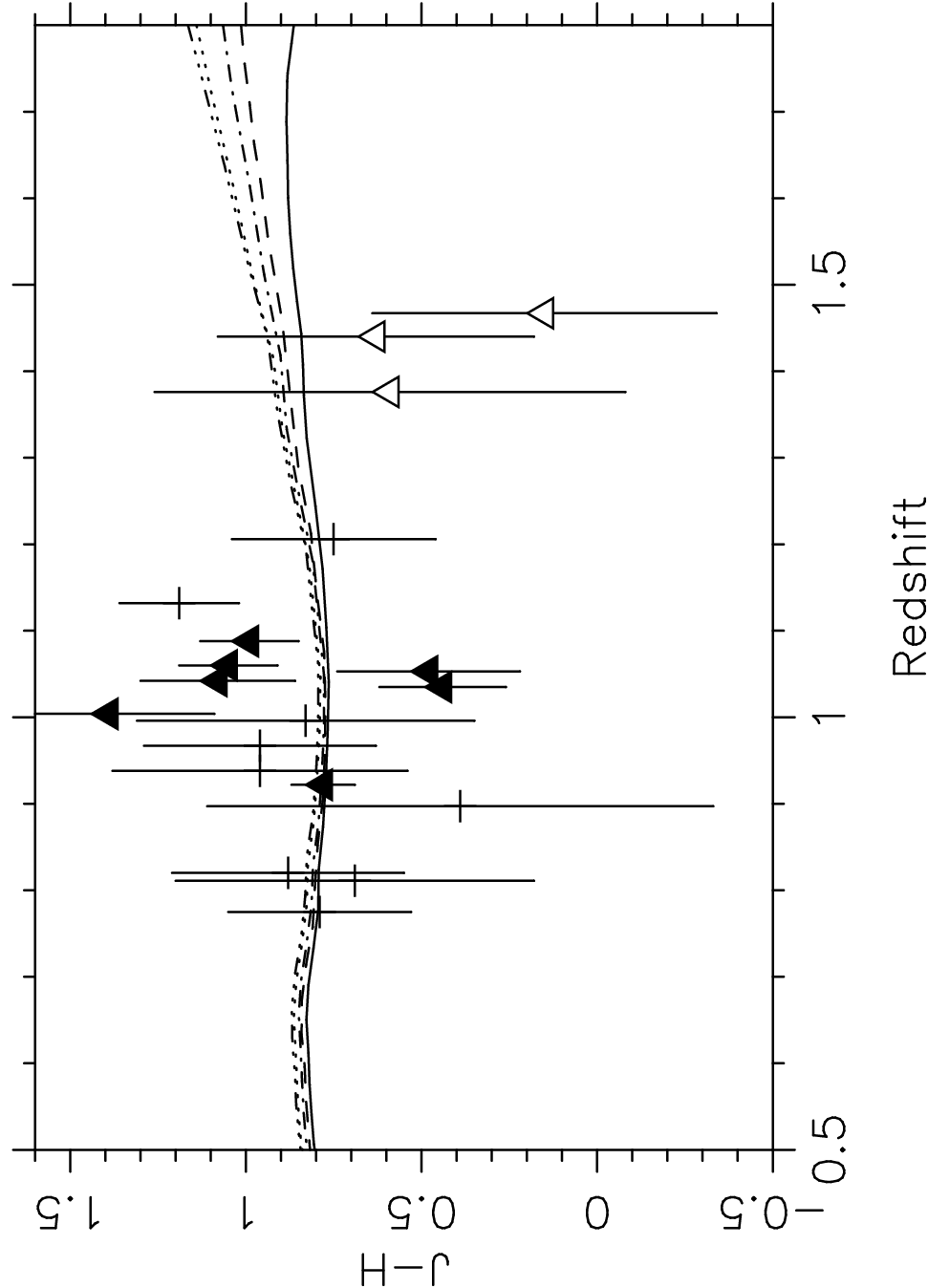}
\end{center}
\caption{Observed (left) and corrected (right) 4\arcsec\ $J-K$ (top),
  $H-K$ (centre) and $J-H$ (bottom) colours 
  vs. redshift for the 6C and 3CR galaxies at $z 
\sim 1$.    The tracks represent the $J-K$
colours for passively evolving elliptical galaxies, formed at
redshifts of 3 (solid line), 4 (dashed line), 5 (dot-dashed line), 
10 (dotted line) and 20 (dot-dot-dot-dashed line). Symbols are as
in Fig.~\ref{Fig: 7_3}. 3C41, which is thought to have a
point-source quasar contribution to its $K$--band emission (Best,
Longair \& R\"{o}ttgering 1997b), has an extremely red $J-K$ colour of 2.89 and lies
off the plot. 
\label{Fig: 7_1}}
\end{figure*}

\begin{figure}
\vspace{3.7 in}
\begin{center}
\includegraphics{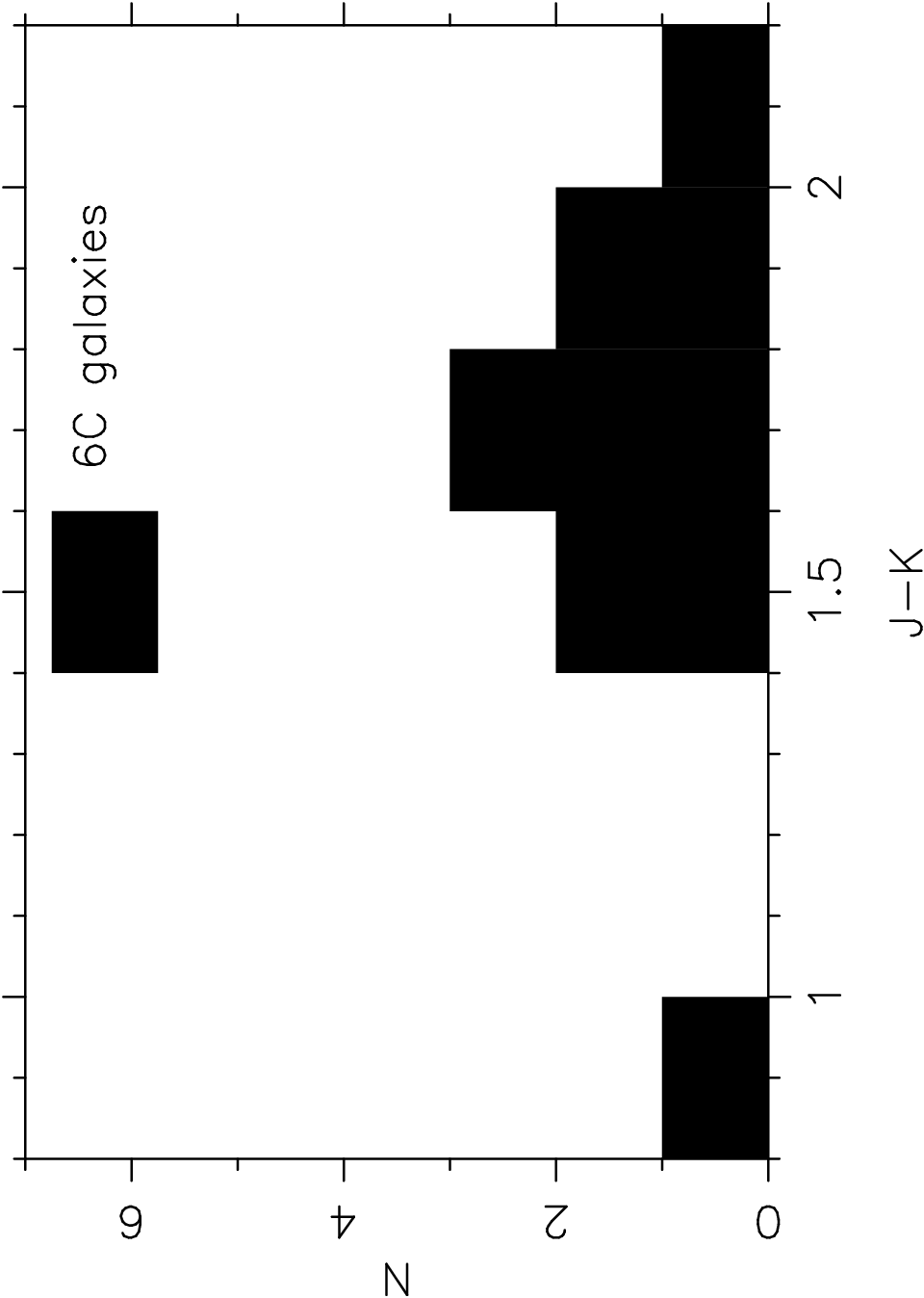}
\includegraphics{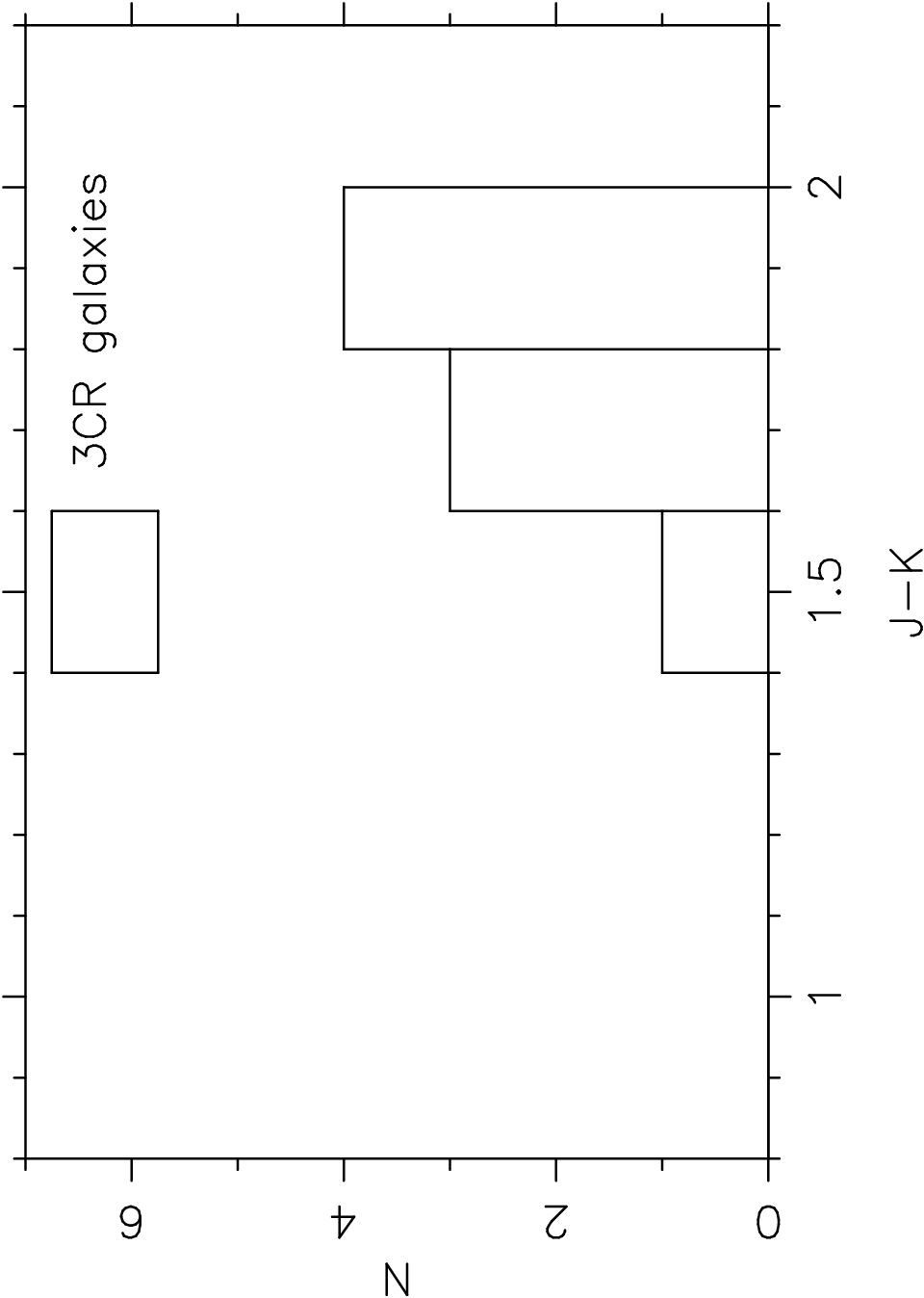}
\includegraphics{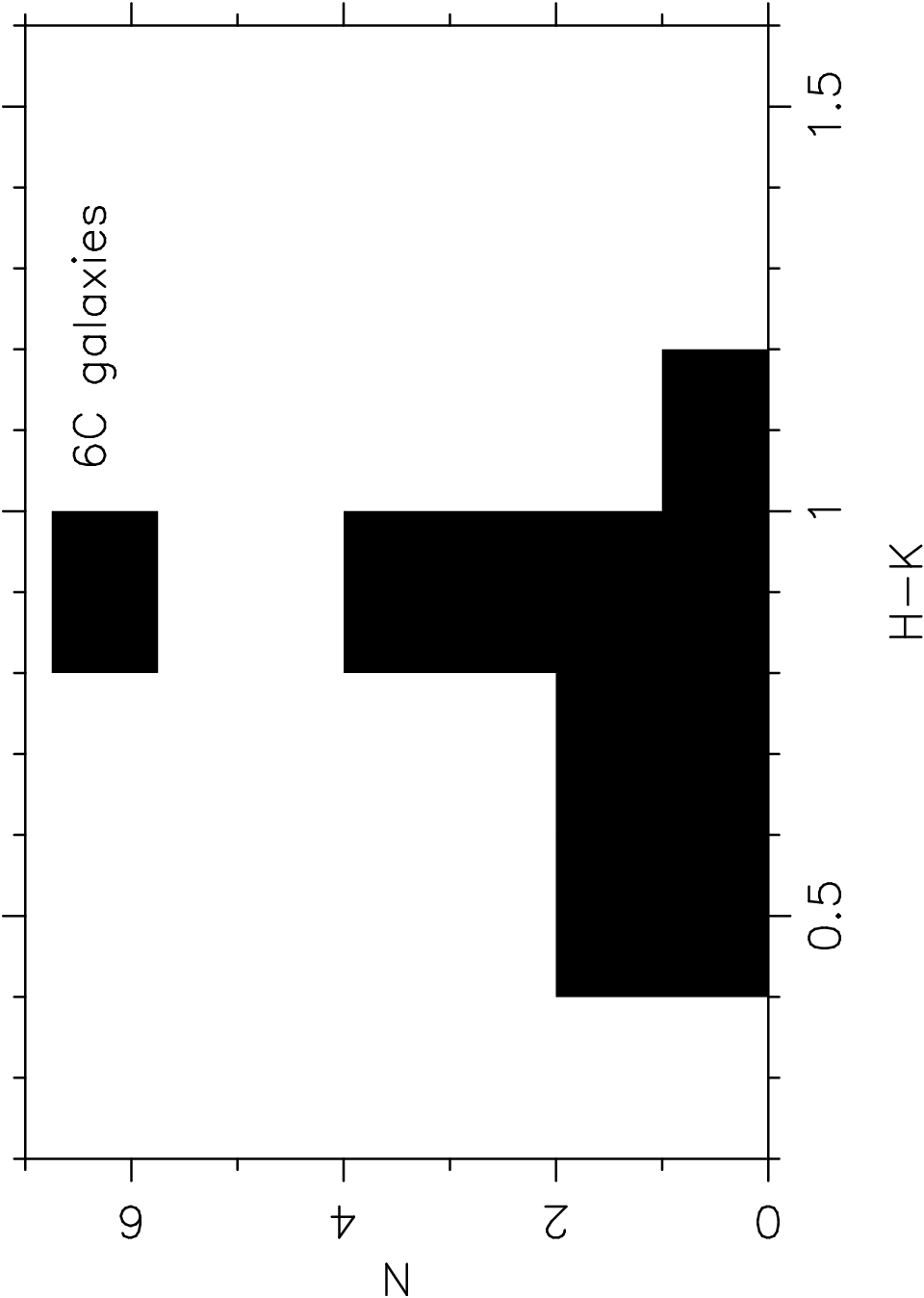}
\includegraphics{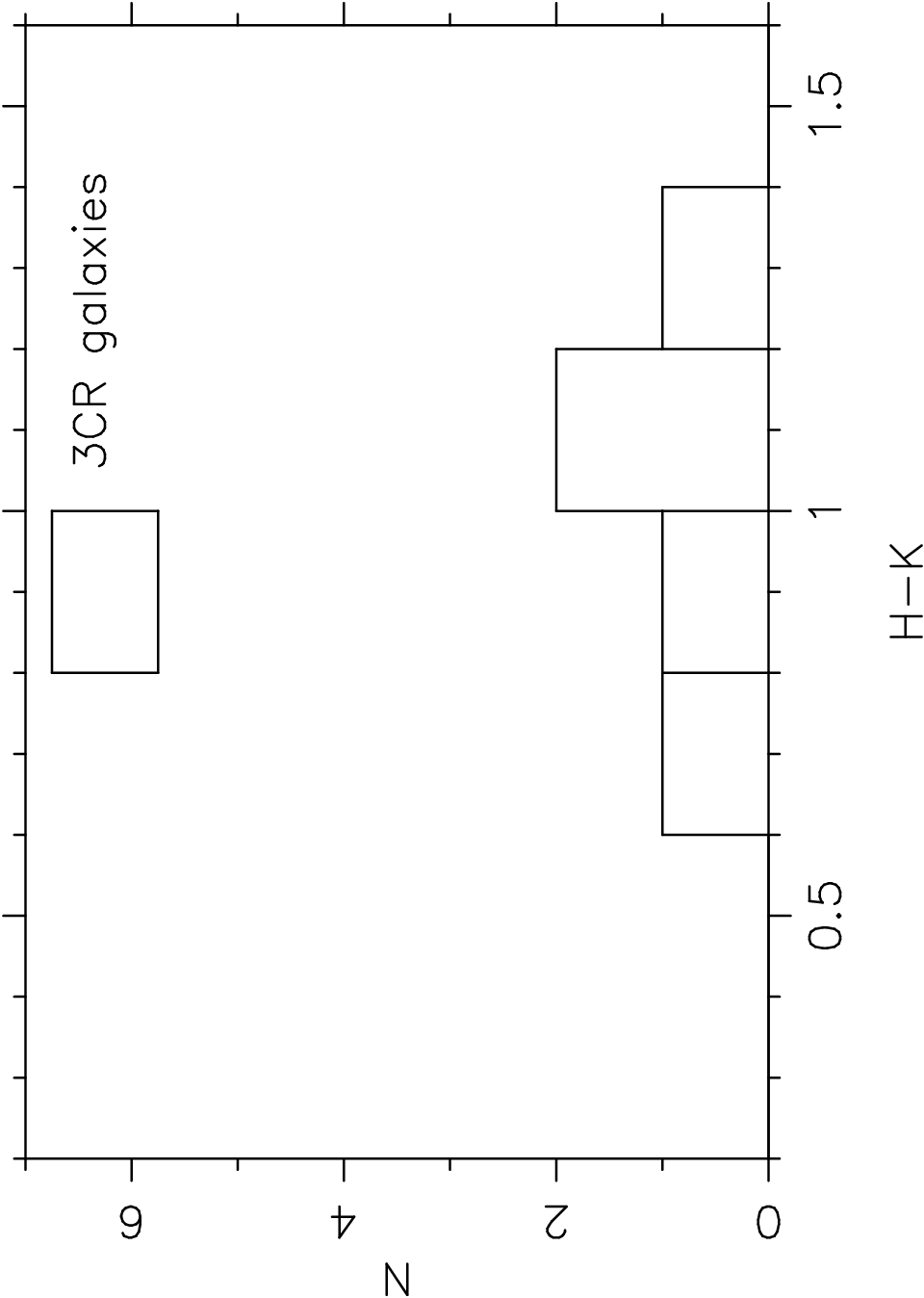}
\includegraphics{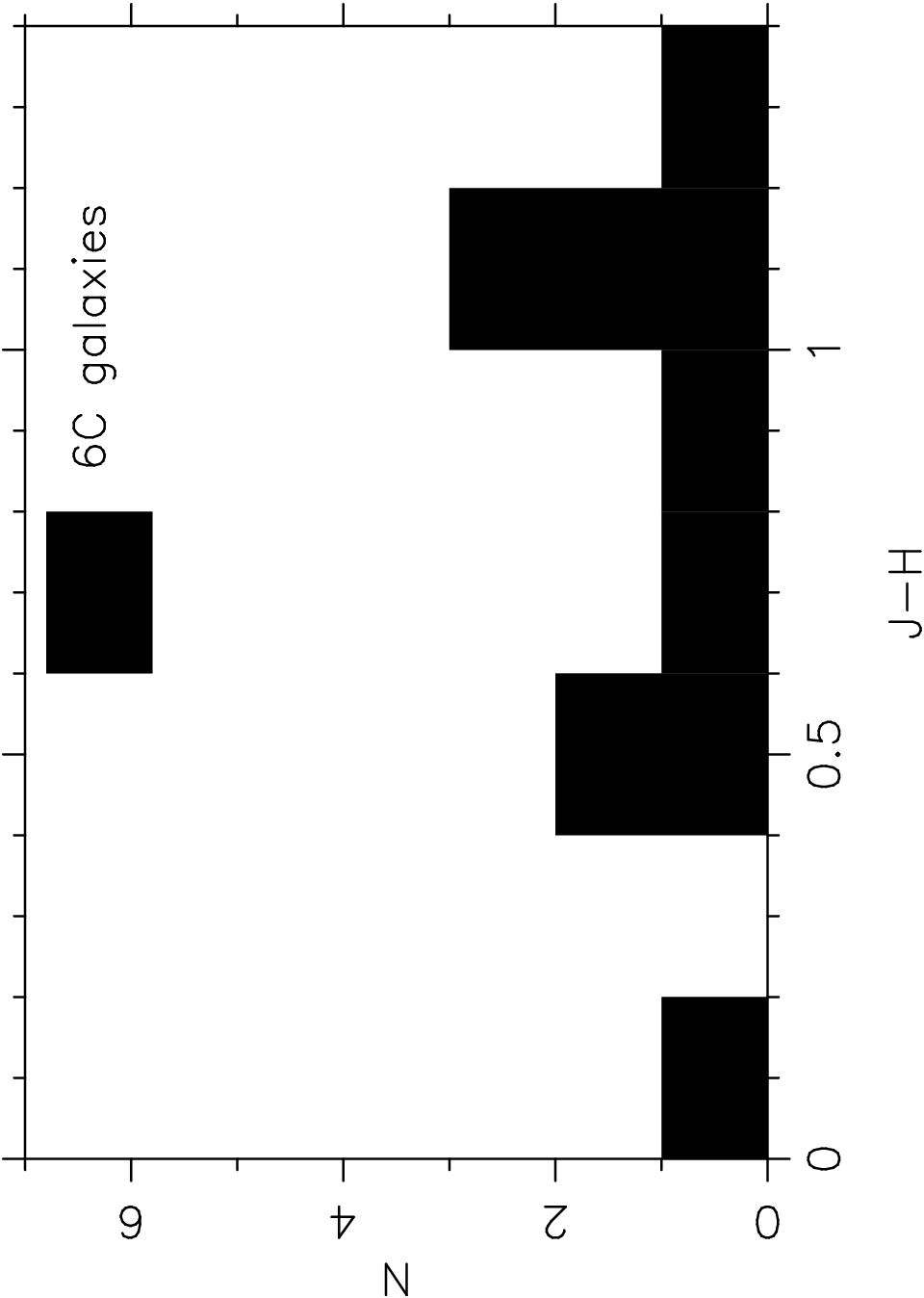}
\includegraphics{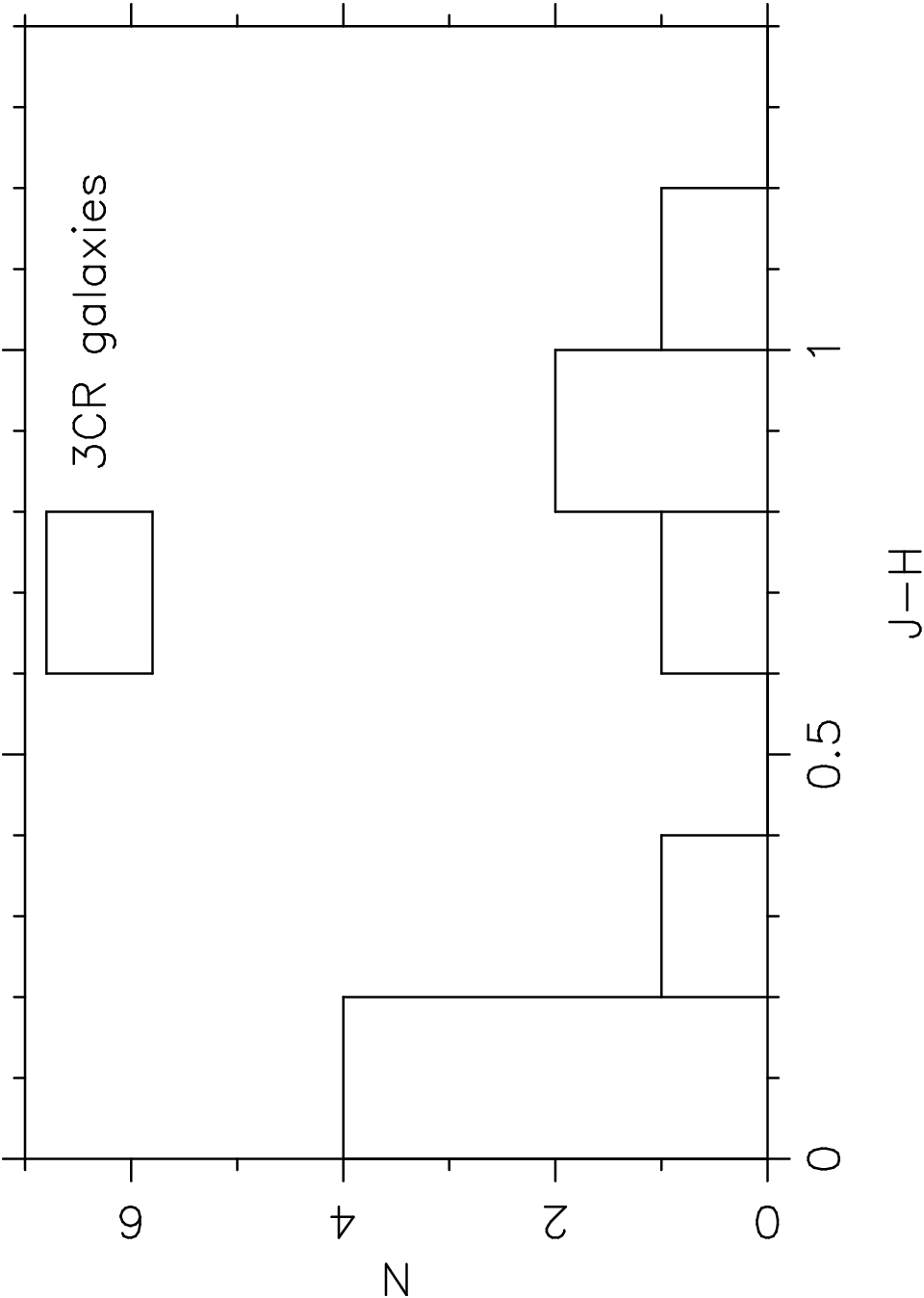}
\end{center}
\caption{Histogram of observed 4\arcsec\ $J-K$ (top) and $H-K$ (bottom) colours for the 6C (left) and 3CR (right)
galaxies in the redshift range $0.85 < z < 1.5$.  The 6C sources are
represented by the black boxes, and the 3CR sources by the white
boxes. Both subsamples display a similar range of colours.  
\label{Fig: 7_2}}
\end{figure}

In addition to our 6C data, $J$-band magnitudes are available for
twenty of the twenty--eight 3CR galaxies in the $z \sim 1$ subsample,
and $H$-band data is also available for 9 of these sources.  Figure
\ref{Fig: 7_1} displays the observed $J-K$, $H-K$ and $J-H$ colours
for the 6C and 3CR galaxies as a function of redshift together with
the tracks for passively evolving elliptical galaxies. We also display
the same data after correcting for the (estimated) emission line contribution.
Histograms displaying the variation in colour for each sample over the
redshift range of the 6C data ($0.85 < z < 1.5$) are displayed in
Fig.~\ref{Fig: 7_2}.  The data for both the 6C and 3CR subsamples
are clearly seen to span the same range of colours.  

The majority of infrared emission for galaxies at these redshifts
(excluding line emission) is believed to be
due to the old stellar populations of the host galaxies, and most
sources do indeed lie on or close to the tracks for passively 
evolving old stellar populations, as would be expected.  There is no clear difference between the 6C and 3CR colours, either
before or after emission line correction.  However, 
emission line correction does move the sources with the bluest
observed colours back towards the predictions of passive evolution
models.   Assuming that the stellar populations of the
radio source host galaxies all 
formed at approximately the same cosmic epoch, there is little
difference in the expected spectral energy distribution (SED) between
galaxies at $z \sim 1$ and those at $z \sim 1.5$ except that due to
their younger age, the more distant sources are expected to be 
intrinsically slightly brighter.  However, the change in rest--frame
wavelength for galaxies observed at these epochs is such that the more
distant sources will be observed to have a bluer $J-K$ colour, due to
an additional small contribution of aligned emission to the $J$-band
magnitudes of the higher redshift sources.  This is indeed the case
for our $J-K$ and $J-H$ colours at $z > 1.2$.  The long-wavelength tail
of the alignment effect will not contribute so strongly (if at all)
for emission observed in the longer wavelength $H$ and $K$ filters.

\section{Interpreting the galaxy colours}

Although the colours of the galaxies in the 6C and 3CR $z \sim 1$
samples are very similar, this certainly does not imply that they are
similarly luminous in any given waveband.  Indeed, it is clear that
this is not the case. In the $K-$band (which at $z \sim 1$ is
dominated by emission from the old stellar populations of the host
galaxies), the 6C sources are $\sim 0.6$ magnitudes fainter than the
3CR sources at the same redshift (Inskip et al 2002a).  Similar trends
are apparent from our shorter wavelength HST observations (Inskip et
al 2003).  The similarity in observed infrared colours, regardless of
luminosity, is to be expected: at these wavelengths, the emission from
the galaxies in both samples is dominated by their passively evolving
old stellar populations.  However, their observed optical colours are also
statistically identical. This result is more surprising, as the more
powerful 3CR sources might be expected to display a stronger alignment
effect (and therefore bluer colours).  There are two possible
explanations: either the alignment effect is not radio power
dependent, or some additional process is working to counterbalance the
expected bluer colours of the 3CR sources.  In this section, we will
investigate the likely causes of this result.

When considering the differences between different galaxy populations,
it is of essential importance that all possible sources of bias are
considered.  This is particularly true for radio galaxies, where the
properties of the radio source are often closely tied to those of its
host galaxy.  Environmental effects
may be an important factor in determining the overall size and luminosity
of a radio source of a given power. Observational evidence suggests
that the very largest ($>1$Mpc) radio sources are expanding more
rapidly than average into a below-average density IGM (e.g. Cotter
1998, Mack et al 1998), and may be of similar age to more typical
sources of several hundred kpc in extent.  Evidence for intermittency
in the radio source activity for these largest radio sources (which, given their age, may be most
likely to display it) is rare (e.g. Lara et al 2004).  For
radio sizes typical of those in our $z \sim 1$ subsamples, spectral
and dynamical ages are in good agreement up to ages of order of $10^7$
years (e.g. Blundell \& Rawlings 2000), but the argument that a larger
source size corresponds to a greater age, while not a 1:1
correspondence, at least holds true as a solid first
approximation for sources over the full range of sizes.  At the
smallest end of the scale, whilst not 
all compact radio sources (i.e. CSS/GPS sources) will necessarily evolve into larger FRII
radio sources, it is most likely that these are genuinely young
objects (e.g. Fanti et al 2000, Tzioumis et al 2002, Ojha et al 2004)
rather than older sources ``frustrated'' by a dense ISM 
 (O'Dea 1998 and references therein). 

Within a single flux limited sample, the higher redshift 
sources are not only more powerful (due to Malmquist bias), but also
have a smaller average size (and therefore younger average age).  This
is due to the fact that radio luminosity decreases with source size;
sources which meet the selection criteria for the sample when small
will not necessarily do so at larger radio sizes (Neeser et al 1995;
Kaiser, Dennett-Thorpe \& Alexander 1997; Blundell, Rawlings \&
Willott 1999).  As a consequence of this, the largest radio sources in
any given redshift range will host (on average) intrinsically more
powerful AGN than their smaller counterparts. We bear these issues in
mind as we consider the variations in galaxy colours as a function of
radio source size and power.

\subsection{The importance of radio source size}
\begin{figure}
\vspace{6.0 in}
\begin{center}
\includegraphics{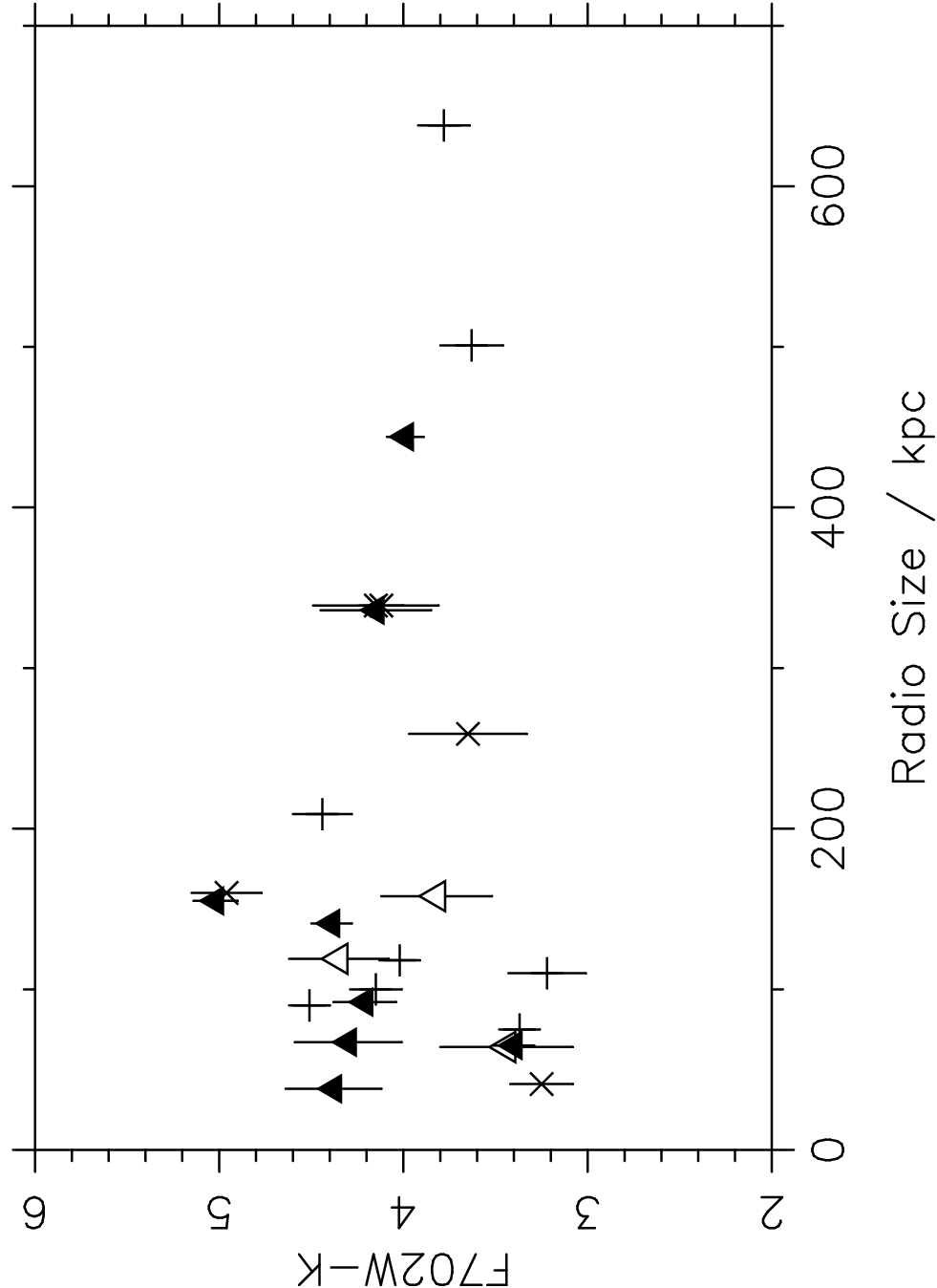}
\includegraphics{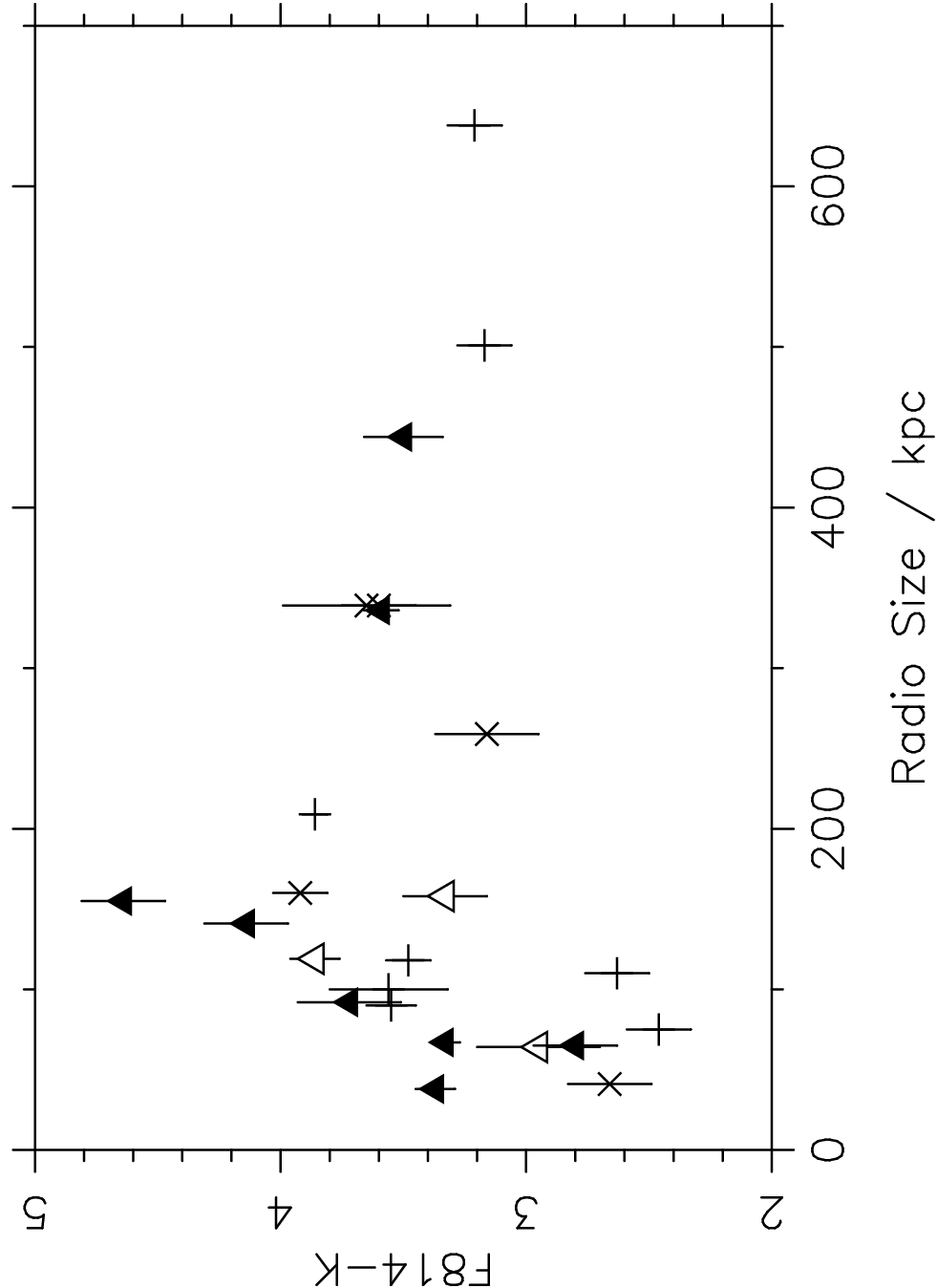}
\includegraphics{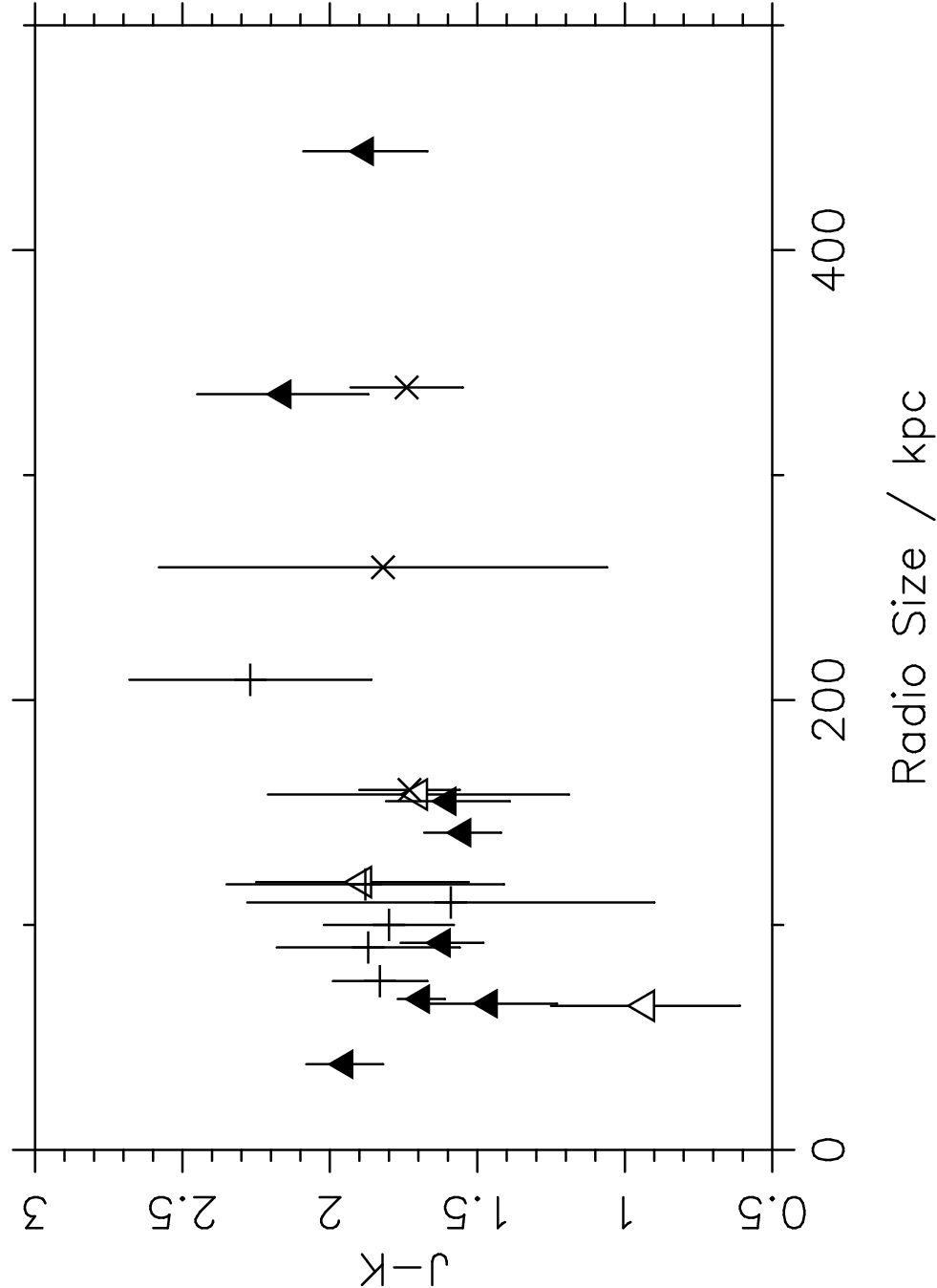}
\end{center}
\caption{4\arcsec\ F702W$-K$ (a --top), F814W$-K$ (b -- 
centre) and $J-K$ (c --bottom) colours (corrected for line emission)  vs. radio size for 6C and 3CR
radio galaxies in the redshift range $0.85 < z < 1.5$. Symbols are as
in Fig.~\ref{Fig: 7_3}. 
\label{Fig: 7_5}}
\end{figure}

A possible explanation for the unexpectedly blue F702W$-K$ and
F814W$-K$ colours displayed by the 6C sources is that the average
projected radio size of the sources in the 6C sample ($\sim 150$kpc) is less
than that for the 3CR sources in the same redshift 
range ($\sim 230$kpc).  Given the observed trend for smaller radio
sources to display a more extreme alignment effect (e.g. Best, Longair
\& R\"{o}ttgering 1996; Inskip et al 2004), any correlation
between radio size and galaxy 
colour could then lead to a bluer average colour for the 6C sources
than would be expected if both samples had the same mean radio size.
The variation in these colours with radio source size is an essential
consideration; plots of the corrected colours vs radio size are
displayed in Fig.~\ref{Fig: 7_5}.
Also displayed in Fig.~\ref{Fig: 7_5} is the distribution of $J-K$
colours with radio size for the two samples.  These data do not show any
signifcant correlation with radio size; neither do our data in the other
IR colours. In general this is as expected,
as radio size should not have any influence on the properties of the
old stellar population. 

As already noted, the distribution of radio sizes within the two $z
\sim 1$ samples is quite different. Therefore, for a fair comparison
between the sources in the 6C and 3CR samples in this redshift range, it is
justified to first consider only the smaller radio galaxies, i.e. those with
a radio size $< 200$kpc, for which the two samples display a similar
distribution of projected sizes. There is no difference between the mean F702W$-K$
and F814W$-K$ colours in each sample over this range of radio
sizes. In this size range, excluding the
sources with large point source contributions, the data sets for both
colours (corrected for emission line contamination) are correlated
with radio source size at significance levels 
of $\sim97.5\%$ and $\sim 95\%$ in a
Spearman Rank correlation test for the F814W$-K$ and F702W$-K$
colours respectively.   The strong correlation between
galaxy colour and projected radio source size for these data is
matched by the behaviour of many other properties of these systems,
which display the greatest variety over this range of radio sizes
(Inskip et al 2002b, 2004).

However, a visual inspection  of the plots suggests that this trend
does not hold true out to the largest radio sizes within our sample,
and has certainly broken down for sources larger than $\sim 400kpc$.
Considering sources up to 400kpc in size, the corrected F814W$-K$
colours are found to be correlated with 
radio size at a reduced significance level of $> 97.5\%$, and no
statistically significant correlation is observed for F702W$-K$.
For the full sample, the correlation for the corrected F814W$-K$
colours is further weakened (with a significance level of $\sim 95\%$). 
It is also worth noting that the radio size 
correlations have an increased significance prior to emission line
correction; this is due to the known strong correlation between
emission line flux and radio source size in these sources (Inskip et
al 2002c). 

These results are generally exactly as expected;  smaller
radio sources typically display brighter, more extensive aligned
emission (Best R\"{o}ttgering \& Longair 2000b, Inskip et al 2003). 
As illustrated by the case of 6C1257+36 (Inskip et al 2003; {\it Paper
 I}), at the wavelengths of the HST observations the aligned emission
for sources of all radio source sizes is 
generally bluer than the host galaxy itself. 

There are no obvious reasons why the very largest sources ($D_{\rm
  rad}>400$kpc) in both $z \sim 1$ samples should have 
colours as blue as those observed, which goes against the strong trend
with radio size observed for the smaller sources.  One possible
  explanation is that these largest 
  sources may exist in a very much less dense IGM, and may have ages
  (and therefore colours) more comparable to the smallest sources in
  our sample.
 However, two of
these three sources (6C1011+36 and 3C356) appear to have one or more
nearby companions (Best 1996; Inskip et al 2003), as do several other
smaller radio sources which also display a more pronounced alignment
effect (e.g. 6C1129+37).  These factors could potentially lead to an
increase in the interactions between the radio source and the
surrounding IGM, above the level expected for the other large sources
($D_{rad} > 120$kpc) in the sample. Indeed, it has been suggested
  (Simpson \& Rawlings 2002) that the triggering of 3C356 may have
  been caused by a violent
  interaction with a nearby object. It can also be argued that in
  order for a radio source to survive at high luminosity to such large
  sizes, it may require a slightly denser IGM (e.g. Kaiser \&
  Alexander 1999) and therefore may be more likely
  to reside in a richer cluster environment, where interactions are 
  more frequent.  Whilst not the norm,
interactions are seen in many different sources, over a wide range of
redshifts and radio powers (e.g. Smith \& Heckman 1989; Roche \& Eales
2000; Lacy et al 1999), often resulting in bluer colours for the host
  galaxy than would be expected
given the predictions of passive evolution models.   The faint
companion objects nearby many of these blue/interacting systems often
seem to display a preference for lying close to the radio source axis
(e.g. Roche \& Eales 2000; Pentericci et al 2001; van Breugel et al
1998).  Rigler et al (1992) also find that small red companion objects
are often observed nearby powerful radio sources; although they find no
preferential location for these objects, they are generally associated
with the radio sources which display the bluest colours and the most
extreme alignment effects.   Although the most extreme star forming
features are only frequently observed at the highest redshifts, it is
certainly plausible that any merger related activity at lower
redshifts could easily boost any observed alignment effect above the
norm for the cosmic epoch of the source in question.  It does seem
likely that the anomalous sources within our own two samples also fit
this pattern, and suggest that environmental factors are of
  importance.  

\subsection{Variations with radio source power.}

By comparing galaxies of a similar radio size within a fixed redshift
range, it should be possible to cleanly deduce the influence of radio
source/AGN power on the observed properties of these complex systems.
The most plausible option for explaining the greater
luminosity of the more powerful 3CR sources is a greater mass for their host
galaxies.  Given the known correlation between black hole mass and
galaxy bulge mass (Kormendy \& Richstone 1995), a further correlation
between galaxy mass and radio source power might be expected, if the
black holes were being fuelled at the Eddington limit.  However,
whilst variations in mass may account for much of the scatter within
the two samples, analysis of the galaxy morphologies suggests that
radio galaxies are hosted by similar sized galaxies over a wide range
of redshifts and radio powers (Inskip et al 2005 and references 
therein).   

There are several other options by which this luminosity difference
between the samples may be explained.  Firstly, the more
powerful 3CR sources might be expected to display greater AGN
contamination in their broad band fluxes.  However, the average
unresolved point 
source contributions for the $z \sim 1$ 6C and 3CR samples are statistically
indistinguishable (Inskip et al 2005, paper 2).  Although this result is somewhat surprising, the
radio core flux fraction of the 6C data is higher on average 
than for the 3CR sources at the same redshift (Best et al 1999, 1997), suggesting that the
sources in our 6C subsample are typically observed at an angle closer
to the line of 
sight than the 3CR sources.  This would imply less obscuration of the
6C AGN, and therefore bluer colours/brighter emission than might
otherwise be expected. 
Overall however, AGN contamination cannot account for the luminosity
difference between the two samples, nor is it sufficient to explain
the similarity 
in colours between the 6C and 3CR $z \sim 1$ sources.

A second factor is the alignment effect itself, which is particularly
luminous in the case of the $z \sim 1$ 3CR sources.  The long
wavelength tail of the alignment effect is clearly visible in  
the $J$--band imaging observations of the 3CR data, which reveal a
definite contribution of emission from the extended aligned
features.   At a much lower level, the $K$--band emission of the 3CR
sources also displays some extension along the radio source axis. 
Such features are not observed in our images of the 6C sources,
although this could be due to the lower signal--to--noise levels of
these observations (Best et al 1996, 1997b, Inskip et al 2004).  
Whilst an alignment effect is only observed in the IR for the more powerful
3CR $z \sim 1$ sources, it is not likely to contribute more than $\sim
10$\% of the total $K$-band flux for these sources (Best et al 1998,
Rigler et al 1992, Zirm et al 2003). 
Therefore, excess emission from the long-wavelength alignment
effect cannot reconcile the differences in $K$-band emission for the
two samples, nor can it account for the fact that high redshift radio
galaxies are on average more luminous ($\sim 0.5$ magnitudes at $z
\sim 1$) than expected based on the predictions of
passive evolution models (Inskip et al 2002a).  The only
plausible options remaining are to also consider the effects of
variations in the radio source environment, and the host galaxy
stellar populations.  We will now discuss
the effects of each of these in turn.

\subsubsection{Host galaxy environment}

The environments of radio sources vary quite markedly with redshift.
Low redshift radio galaxies are predominantly found in small groups and
avoid richer cluster environments, whilst at higher redshifts, a wider
range of environments is observed, including richer groups and
clusters (e.g. Hill \& Lilly 1991; Allington-Smith et al
1993; Zirbel 1997).  The $z \sim 1$ 3CR sources in particular lie in
regions typically consistent with Abell class 0-1 clusters (Best 2000).
If the density of the IGM surrounding a radio source is correlated
with the richness of the local galaxy environment, this has
implications for the other observed properties of the $z \sim 1$ 3CR radio sources. 
Under the conditions of a higher density IGM, the radio source will be
subject to a somewhat greater level of confinement, boosting its
observed luminosity and potentially also restricting the rapidity of
its growth (e.g. Kaiser, Dennett-Thorpe \& Alexander 1997; Kaiser \&
Alexander 1999).  The extreme radio luminosity of the high redshift 3CR
sources could possibly be caused by a richer environment than average.
Given the limited differences between the properties of
the $z \sim 1$ 3CR and 6C host galaxies, it is worth hypothesising
that the factor of
six difference in their radio luminosities may in part be due to
environmental effects, in addition to other factors such as the
fuelling rate of the AGN.  Analysis of the environments of the 6C
sources at $z \sim 1$ suggests that they lie in Abell class 0
environments (Roche, Eales \& Hippelein 1998), somewhat less rich on
average than those of the 3CR sources at the same redshift. 

Other environmental factors may also be important. It is now becoming
increasingly likely that galaxy mergers/interactions are responsible
for the triggering of most (if not all) radio sources at some level.
The levels of gas involved in such a scenario, and therefore the
subsequent fuelling of the AGN, star formation in the host galaxy and
the density of the disturbed IGM surrounding the radio source, may all
be strongly dependent on the properties of the local environment and
the nature of the galaxies involved.  As was noted in the preceding
section, the emission from several of the larger 6C and 3CR sources at
$z \sim 1$ is particularly blue in colour, and comparable to that of
smaller sources which display a more extreme alignment effect.  It is
noteworthy that many of these sources also either display signs of
interactions, or that they have a larger than average number of
companion objects at a close projected distance; this may very well be
an example of a richer environment leading to the boosting of activity
in and around the radio galaxy, as well as maintaining relatively high
radio luminosities out to large radio sizes. Scattered emission, jet-cloud
interactions and jet induced star formation could all take place more
readily for a radio source in rich surroundings, leading to a stronger
alignment effect; this could help explain the differences in the
luminosity of the aligned emission observed between the powerful 3CR
sources at $z \sim 1$ and the less powerful 6C sources at the same
redshift.  These factors may also help explain why the alignment
effect is so much less extreme at lower redshifts for the sources in
both the 6C and 3CR samples: low redshift radio sources are
predominantly found in less rich environments, and the merger history
of the galaxies in question is likely to have been substantially
different to that of higher redshift sources.   

\subsubsection{Stellar populations of the host galaxies, and
  deviations from passive evolution}

\begin{figure}
\vspace{1.8 in}
\begin{center}
\includegraphics{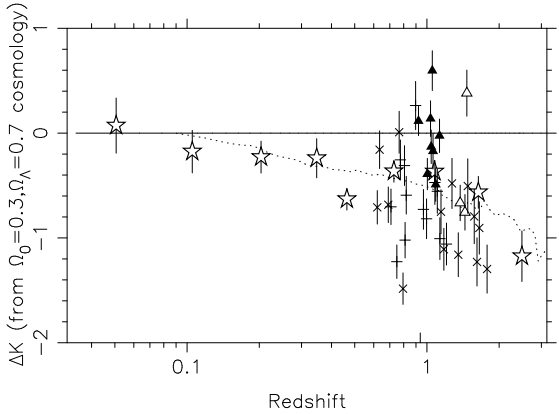}
\end{center}
\caption{Deviations from the predictions of passive evolution models.
  The solid line represents 
  the track for a passively evolving elliptical galaxy formed at
  $z=10$ in a cosmological model with $\Omega_0=0.3$,
  $\Omega_{\Lambda}=0.7$ and $H_0 = 65 \rm{km\,s^{-1}\,Mpc^{-1}}$, and
  the dotted line represents the same model with  $\Omega_0=1.0$,
  $\Omega_{\Lambda}=0.0$ and $H_0 = 50 \rm{km\,s^{-1}\,Mpc^{-1}}$. The
  models assume a fixed galaxy mass, normalised to fit the low
  redshift data.  The
  data points represent the magnitude difference between the passive
  evolution model and the observed data (such that negative values
  imply the galaxies are brighter than the model predictions): large
  stars represent the average $K$-band magnitudes for 57 6C galaxies
  with $z < 3.4$ and 72 3CR galaxies with $z < 1.8$, and the other
  symbols represent the data in the 6C and 3CR $z \sim 1$ subsamples
  as in  Fig.~\ref{Fig: 7_3}.  All 
  data are evaluated in a 63.9kpc aperture. There would be little
  change to this plot for a slightly later formation redshift of
  $z=5$. The discrepancy between models and data would be
  subtsantially less for a formation redshift of $z=3$, but other
  observational factors (e.g. red colours and the presence of radio
  sources at $z \gta 5$) disfavour this scenario.
\label{Fig: 7_6}}
\end{figure}

The final factor which requires consideration is the nature of the
host galaxy stellar populations.  Although the radio source host
galaxies appear to be well-behaved de Vaucouleurs ellipticals (Inskip
et al 2005), with $K-$band emission dominated by an old passively
evolving stellar population, this may not be a complete picture.
There is increasing evidence that the host galaxy population for
powerful radio galaxies evolves considerably with redshift.  In the
currently favoured cosmological model, the infrared
magnitude--redshift relation for 6C and 3CR sources (Inskip, Best,
Longair \& MacKay 2002a) suggests that at high redshifts, the
most powerful 3CR radio galaxies are not only systematically more luminous in
the $K-$band than their less powerful 6C counterparts, but that both
samples are more luminous than the predictions of passive evolution
(assuming the same average mass at all redshifts).
Fig.~\ref{Fig: 7_6} displays this deviation from passive evolution for
the complete 6CER (Rawlings, Eales \& Lacy 2001) and 3CR (Laing, Riley
\& Longair 1983) samples, plus the individual data for our $z
\sim 1$ subsamples, clearly showing the increased luminosity for the
higher redshift and higher radio power systems.  
The most obvious mechanism by which the 
host galaxy can deviate from a simple passively evolving old stellar
population is via an additional burst of star formation at some point
in the galaxy's past, plausibly associated with a merger event which
may also be responsible for triggering the production of a powerful radio source.  It
is certainly worthwhile investigating the effects of recent star
formation in more detail, paying close attention to the constraints
provided by the observational data.  Our consideration of the
influence of the host galaxy stellar populations on the observed
galaxy colours is presented in the following section.

\section{The influence of the host galaxy stellar populations}

Variations in the stellar populations of the $z \sim 1$ 6C and 3CR
host galaxies are likely to provide the best explanation for both their
observed deviation from passive evolution models and their strikingly
similar colours.  A slightly later formation redshift for one of the
samples is one option: this would lead to younger, more luminous,
bluer host galaxies. 
The shapes of the spectral energy distributions for galaxies with ages
between 2.5 and 5 Gyr (which encompasses the range of ages expected
for a galaxy at $z \sim 1$, whose stellar populations formed at a
redshift of 3-5) are very similar (e.g. Inskip, Best, Longair \& McKay
2002a).   F702W$-K$ and F814W$-K$ evolutionary tracks for galaxies
formed at between redshifts of 3 to 10 are displayed on Fig.~\ref{Fig:
7_3}; tracks for the predicted $J-K$ colours are displayed on
Fig.~\ref{Fig: 7_1}.  At $z \sim 1$, the predicted $J-K$ colours for
different star formation redshifts vary by less than $\sim 0.1$
magnitudes; changes in the observed F702W$-K$ and F814W$-K$ colours at
$z \sim 1$ are not much greater, at $\lta 0.3$mag.  Younger 6C host
galaxies could therefore provide bluer F702W$-K$ and F814W$-K$ colours
than would otherwise be expected; this bluening could balance the
blue 3CR colours caused by the stronger alignment
effect of the 3CR sources, leading to similar overall colours.  It
could similarly reduce any difference 
between the $J-K$ colours of the two samples. However, the $K-$band
data require that of the two samples, the host galaxies of the more
powerful 3CR sources should be the more luminous, which contradicts
the possibility of younger 6C host galaxies.  Secondly, if the 3CR sources are
hosted by significantly younger galaxies, we would then need to
explain how the 6C sources could match the bluer colours of the
younger 3CR sources, unless the 3CR sources are also heavily
reddened. Finally, there is no real physical justification for the age of
the host galaxy having any impact on the power of a radio source
produced at much later times.

\begin{figure*}
\vspace{3.2 in}
\begin{center}
\includegraphics{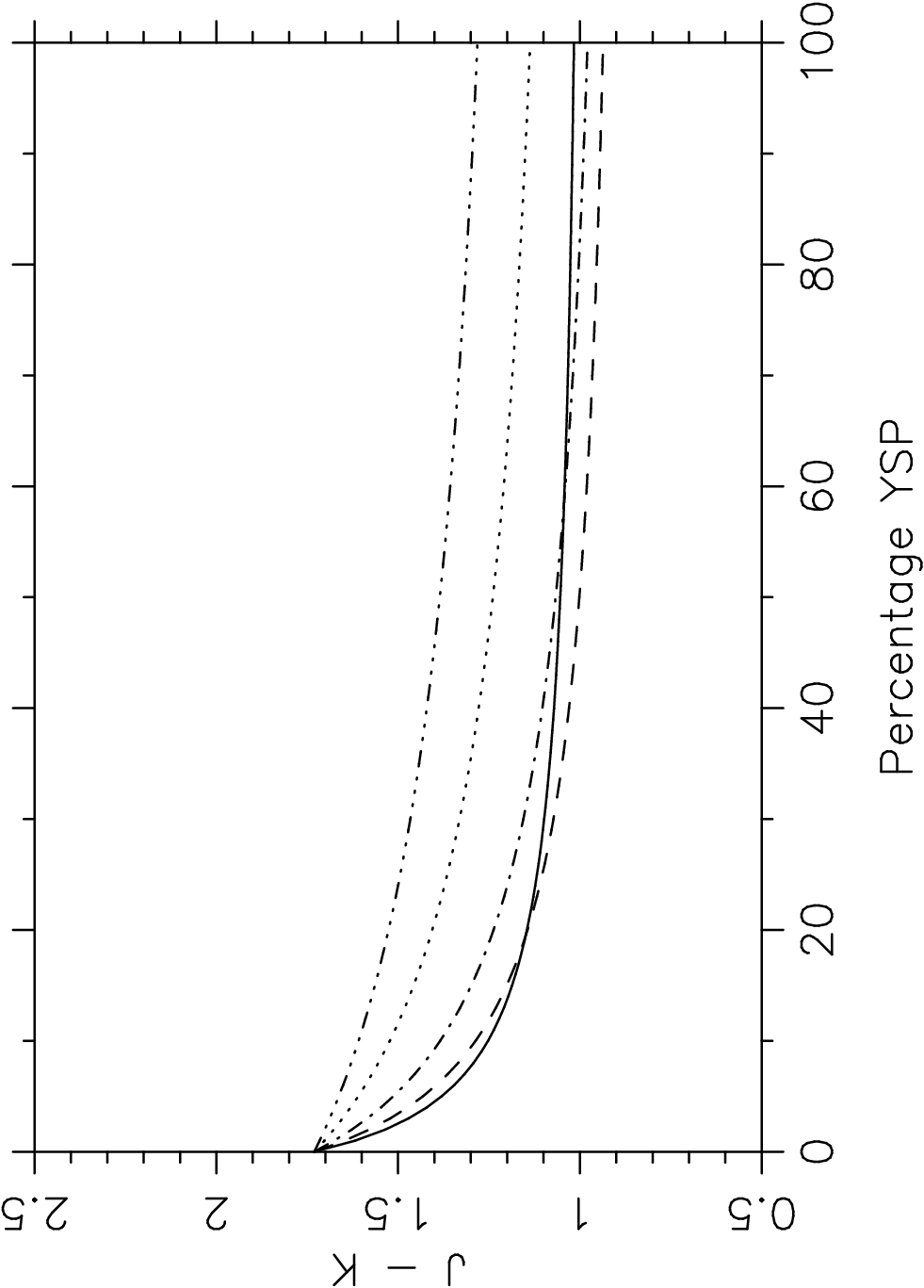}
\includegraphics{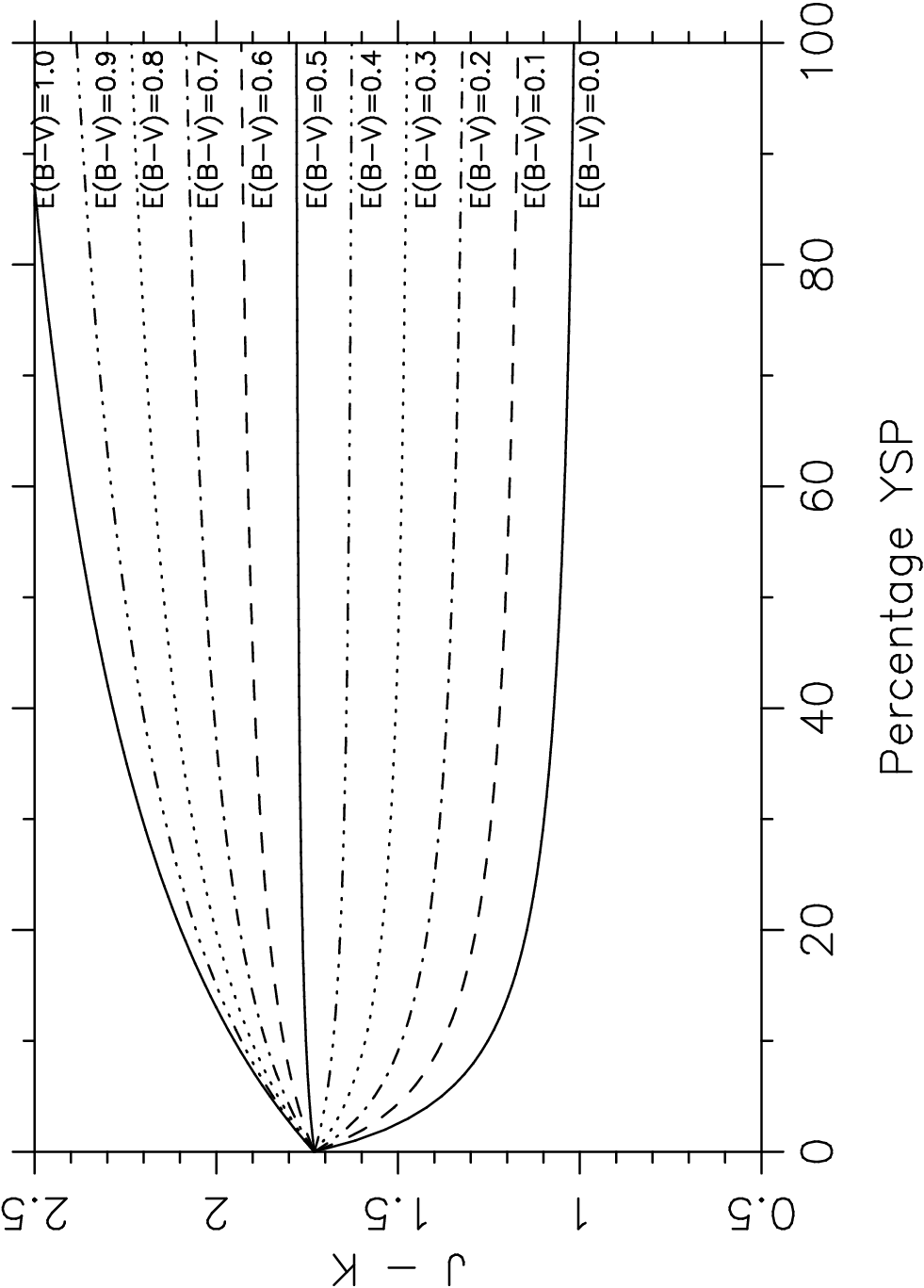}
\includegraphics{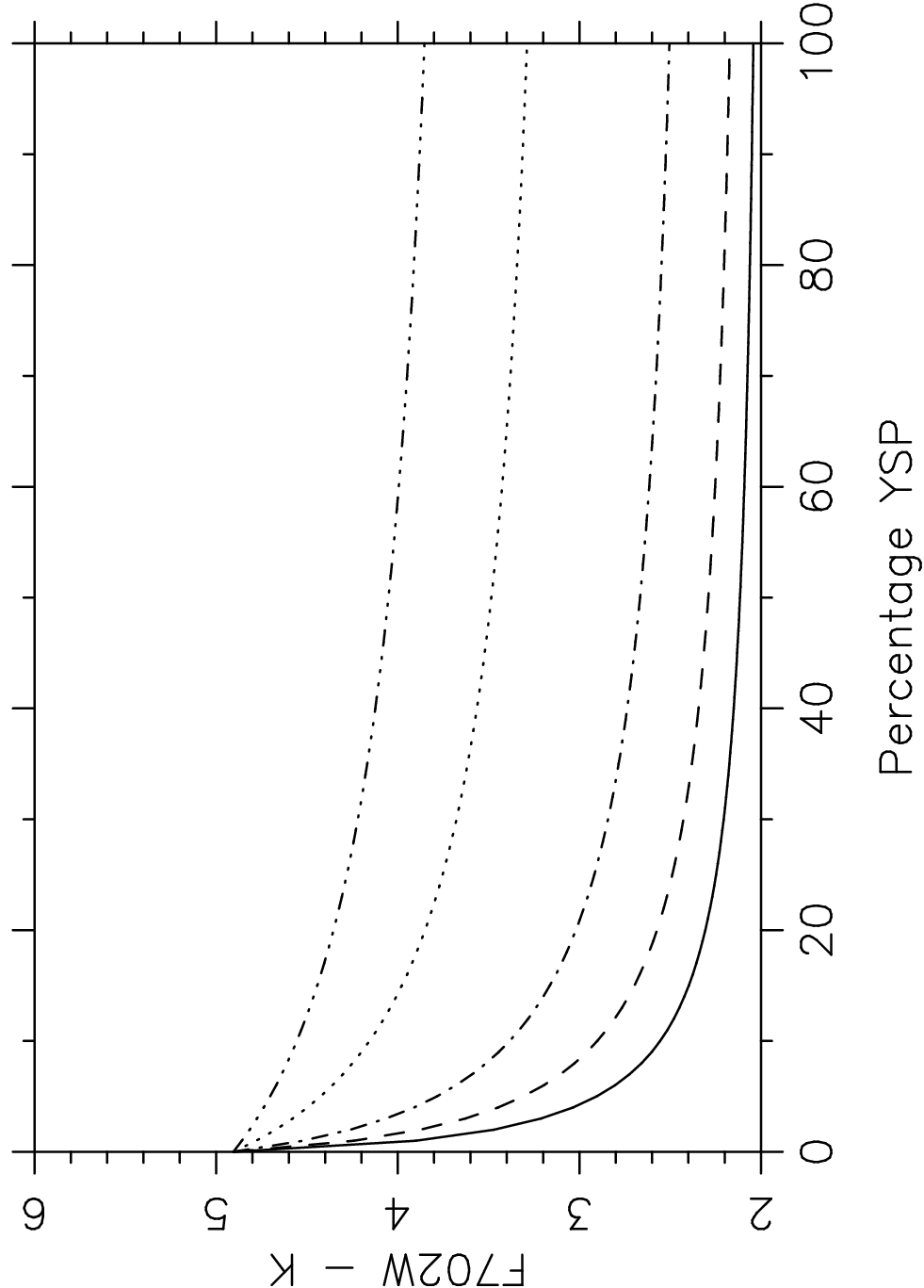}
\includegraphics{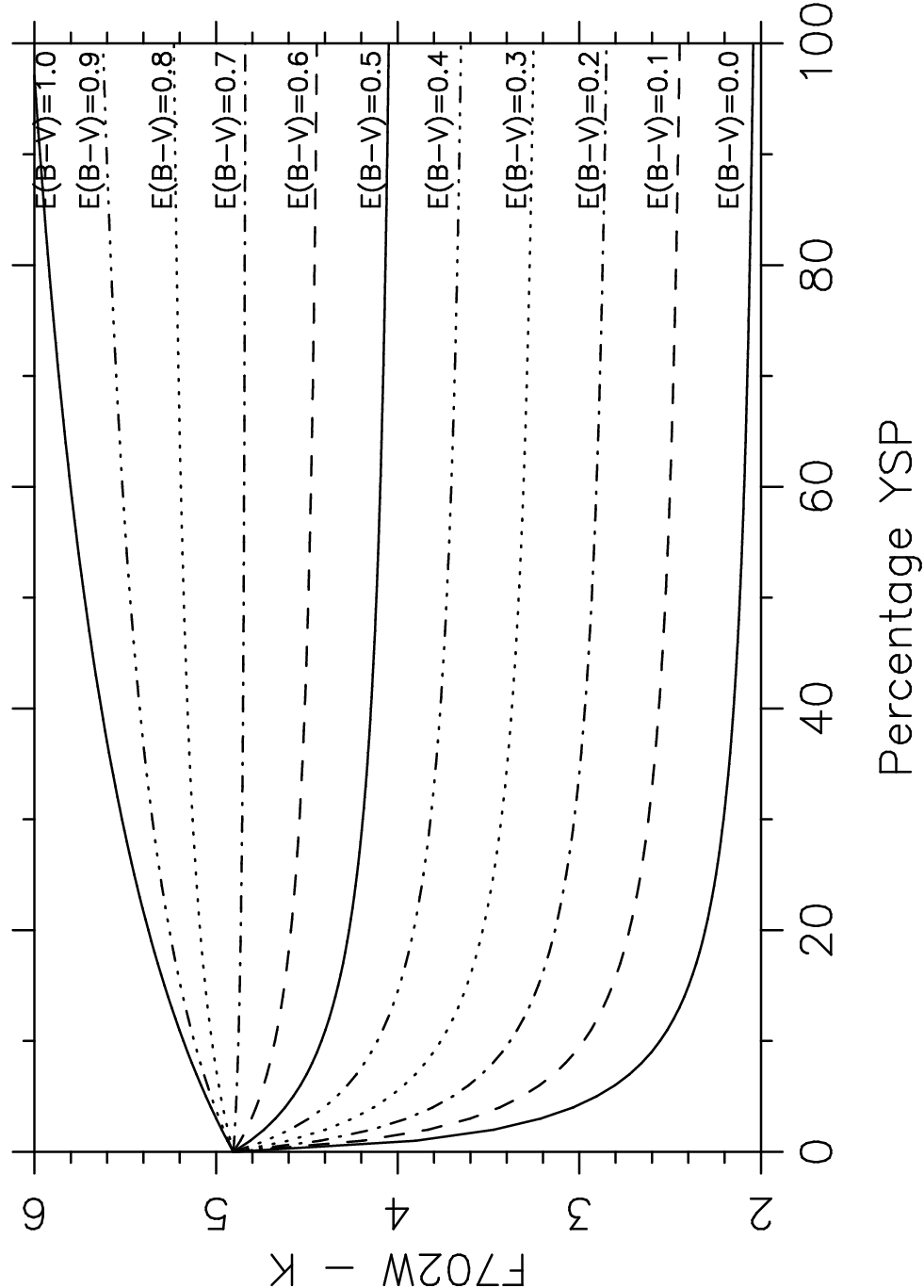}
\includegraphics{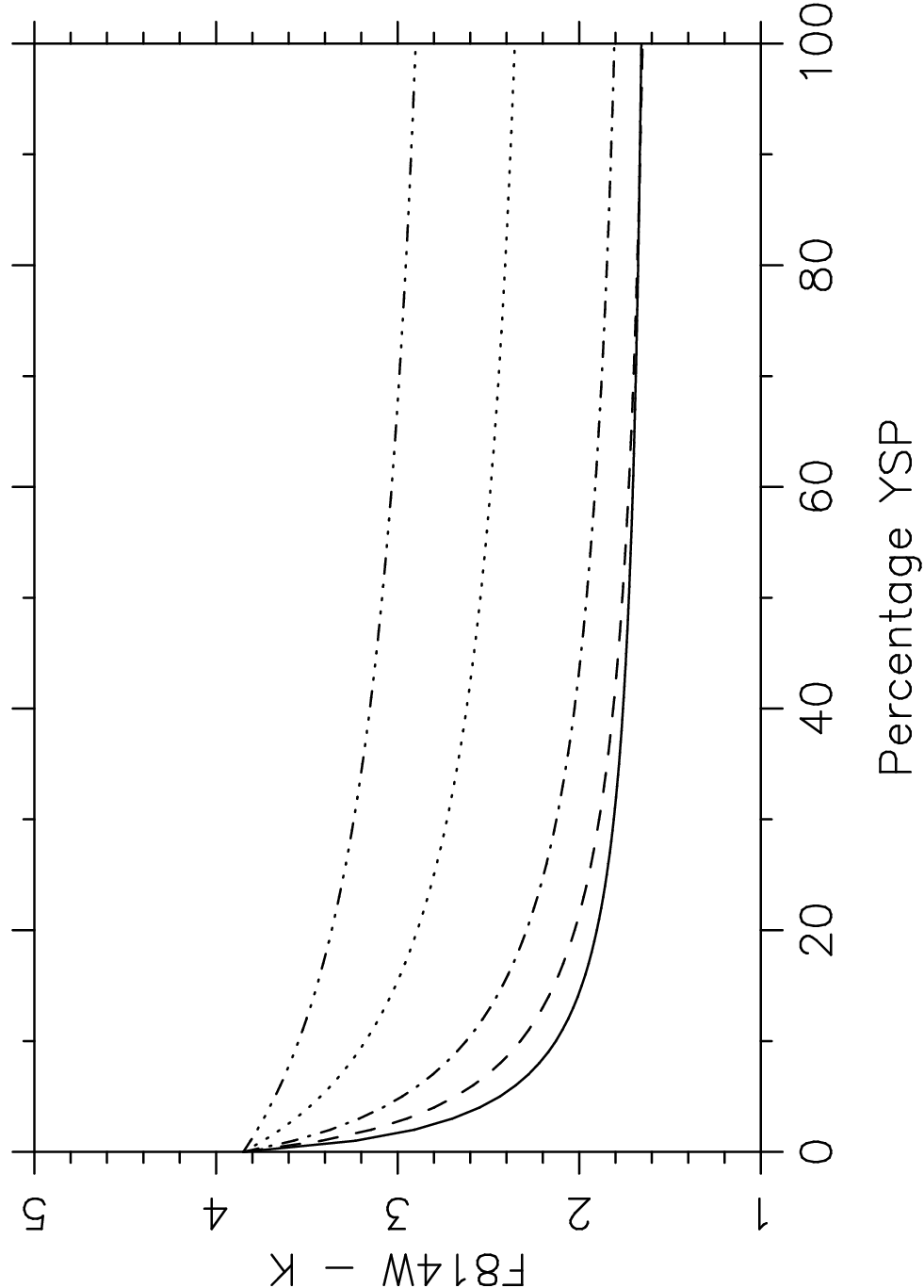}
\includegraphics{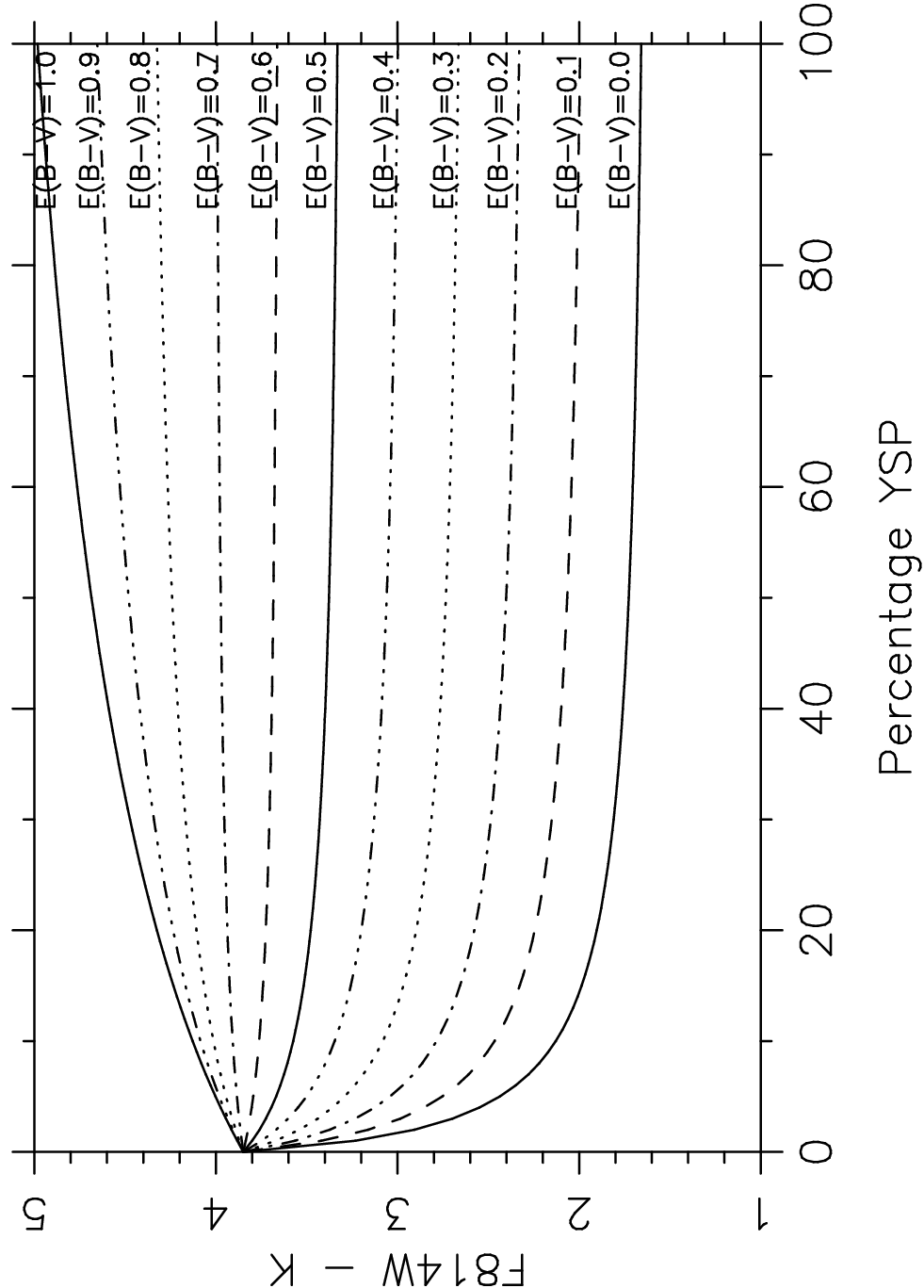}
\end{center}
\caption{Plot displaying the predicted variations in galaxy colours at
  $z \sim 1$ given the inclusion of a young stellar population.  The
  top row presents the predicted $J-K$ (left), F814W$-K$ (centre) \&
  F702W$-K$ (right) colours
  as a function of percentage young stellar population, assuming a
  galaxy composed of old stellar population formed at $z=5$, and a
  young stellar population with an age of either $5 \times 10^7$ years
  (solid track), $1 \times 10^8$ years (dashed track), $2
\times 10^8$ years (dot-dashed track), $5 \times 10^8$ years (dotted
  track) and $1 \times 10^9$ years (double dot-dashed track).  The
  lower row displays the same galaxy colours at $z \sim 1$, with the
  addition of a $5 \times 10^7$ year old young stellar population
  affected by varying levels of reddening (E(B-V) ranging from 0.0 to
  1.0). 
\label{Fig: YSPcols}}
\end{figure*}

The alternative option is a young stellar population (YSP) accounting
for only a small fraction of the total stellar mass of the host
galaxy, formed at some point in its recent history.  In addition to
star formation associated with the passage of the expanding radio
source (van Breugel et al 1985; Best, Longair \& R\"{o}ttgering 1997a),
star formation associated with a galaxy merger and/or 
interaction is also a viable possibility, particularly since the
triggering of the radio source activity may very well be linked with
such processes (e.g. Johnston, Hunstead, Cotter \& Sadler 2005).  There is increasing evidence that the host galaxies
of distant radio sources include a contribution from a relatively
young stellar population (e.g. Aretxaga et al 2001), with an age of the order of that of the
radio source. Whilst the presence of a YSP of such an age is entirely plausible,
recent research (e.g Tadhunter et al 2002; Wills et al 2002; Tadhunter
et al 2005) has identified the clear presence of older populations of
young stars, with ages of between 0.1-2 Gyr.  This scenario is in good
agreement with recent results of the Sloan Digital Sky Survey
(Kauffmann et al 2003), which suggested that the host galaxies of the
highest luminosity AGN contain similarly young stellar populations.  
Although necessarily involving a greater proportion of the host
galaxy stellar mass than a younger population of stars, such
populations could easily account for an increased observed $K-$band
luminosity without causing excessive boosting of the rest-frame UV
flux.  

\begin{figure}
\vspace{3.8 in}
\begin{center}
\includegraphics{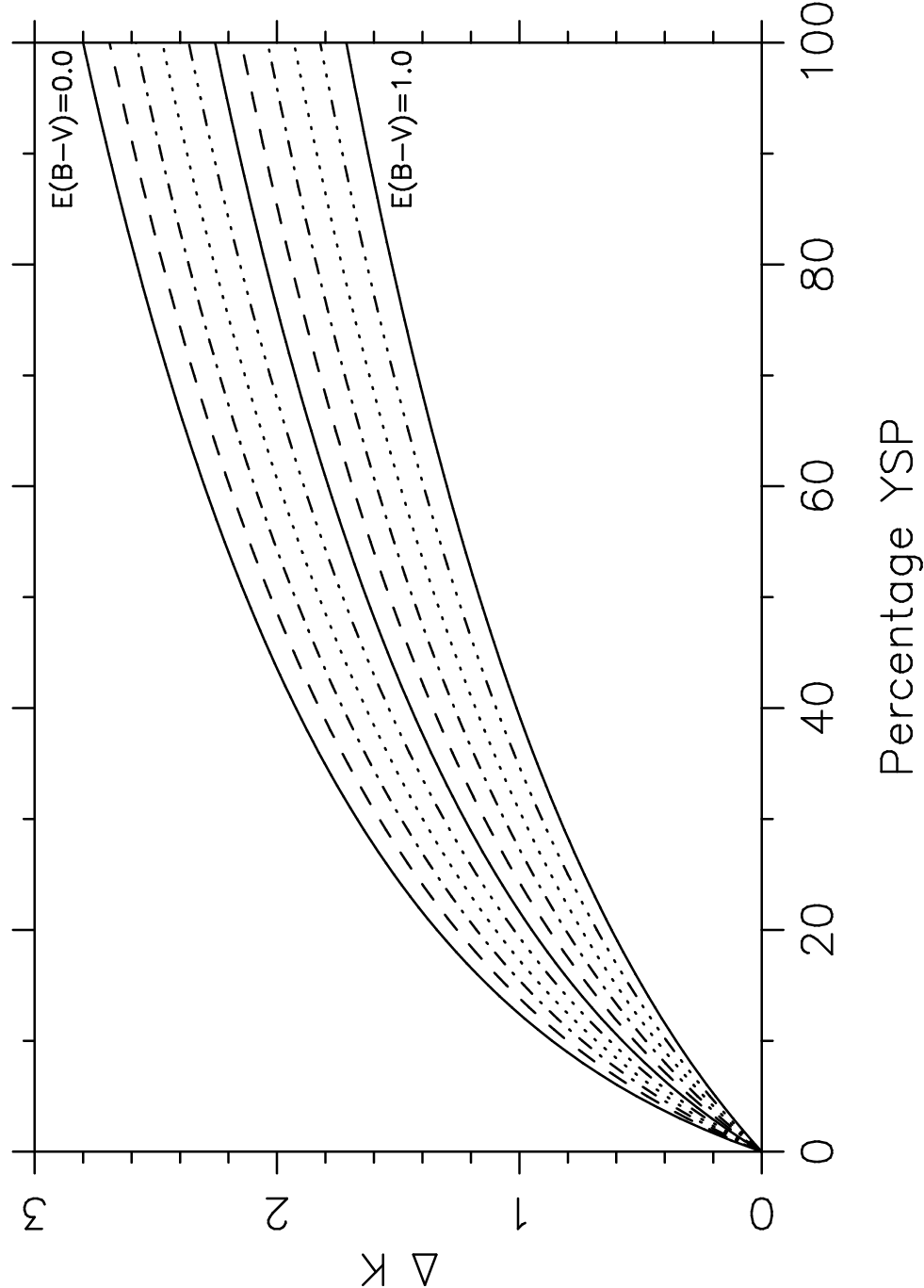}
\includegraphics{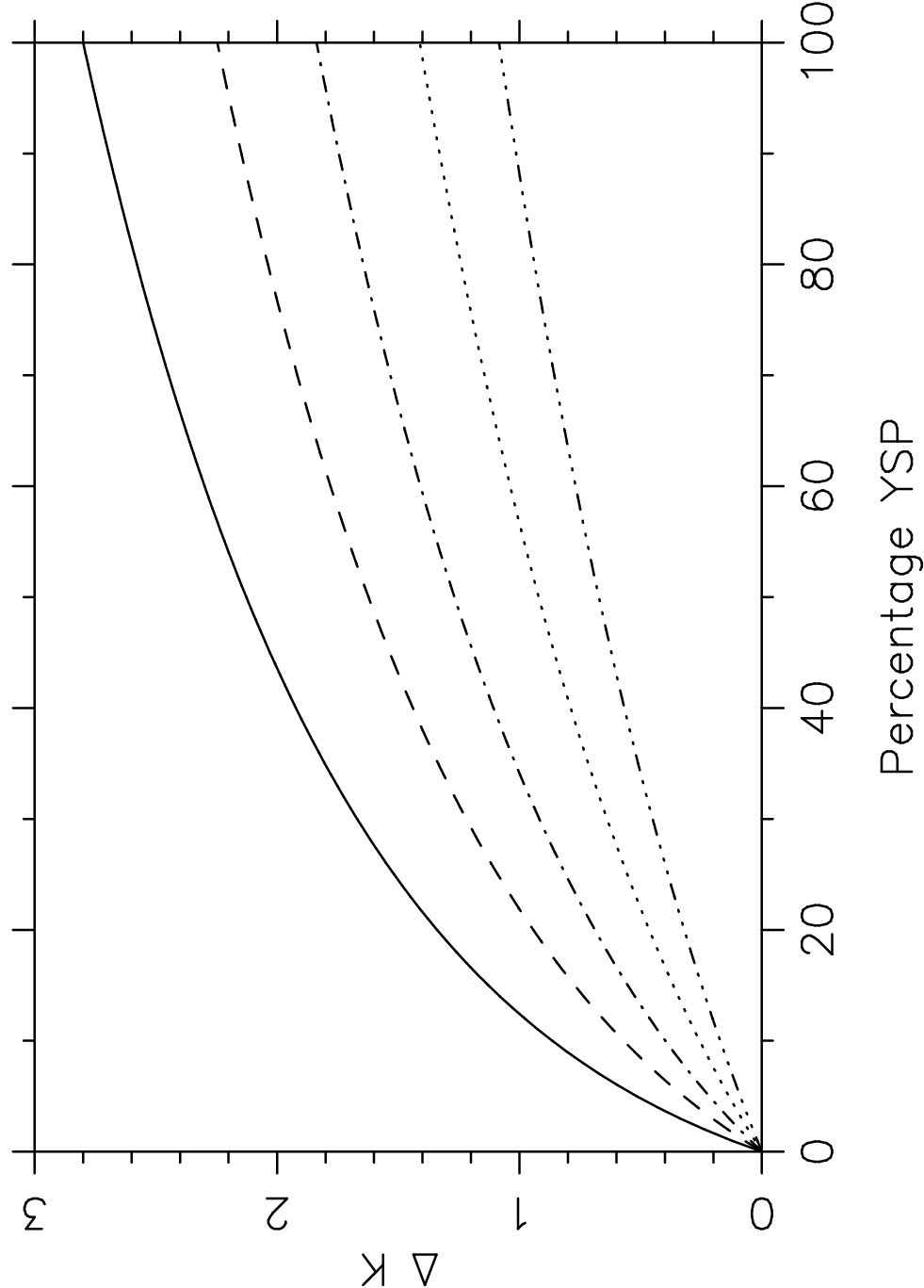}
\end{center}
\caption{Plots displaying tracks for predicted variation in K-band
magnitude produced by the addition of a young stellar population (YSP) to a
$5 \times 10^9$ year old stellar population (consistent with a $z \sim
1$ galaxy with a star formation redshift of $z \sim 5$).  (Top) Change
in $K-$band magnitude with the addition of model YSP
ages of $5 \times 10^7$ years (solid track), $1 \times 10^8$ years
(dashed track), $2 \times 10^8$ years (dot-dashed track), $5 \times
10^8$ years (dotted track) and $1 \times 10^9$ years (double
dot-dashed track). (Bottom) Change in $K-$band magnitude with the
addition of a
reddened $5 \times 10^7$  year old young stellar population, with
E(B-V) values ranging from 0.0 to 1.0.    
\label{Fig: YSP_K}}
\end{figure}

\begin{figure}
\vspace{4.3 in}
\begin{center}
\includegraphics{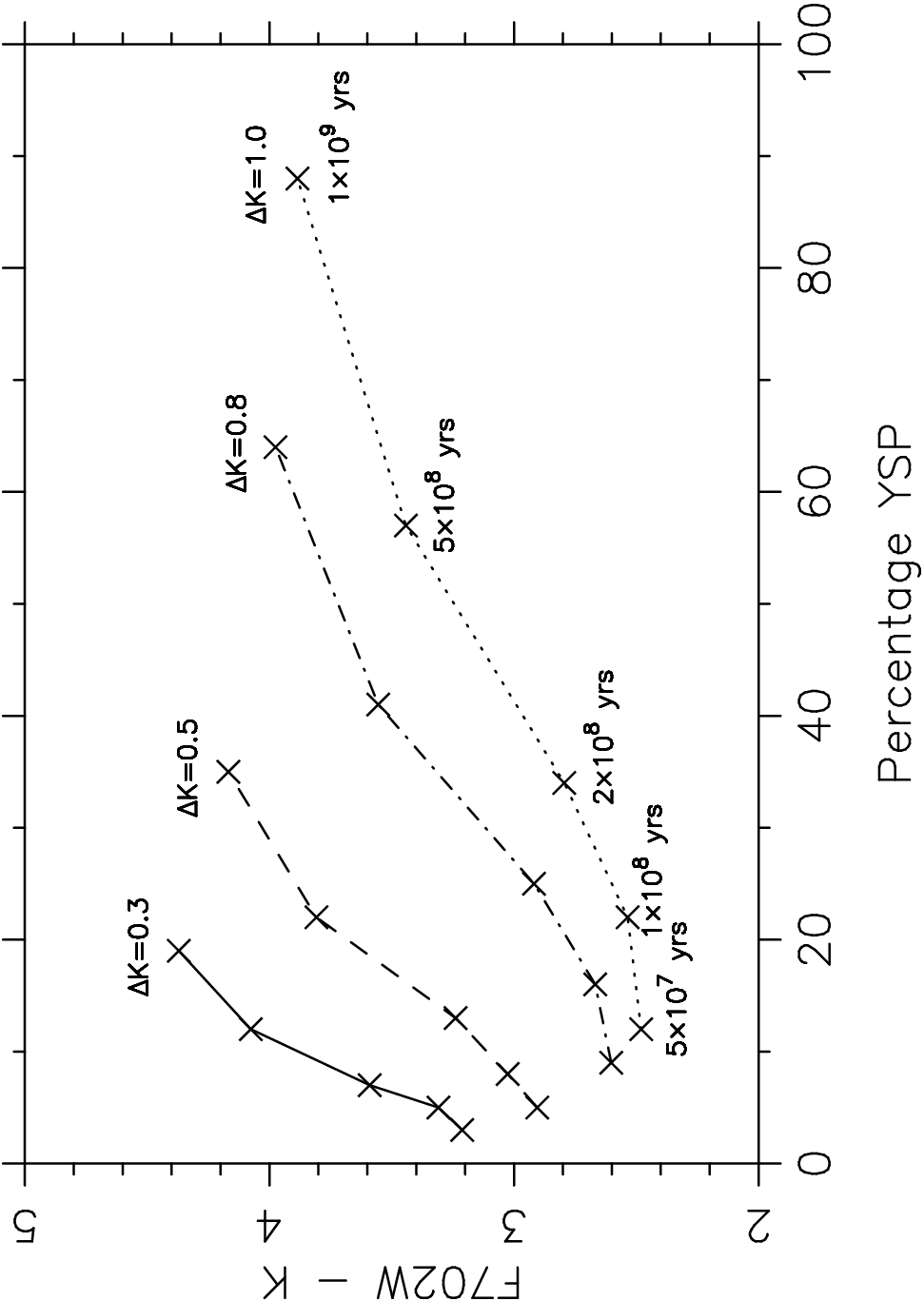}
\includegraphics{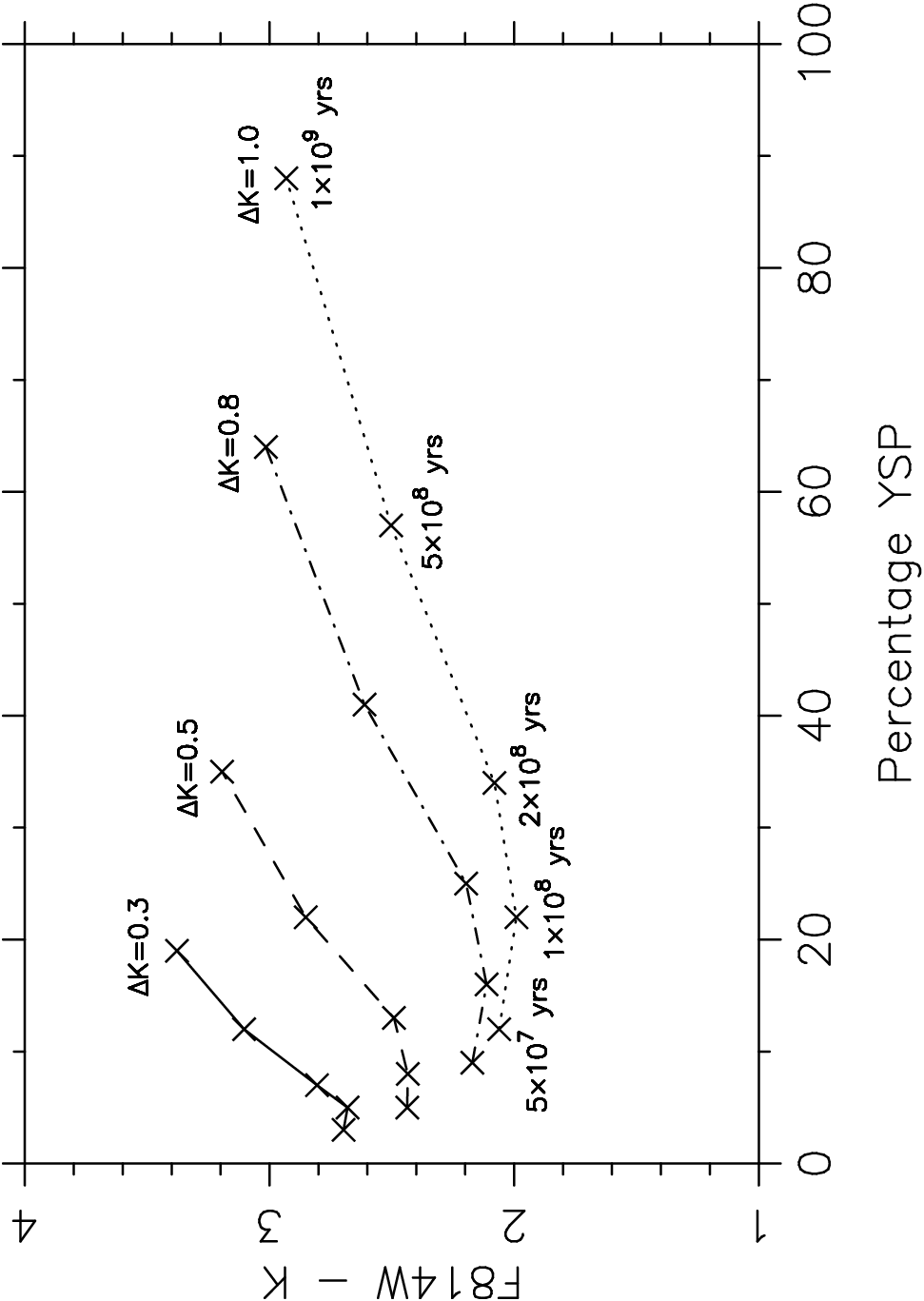}
\end{center}
\caption{Plot displaying the tracks for predicted F814W$-K$ (top) and
F702W$-K$ (bottom) colours, and
the percentage contributions of young stellar populations which would
be required in order to increase the $K$-band magnitude of the host
galaxy by 0.3, 0.5, 0.8 or 1.0 magnitudes.  The models consider young
stellar populations with ages of $5 \times 10^7$, $1 \times 10^8$, $2
\times 10^8$, $5 \times 10^8$ and $1 \times 10^9$ years, in
conjunction with a $5
\times 10^9$ year old stellar population (consistent with a $z \sim 1$
galaxy with a star formation redshift of $z \sim 5$).  
\label{Fig: YSP}}
\end{figure}

We have used the \textsc{GISSEL} spectral synthesis models of Bruzual \&
Charlot (2003) to investigate the effects of a YSP on the host
galaxy emission at different wavelengths.  The formation of the youngest YSPs
which we consider is assumed to be simultaneous with the triggering 
of the radio source itself, i.e. implying an age of $\lta 10^8$years,
whilst we also consider older ages, up to a Gyr.
Fig.~\ref{Fig: YSPcols} displays the predicted $J-K$, F814W$-K$ and
F702W$-K$ colours for a $\sim5$Gyr old galaxy at $z=1$, including a contribution
from young stellar populations (YSP) of varying ages.  Also displayed
are the predicted colours for a galaxy composed of a mixture of old
stellar population and a reddenned $5 \times 10^7$ year old YSP.  The
resulting increase in the observed $K-$band emission for these models
is displayed in Fig.~\ref{Fig: YSP_K}.  Fig.~\ref{Fig: YSP} displays
the variations in F814W$-K$ and F702W$-K$ colours induced by
the presence of these YSPs of different ages, given the requirement that the
observed $K-$band emission is increased by a specified level (of
between 0.3 to 1.0 magnitudes).   

Our observations (Fig.~\ref{Fig: 7_3}, which
have F814W$-K \gta$ 2-3 and F702W$-K \gta$ 3-4) allow us to imply
limits on the YSP mass fraction.  A
$5 \times 10^7$ year YSP must account for $< 5$\% of the total
stellar mass of the galaxy if it is to lead to an increase in the
observed $K$-band luminosity of $< 0.5$mag.  The corresponding limit for a $10^8$
year old YSP is $\lta 9$\% of the galaxy stellar mass.  Smaller bursts
of star formation triggered by the expanding radio source would
produce colours in good agreement with those observed for the aligned
emission in these systems.  These limits can be relaxed with the
inclusion of reddening; with an E(B-V) of 1.0, a $5 \times 10^7$ year old YSP accounting for
15\% of the total galaxy mass will not produce either excessively blue
colours, or an excessively large increase in the observed $K-$band flux.

However, in order for a YSP to produce any difference between 3C and
6C colours, we would need to consider the radio power dependence of
the properties of any YSP.   If we assume that the rest-frame UV
excess due to the alignment effect for the 6C sources is equal to
or weaker than that of the 3CR sources, then the 6C host galaxies must
be bluer on average than those of the 3CR sample, in order to obtain comparable
colours for the two samples.  Whilst this could be
achieved with a stronger YSP contribution for the 6C sources than the
3CR sources, the luminosity differences between the samples rule this
option out. The alternative is to include reddening of
both the old and young stellar populations of
the 3CR host galaxies; however, it is not clear that the levels of reddening should
depend in any way on the radio power of the source in question.

The observed YSP ages are in many cases significantly older than the radio source
itself, and it is therefore difficult to relate the properties of a
YSP of such an age with the power of the radio source (although links
could possibly be found by considering the exact details of any merger event
associated with the formation of a YSP, and the much later triggering
of the radio source/AGN activity; both factors may also be related to
the host galaxy environment). This implies that it is unlikely that
the presence of older YSPs can account for the observed differences
between the 3CR and 6C host galaxies at $z \sim 1$.  
However, the presence of an `old' YSP may explain why radio source
host galaxies are bluer than expected given the predictions of passive
evolution models. In order to increase the $K$-band emission by 0.5mag
(i.e. the observed difference between radio galaxies and passive evolution
models at $z \sim 1$; Inskip et al 2002a), a young stellar population
$\gta 0.5$Gyr in age and containing $\gta 20$\% of the host galaxy
mass is required; this is in excellent agreement with the findings of
Tadhunter et al (2005). 

\section{Conclusions}
The major results of our analysis of the $z \sim 1$ galaxy colours can
be summarised as follows. 
\begin{enumerate}
\item[$\bullet$] The observed infrared colours of the 6C and 3CR
  sources are indistinguishable, and well explained by passively
  evolving galaxies up to redshifts of $\sim 1.3$.
\item[$\bullet$] The galaxy colours begin to deviate from passive old stellar
  populations at higher redshifts ($z \gta 1.3$), as increasing amounts of redshifted
  aligned emission lie in the wavelength range of the infrared filters.
\item[$\bullet$] The observed F702W$-K$ and F814W$-K$ colours 
become increasingly blue out to redshifts of $z \approx 
1.1$, and then redden again at higher redshifts. This is linked to the
fact that the excess rest-frame UV emission produced by the aligned structures
becomes increasingly important between redshifts of 0.6 and 1, and
that as redshift increases beyond that, the underlying old stellar
population of elliptical galaxies can be expected to have increasingly 
red colours in these observed wavebands.   
\item[$\bullet$] The observed F702W$-K$ and F814W$-K$ colours 
for both galaxy
samples are also statistically indistinguishable, suggesting that
either the
predominant alignment effect mechanisms do not scale strongly with
radio power, or that some additional effect works to counterbalance this. 
\item[$\bullet$] Just
as the most extreme rest-frame UV morphologies are generally associated with 
the smaller radio sources in the sample, these sources were also
observed to display bluer colours due to the increased excess
rest-frame UV
emission observed on the 
HST/WFPC2 images. We see some signs that galaxy environment affects
the strength of the observed alignment effect.
\end{enumerate}
\vspace{-3pt}

The interpretation of these results is not totally straightforward.
Whilst the overall redshift evolution of the galaxy colours can be
readily understood, the (lack of) variation in the observed optical-IR colours between the
two $z \sim 1$ subsamples is less easily accounted for.   The 3CR
radio sources are $\sim 6$ times more powerful than those of the 6C
subsample at $z \sim 1$, but the difference in mean $K$-band flux
between the two data sets is much lower, at a factor of $\lta 2$.   
Given the fact that the emission produced by many alignment effect
mechanisms should scale with radio power one might expect the more
powerful sources to display bluer colours. Several mechanisms for
producing the excess rest-frame UV aligned emission will 
not necessarily produce more emission in the presence of a more
luminous radio source.  Star formation triggered by the expanding
radio source through the surrounding IGM could potentially be more 
efficient for the 6C sources.  The passage of a less powerful radio
source may lead to lower levels of cloud shredding, and thus allow
greater amounts of star formation.  Additionally, the AGN of the more 
powerful 3CR galaxies are likely to heat and ionize the gas clouds to
a greater extent, impeding the star formation rate within such clouds.
However, observations of the mJy radio sources LBDS 53W091 \& LBDS
53W069 at $z \sim 1.5$ (Dunlop 1999; Spinrad et al 1997) would seem to
contradict this hypothesis.  These very low power radio galaxies have
very red stellar populations consistent with ages of over 3.5Gyr.  The
observations of both galaxies are consistent with a de Vaucouleurs law
luminosity profile, and neither display any aligned emission. But,
this may be a reflection of the radio properties of these sources (largest
angular sizes are 4.2 arcsec and $<5.1$ arcsec respectively).  The
disturbance to the IGM caused by such small, low power radio galaxies
may not be sufficient to trigger any extra star formation at all.
 
It should be noted that evidence for star formation is not restricted
to sources displaying a strong alignment effect, and that a large
variety of stellar populations are observed in different systems.  UV HST observations
of very low redshift 3CR sources (Allen et al 2002) show that a high
proportion of sources display some level of recent or ongoing star
formation, without any need for a large-scale alignment effect.  Very
young (several Myr old) stellar populations accounting for $\lta 1\%$ of
the total stellar mass are observed in the host galaxies of the intermediate redshift
compact radio sources PKS1345+12 (Rodriguez et al, {\it priv. comm.}) and
9C1503+4528 (Inskip et al, {\it in prep}), whilst a wider study of 2Jy
and 3CR radio galaxies (Tadhunter et al 2005, Wills et al 2002, Holt
{\it priv. comm.})
revealed evidence for older (0.05 to 2 Gyr), more massive (up to
50\% of the total stellar mass) stellar populations.  Such varied
stellar populations are likely to account for much of the scatter
within different samples.  However, this does not prevent us from drawing
conclusions regarding the more general effects of properties such as
radio power on the samples as a whole.

One important consideration is whether star formation induced by an expanding radio
source may instead depend more strongly on the mass of available gas,
rather than any other parameters.  Scattering processes are also
likely to depend on the available mass of gas and dust as well as on
the power of the rest-frame UV emission from the AGN.  If true, the fact that the
6C and 3CR host galaxies at $z \sim 1$ are of comparable size would be
a point in favour of similar masses and hence observed colours.  However, the total mass of 
cool clouds in the regions of the IGM surrounding the host galaxy may
or may not scale with galaxy mass.  If, for example, the radio source
is triggered by a galaxy merger, the scattering processes may depend
most strongly on the amount of dust and gas brought in by that
merger. A further consideration is whether the 3CR sources lie in
particularly rich environments, which may boost both their observed
radio luminosity and also the availability of gas in the surrounding IGM.

Two main mechanisms are known/expected to vary with radio source
size (and plausibly age).  Emission line flux scales strongly with radio source size
as well as radio power, due to the increased importance of shock
excitation in smaller, younger radio sources.  Blue galaxy colours due to jet
induced star formation should also weaken at larger radio sizes, due
to the rapid aging of the recently formed young stellar population.
Although radio power does not appear to strongly influence the
observed galaxy colours, it is noticeable that the trends observed with radio
source size in the case of the more powerful 3CR data are considerably
weaker (or absent) at the lower powers of the $z \sim 1$ 6C sample,
particularly after the removal of emission line contamination.
 The lack of any clear cut trend in these data does
not rule out jet induced star formation as an important mechanism for
producing the alignment effect, but rather indicates that
the high level of scatter, due to other processes and the influence of
the local environment/IGM, has swamped any underlying evolution of the
alignment effect with source age/size.
Given the comparable colours between
the samples and the distinct lack 
of discrete blue components at any great distance from the 6C host 
galaxies, it is certainly clear that the bulk of the excess rest-frame
UV emission lies closer to the host galaxies (or indeed within them) in
the case of the 6C sources.   

%
%
%


Finally, it seems likely that the most important factor in
explaining these data is not variations in the origin or nature of the
rest-frame UV excess in each sample, but the longer wavelength emission of our
$K$-band data.  The $K-$band emission from the more powerful 3CR
sources is increased over the levels expected from passive evolution
scenarios, suggesting that the presence of a young  stellar population
(or at least one more youthful than the majority old stellar
population) may be skewing the galaxy colours.    The presence of either
a reddened young stellar population of similar age to the radio source
itself (i.e $\gta 10^7$years), or an older population of up to a
Gyr or so in age, can account for the excess $K-$band emission from the more
powerful sources without leading to substantially bluer emission from the host
galaxy.  However, given that the rest-frame UV excess must be balanced by any
long-wavelength emission from an additional stellar population (in order
that the lack of any colour difference 
between the samples is thereby maintained), it seems certain that
reddening effects within the host 
galaxy are also an important factor.

\section*{Acknowledgements}

KJI acknowledges the support of a
Lloyds Tercentenary Foundation Research Fellowship and a PPARC
postdoctoral research fellowship.  PNB is grateful
for the generous support offered by a Royal Society Research 
Fellowship.  The United Kingdom Infrared Telescope is operated by the
Joint Astronomy Centre on behalf of the U.K. Particle Physics and
Astronomy Research Council. Some of the data reported here were
obtained as part of the UKIRT Service Programme.
Parts of this research are based on observations made with the
NASA/ESA Hubble Space Telescope, obtained at the Space Telescope
Science Institute, which is operated by the Association of
Universities for Research in Astronomy, Inc., under NASA contract NAS
5-26555. These observations are associated with proposals \#6684 and
\#8173.

\label{lastpage}

\end{document}